\documentclass[aps,pra,reprint,nofootinbib]{revtex4-2}
\usepackage{endnotes}

\usepackage{graphicx}
\usepackage{dcolumn}
\usepackage{lipsum}
\usepackage{bm}
\usepackage{amsmath, amsthm, amssymb, amsfonts, mathtools}
\numberwithin{equation}{section}
\usepackage{etoolbox} 

\makeatletter

\@ifclassloaded{revtex4-2}{
  \appto\appendix{%
  }
}{}
\makeatother
\usepackage{dsfont, float}
\usepackage{enumitem}
\usepackage{hyperref}
\usepackage{physics}
\usepackage{xcolor}
\definecolor{DarkBlue}{rgb}{0.0, 0.0, 0.5}
\usepackage{tcolorbox}

\usepackage{pifont}
\usepackage{lettrine}
\usepackage{tikz}
\usepackage{pagecolor}
\usepackage{lipsum}
\usepackage{enumitem}
\usepackage{tikz}
\usetikzlibrary{fadings}
\usepackage[bottom]{footmisc}
\usepackage{graphicx}
\usepackage{setspace}
\usepackage{subfigure}
\usepackage{makecell}
\usepackage{nicematrix}
\usepackage{amssymb}  
\usepackage{colortbl}
\usepackage{array}
\usepackage{booktabs}
\usepackage{cancel}
\usepackage{eso-pic}
\usepackage[absolute,overlay]{textpos}
\usepackage{makecell}
\usepackage[bb=boondox]{mathalfa}
\usepackage{float}
\usepackage[utf8]{inputenc}
\usepackage[english]{babel}
\usepackage{empheq}
\usepackage{framed}
\newlist{todolist}{itemize}{2}
\setlist[todolist]{label=$\square$}
\usepackage{pifont}
\usepackage{tcolorbox} 
\definecolor{lightgreen}{HTML}{90EE90}

\newcommand{\xmark}{\ding{55}}%

\usepackage{amsmath}

\DeclarePairedDelimiter\floor{\lfloor}{\rfloor}

\tcolorboxenvironment{principle}{colback=LightGreen}
\usepackage{amssymb}
\usepackage{blindtext}

\begin{document}
\definecolor{darkgreen}{RGB}{0,150,0}  

\preprint{APS/123-QED}

\title{Assessing the dynamical assumptions in Tsirelson inequality\\tests of non-classicality in harmonic oscillators}

\author{Arush Garg}
 \email{ag2515@cam.ac.uk}
\author{Jonathan J. Halliwell}%
 \email{j.halliwell@imperial.ac.uk}
\author{Taejas Venkataraman}%
 \email{tv296@cam.ac.uk}
\affiliation{Blackett Laboratory, Imperial College, London SW7 2BZ, United Kingdom}

\date{\today}

\begin{abstract} ``Macrorealism" posits that a system possesses definite properties at all times and that we can discover these properties, in principle, without disturbing the system's subsequent behaviour. The Leggett–Garg inequalities are derived under these assumptions and are readily violated by standard quantum mechanics, thereby providing a scheme to test whether demonstrably macroscopic systems can exhibit quantum coherence. Unfortunately, Leggett–Garg tests suffer from the difficult to avoid clumsiness loophole—the difficulty of proving that sequential measurements have not inadvertently disturbed the system. The recently uncovered Tsirelson inequality is derived from the simple dynamical assumption of uniform precession, obeyed by many classical systems, and requires only single-time measurements. However, Tsirelson inequality violations could be explained by a macrorealistic system that merely breaks the dynamical assumption, rather than genuine quantum behaviour. By carrying out a quantum-mechanical analysis of the Tsirelson inequality in the harmonic oscillator, we develop a protocol to rule out this possibility by assessing generalised conditions of uniform precession. We show that various measures of uniform precession, some of which are related to Leggett–Garg quantities, are satisfied well enough that the presence of quantum-mechanical interference terms must be implied. We derive several incidental mathematical results relating to violating states of Tsirelson's inequality, concerning dwell time, crossing number and probability currents, and also consider a group theoretic analysis of the Tsirelson operator.
\end{abstract}

\maketitle


\section{Introduction} \label{Introduction}

\subsection{The Tsirelson inequality}

In 2006, Boris Tsirelson wrote a paper with a title that asks an endearingly straightforward question: ``How often is the coordinate of a harmonic oscillator positive?" \cite{tsirelson_how_2006}. Tsirelson demonstrated that a simple dynamical assumption about classical oscillators can give rise to a surprising disparity between classical and quantum expectations. Violations of the ``Tsirelson inequality" \cite{plavala_tsirelson_2024} are associated inherently with the negativity of the Wigner function \cite{Wigner_function} and present an alluring scheme to assess the ``quantumness" of a system using only single-time measurements on an ensemble of systems.

This work went virtually unnoticed for more than a decade and a half until it was rediscovered by Zaw \textit{et al.}, who in their 2022 paper extended Tsirelson's work to a broader family of systems undergoing uniform precession, such as finite-dimensional spins in real space \cite{precessions}. Tsirelson's protocol has been generalised as a viable entanglement witness for coupled harmonic oscillator systems \cite{PhysRevLett.130.160201,PhysRevA.108.022211} and entangled spin ensembles \cite{PhysRevA.109.042402}. The most recent work in the field has focussed primarily upon further generalisations of the protocol \cite{PhysRevA.110.062408} and computing stricter bounds on maximal quantum violations \cite{zaw2024threeanglevariantstsirelsonsprecession,PhysRevLett.134.190201}. A recent triumph has been the first experimental realisation (a so-called quantumness certification) of Tsirelson violations in precessing nuclear spins \cite{vaartjes_certifying_2025}.

\begin{figure}
    \centering
    \includegraphics[width=0.7\linewidth]{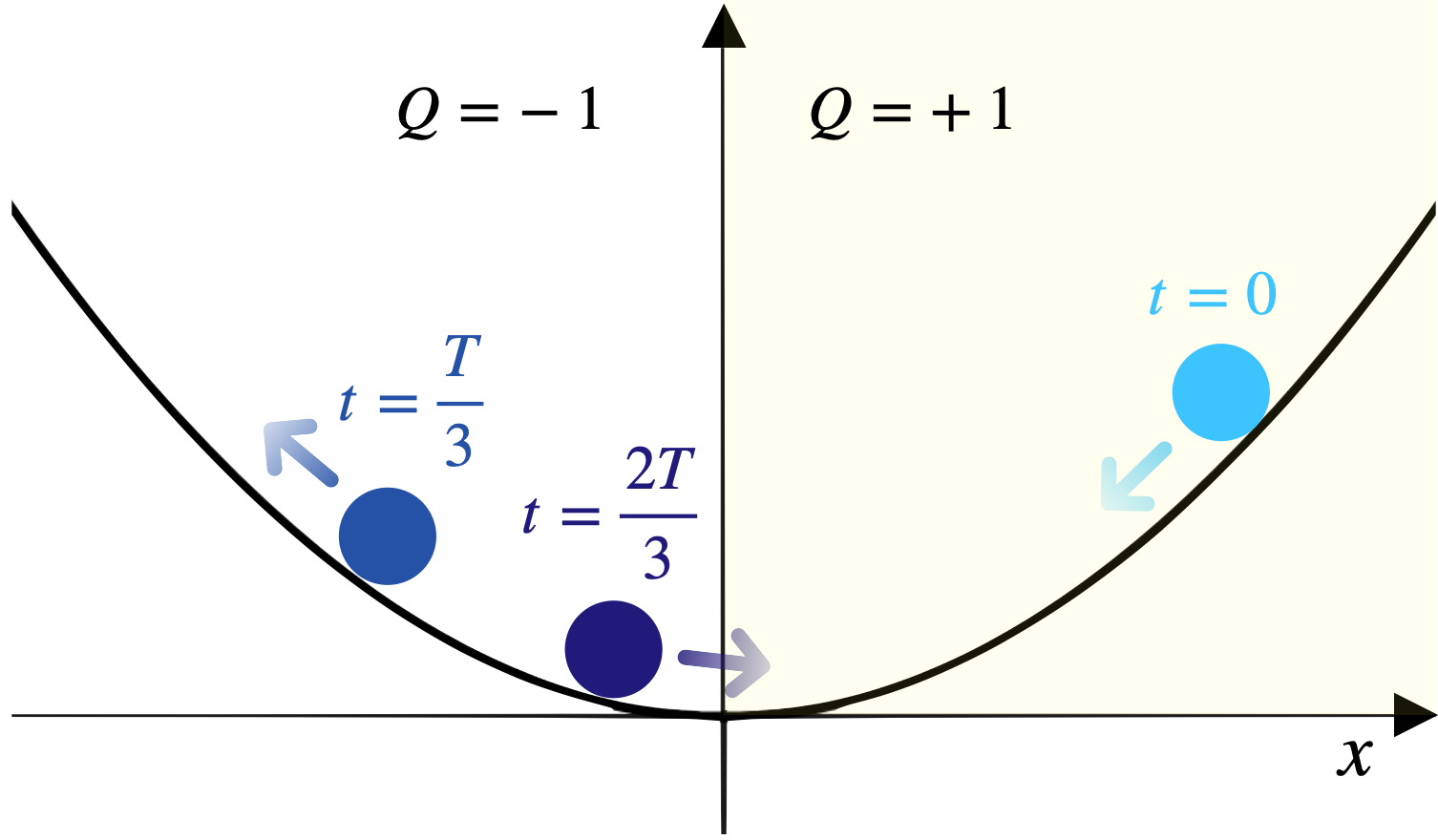}
    \caption{A classical particle oscillating in a quadratic potential is one example of a system undergoing uniform precession. The dynamical assumption is that $Q$ cannot have the same sign at all three times, irrespective of where in its trajectory the particle starts at time $t=0$.}
    \label{fig:Dynamical assumption illustrated}
\end{figure}

To derive Tsirelson's inequality, we consider a classical harmonic oscillator with time period $T$. Define the dichotomic variable $Q(t)$ to be the sign function $\text{sign}(x(t))$, which is $\pm 1$ depending on the sign of $x$, and consider its value at three equally-spaced times, every one-third of a period ($t_1=0$, $t_2=\frac{1}{3}T$, and $t_3=\frac{2}{3}T$). A simple dynamical assumption about oscillators is this: \textit{the particle will never be found on the same side of the origin three times in a row.} Therefore, the sum $Q(0)+Q(\frac{1}{3}T)+Q(\frac{2}{3}T)$ can only assume the values $+1$ or $-1$ for any given classical trajectory. This is illustrated in Fig. \ref{fig:Dynamical assumption illustrated}.

\begin{figure}
    \centering
    \includegraphics[width=0.5\linewidth]{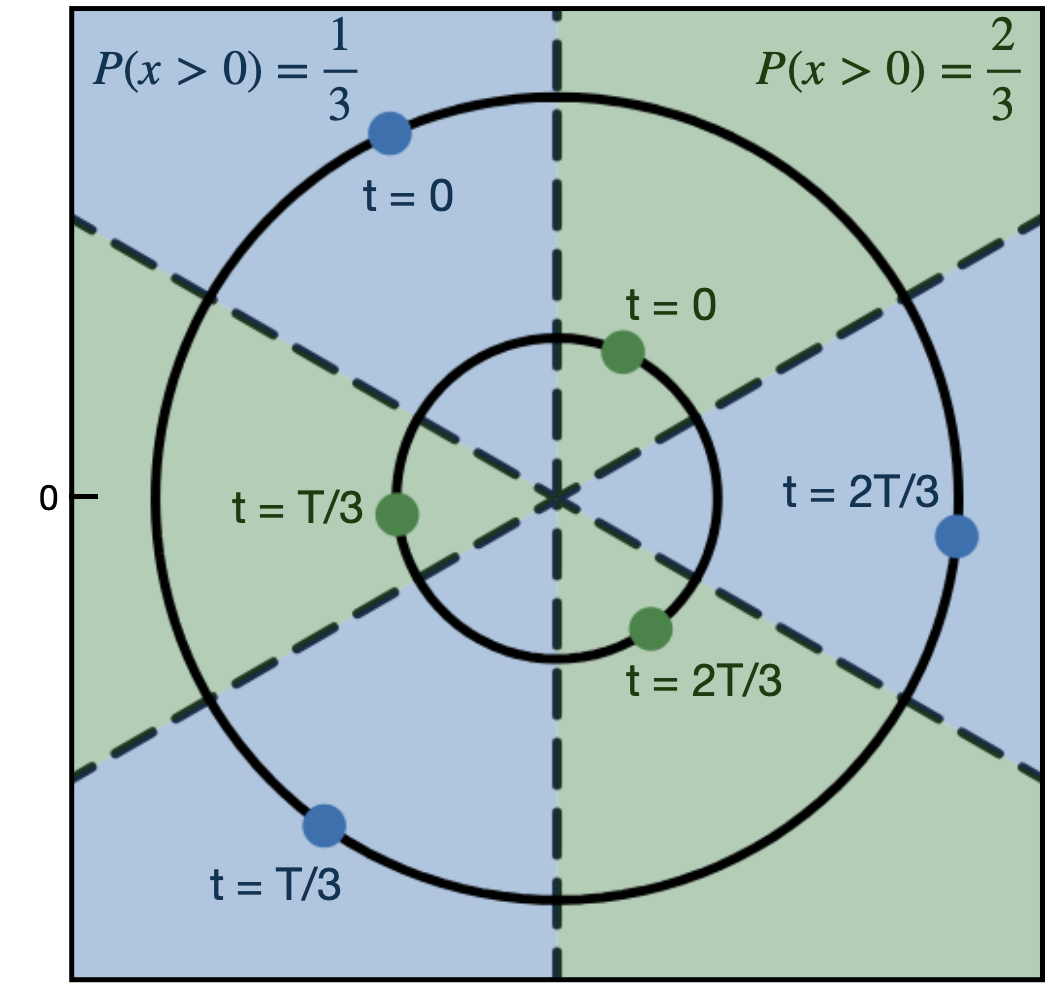}
    \caption{Classical phase-space depiction of the system's dynamics for three possible measurement times. If the system is initialised in the green region for one of the experimental runs, then it will have a probability of $2/3$ of being found on the right side of the origin at a randomly-chosen measurement time $t\in\{0,T/3,2T/3\}$. Conversely, the probability will be $1/3$ if the particle begins in a blue region. As such, the ensemble average, a convex sum of these probabilities, is bounded between $1/3$ and $2/3$. This is equivalent to Eq. \ref{The Tsirelson inequality!} up to a scale factor. (Adapted from Zaw \textit{et al.} \cite{PhysRevA.106.032222}.)}
    \label{fig:Phase plot}
\end{figure}

Defining $A=Q(0)+Q(\frac{1}{3}T)+Q(\frac{2}{3}T)$, it follows that the ensemble average $\langle A \rangle$ over several experimental runs must satisfy the Tsirelson inequality:
\begin{equation} \label{The Tsirelson inequality!}
    -1 \leq \langle A\rangle \leq 1 \ .
\end{equation}
(Tsirelson originally considered the harmonic oscillator's classical phase-space dynamics, illustrated in Fig. \ref{fig:Phase plot}, and the probability that a particle is found on one side of the origin.) As the expectation value of a sum equals the sum of expectation values, $\langle A \rangle$ can instead be written in terms of single-time averages:
\begin{equation} \label{Sum of Q's}
    \langle A \rangle=\langle Q(0) \rangle+\langle Q(\frac{1}{3}T) \rangle+\langle Q(\frac{2}{3}T) \rangle
\end{equation}
Thus, the Tsirelson quantity $\langle A \rangle$ can be measured using only single-time measurements on an ensemble of identical systems. Note, in practice, measuring the dichotomic variable $Q$ may be done using a light beam that illuminates one half of the harmonic well \cite{PhysRevLett.120.210402}.

The Tsirelson quantity $\langle A \rangle$ may be conveniently written in the Wigner representation \cite{Wigner_function,zachos2005quantum,Hillery1984,Tatarskii1983,Case2008,zaw2024threeanglevariantstsirelsonsprecession} in terms of the function $W(p,x,t)$ (and see also Appendix \ref{Wigner function appendix}). It has the form
\begin{equation}
    \langle A \rangle = \iint \text{sign}(x) \sum_i W(p,x,t_i) \ \text{d}x \ \text{d}p \ .
\end{equation}
In the Heisenberg picture, with a time-independent Wigner function, the Tsirelson quantity is
\begin{equation}
    \langle A \rangle = \iint \left[Q(0)+Q(\frac{T}{3})+Q(\frac{2T}{3})\right] W(p,x) \ \text{d}x \ \text{d}p \ ,
\end{equation}
which makes it clear that Tsirelson inequality violations are indeed effected by negativity in the Wigner function.

However, a crucial point which has, so far, not been addressed in the literature is the validity of the dynamical assumption of \textit{uniform precession} in a quantum-mechanical context. If truly the Tsirelson test is to be interpreted as a credible certification of quantum behaviour, it is imperative to demonstrate that violations are indicative of genuine quantum phenomena and not simply of the failure of its founding dynamical assumption.

\begin{figure}
    \centering
    \includegraphics[width=0.7\linewidth]{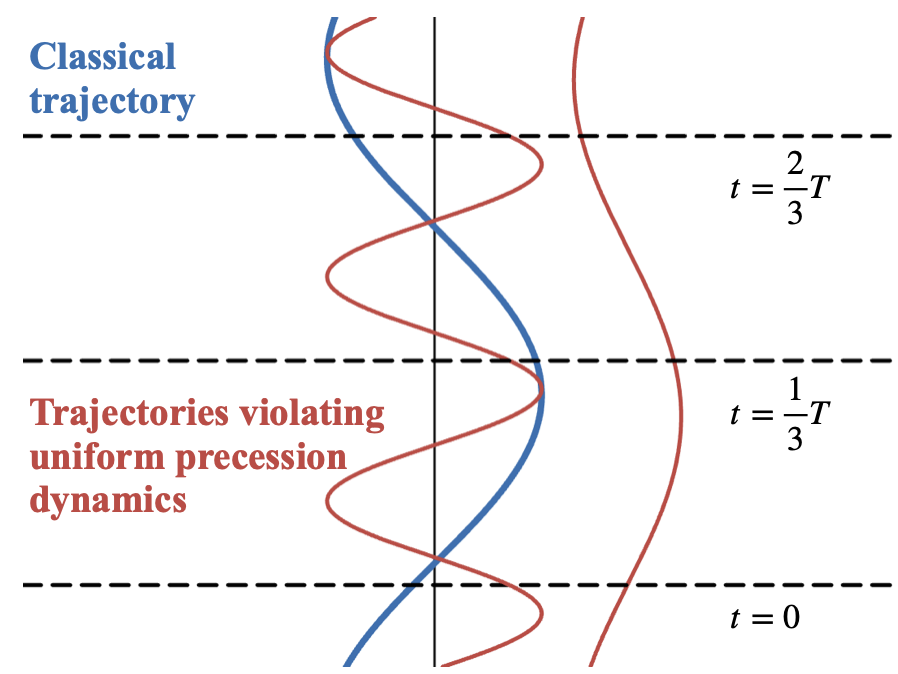}
    \caption{Unlike the blue classical trajectory, which follows the expected sinusoidal dynamics, the erratic red trajectories, still macrorealistic, either dwell for too long on one side of the origin or oscillate too rapidly, and will violate the Tsirelson inequality since they are measured on the same side three times in a row.}
    \label{fig:Erratic classical trajectory}
\end{figure}

A classical system that breaks the dynamical assumption—that is, one which is not precessing uniformly (or not oscillating at all, even)—will easily violate the Tsirelson inequality. Fig. \ref{fig:Erratic classical trajectory} depicts qualitatively that a Tsirelson violation could be caused by a classical particle that does not obey the expected Newtonian dynamics of oscillation induced by a harmonic potential, either lingering too long on one side of the origin or oscillating faster than the classical frequency.

Uniform precession can be defined perfectly well in the case of a classical oscillator by considering, for instance, the phase-space representation of its trajectory. The condition for uniform precession is then simply that the angular variable $\varphi=\tan^{-1}(x/p)$ evolves linearly in time ($\dot{\varphi}=\text{const.}$). However, quantum-mechanically, the corresponding operator, related to the arrival time operator, is conjugate to the Hamiltonian and is not self-adjoint \cite{pauli1958handbuch,Arrival_times,muga2008time,PhysRevA.99.012124}. As such, the eigenstates of the angular operator cannot be orthogonal, and there is some intrinsic indefiniteness in the angular behaviour and hence in the definition of uniform precession; this property is also related to the difficulties of defining a phase operator for the harmonic oscillator \cite{nieto_quantum_1993}. In the Wigner picture, although the Wigner function evolves along the classical path, it is not a probability distribution, since it can be negative, and thus we cannot immediately assert that the evolution in the quantum case respects uniform precession. Also, from an operator perspective, the position and momentum operators of course evolve according to classical equations of motion, so in this sense uniform precession is respected in the dynamics. However, uniform precession involves confirming that the particle is not on the same side of the origin at three different times, which means specifying three non-commuting positions, so we would not expect to see exact classical behaviour. In practice, these features mean that uniform precession can be defined in multiple inequivalent ways in quantum mechanics, as we shall see.

\subsection{This paper}

The aim of this paper is to find precise sets of conditions characterising uniform precession in terms of
measurable three-time probabilities and then to link them to the Tsirelson inequality. This provides a way to make a clear assessment of the degree to which a Tsirelson inequality violation is caused by the presence of genuine quantum effects or merely explained by a violation of uniform precession.

In order to carry out this aim, we embed the Tsirelson scheme within the more comprehensive Leggett–Garg
framework \cite{PhysRevLett.54.857}, the canonical test of non-classicality in temporal correlations; this is done in Section \ref{LG section}. In Section \ref{Tsirelson Violations in the QHO}, we establish the tools required for quantum-mechanical calculations and discuss quantum violations of the Tsirelson inequality. We present our most important results in Section \ref{Assessing UP} on the main protocol to assess uniform precession, broken down into two methods; both of these utilise the Leggett–Garg formalism. In Section \ref{Heuristic definitions}, we investigate two other quantities—the dwell time and crossing number—that are less strictly related to the Tsirelson inequality but provide further heuristic evidence of the extent to which uniform precession is satisfied. We will show that the former has quantum properties identical to the classical case and the latter approximately so. Moreover, we will argue that these results can be used to rule out alternative classical explanations. Finally, we conclude in Section \ref{Conclusion} with a critical assessment of whether the protocol introduced in Section \ref{Assessing UP} ultimately sidesteps the problem of non-invasive measurement. We report lengthy technical results and important mathematical proofs in the appendices.

\section{The Leggett–Garg framework} \label{LG section}

The Leggett–Garg inequalities (LGIs) \cite{PhysRevLett.54.857,Leggett2008}, a temporal analogue to Bell-type inequalities \cite{PhysicsPhysiqueFizika.1.195,PhysRevLett.23.880}, must be satisfied by any theory obeying ``macrorealism", a quintessentially classical world-view that is encapsulated by the following three postulates: \textit{macroscopic realism per se}—a macroscopic system exists in a definite state at all times; \textit{non-invasive measurability}—the state of the system can be determined, in principle, without affecting its subsequent dynamics; and \textit{induction}—the current state does not depend upon future measurements.

Macrorealism pertains to measurements made on a specific data set \cite{PhysRevA.103.062212}, which we take here to be the set of three single-time averages $\langle Q_i \rangle$, where $Q_i \equiv Q(t_i)$, and the three correlators $C_{ij}$, the latter measured in a non-invasive way. With possible measurement outcomes $s_i=\pm1$, where $i=1,2,3$ for measurements taken at times $t_1,t_2,t_3$, respectively, the three two-time correlators are defined as
\begin{equation}
    C_{ij} \equiv \langle Q_iQ_j \rangle = \sum_{s_i,s_j}s_is_jp_{ij}(s_i,s_j) \ ,
\end{equation}
with $-1 \leq C_{ij} \leq 1 $ and $\{i,j\}=\{1,2\},\{2,3\},\{1,3\}$, and where $p_{ij}(s_i,s_j)$ is the proportion of outcomes in the data set where a measurement of $Q$ at time $t_i$ yields $s_i$ and at (a later time) $t_j$ yields $s_j$.

It will be useful to introduce here some mathematical expressions we will require for later analysis. The measured data set fully determines the two-time quasi-probability distribution \cite{PhysRevA.87.022114,klyshko_bell_1996}
\begin{equation} \label{moment expansion 2-time}
    q(s_i,s_j)=\frac{1}{4}\left(1 + s_i\langle Q_i \rangle + s_j\langle Q_j \rangle + s_is_jC_{ij}\right) \  ,
\end{equation}
for $ s_i,s_j = \pm1 \ , \ j>i \in \{1,2,3\}$, and partially determines the three-time quasi-probability
\begin{multline} \label{moment expansion 3-time}
    q(s_1,s_2,s_3) = \frac{1}{8}(1+ s_1\langle Q_1\rangle + s_2\langle Q_2\rangle + s_3\langle Q_3\rangle \\ + s_1s_2C_{12} + s_2s_3C_{23} + s_1s_3C_{13} + s_1s_2s_3D) \ ,
\end{multline}
where $D=\langle Q_1 Q_2 Q_3 \rangle$ is the triple correlator but is not measured in the simplest protocols. These two quasi-probabilities are classically positive. Quantum-mechanically, they are equivalent to the expressions
\begin{equation}
    q(s_1,s_2) = \text{Re}\left\{\text{Tr}\left(P_{s_2}(t_2)P_{s_1}(t_1)\rho \right)\right\} \ ,
\end{equation}
\begin{equation} \label{quasi trace expression}
    q(s_1,s_2,s_3)=\text{Re}\left\{\text{Tr}\left(P_{s_3}(t_3)  P_{s_2}(t_2) P_{s_1}(t_1) \rho  \right) \right\} \ ,
\end{equation}
with projection operators $P_{s_i}(t_i)=\frac{1}{2}(1+s_i\hat{Q}_i)$ (we avoid using hats for quantum operators except when they are needed to avoid ambiguity between classical and quantum variables). We also give here the analogous sequential measurement probabilities and their moment expansions:
\begin{widetext}
\begin{equation} \label{projective 2-time moment expansion}
    p_{12}(s_1,s_2) = \text{Tr}(P_{s_2}(t_2) P_{s_1}(t_1)\rho P_{s_1}(t_1)) = \frac{1}{4}\left(1 + s_1\langle Q_1 \rangle + s_2\langle Q_2^{(1)} \rangle + s_1s_2C_{12}\right) \  ,
\end{equation}
\begin{multline} \label{projective 3-time moment expansion}
    p_{123}(s_1,s_2,s_3) = \text{Tr}(P_{s_3}(t_3)P_{s_2}(t_2)P_{s_1}(t_1)\rho P_{s_1}(t_1)P_{s_2}(t_2)) \\ 
    =  \frac{1}{8}\left( 1+ s_1\langle Q_1\rangle + s_2\langle Q_2^{(1)}\rangle + s_3\langle Q_3^{(12)}\rangle
     + s_1s_2C_{12} + s_2s_3C_{23}^{(1)} + s_1s_3C_{13}^{(2)} + s_1s_2s_3D \right) \ ,
\end{multline}
\end{widetext}
where the superscripts denote the presence of an earlier or intermediate measurement that has been summed out. More specifically, the quantity $\langle Q_2^{(1)} \rangle$ is defined as \cite{halliwell_quasi}
\begin{equation} \label{Q_2 with an earlier 1}
    \langle Q_2^{(1)} \rangle = \sum_{s_1,s_2} s_2 p_{12}(s_1,s_2) \ ,
\end{equation}
and similarly for $Q_3^{(12)}$, $C_{13}^{(2)}$, and $C_{23}^{(1)}$. This is therefore the average of the operator $ \sum_{s_1} P_{s_1}(t_1)  {\hat Q}_2 P_{s_1} (t_1)$. Note that there are two more candidate three-time quasi-probabilities that we will explore, obtained from Eq. \ref{moment expansion 3-time} by replacing $C_{23}$ with $C_{23}^{(1)}$, or replacing $C_{13}$ with $C_{23}^{(2)}$. See Appendix \ref{Method 2 appendix}.

Turning now to the conditions for macrorealism, the three postulates of macrorealism imply that there exists an underlying joint probability for the given data set, which implies that a set of twelve two-time and four three-time LG inequalities hold. Specifically, the twelve quasi-probabilities Eq. \ref{moment expansion 2-time}, for time pairs $(t_1,t_2)$, $(t_1,t_3)$, $(t_2,t_3)$, are non-negative (LG2s) and the following four LG3 inequalities are satisfied: 
\begin{equation} \label{L1 A}
    L_1 =\frac{1}{4}(1 + C_{12} + C_{23} + C_{13}) \geq 0  
\end{equation}
\begin{equation}
    L_2 = \frac{1}{4}(1- C_{12} - C_{23} + C_{13}) \geq 0
\end{equation}
\begin{equation}
    L_3 = \frac{1}{4}(1 - C_{12} + C_{23} - C_{13}) \geq 0 
\end{equation}
\begin{equation} \label{L4 A}
    L_4 = \frac{1}{4}(1+C_{12} - C_{23} - C_{13}) \geq 0 \ .
\end{equation}
These sixteen conditions are not only necessary for macrorealism but are also sufficient, as is shown by the LG version of Fine’s theorem \cite{halliwell_comparing_2017,halliwell_necessary_2019,PhysRevA.100.042103}. In quantum mechanics, the LG2 and LG3 inequalities can be violated up to the value $-\frac{1}{8}$, the L{\"u}ders bound.\footnote{This is for so-called L{\"u}ders measurements, which simply measure the value of $Q$. More negative values can be obtained by measuring finer-grained quantities (von Neumann measurements) and then coarse-graining their probabilities to find probabilities of $Q$ \cite{Emary_2013,budroni2014temporal}.} 

We also note that the four LG3 quantities may be written as
\begin{equation} \label{L1 B}
    L_1 = q(+,+,+) + q(-,-,-) 
\end{equation}
\begin{equation}
    L_2 = q(+,-,+) + q(-,+,-) 
\end{equation}
\begin{equation}
    L_3 = q(+,-,-) + q(-,+,+)
\end{equation}
\begin{equation} \label{L4 B}
    L_4 = q(+,+,-) + q(-,-,+) \ ,
\end{equation}
from which we see that, when non-negative, the LG3s describe a probability distribution for \textit{changes of sign}. Eqs. \ref{L1 B}–\ref{L4 B} are, respectively, the probabilities for zero or two sign changes and the two different possibilities for a single sign change.

This means that if the LG3s are satisfied but the LG2s are not, then even though MR is violated for the full data set, it can still be satisfied for the data subset of the sign-change variables $\chi_{1}=s_1 s_2$ and $\chi_{2}=s_2 s_3$, which are $\pm 1$ depending on whether or not the sign of $Q$ has flipped between measurement times. We will make extensive use of this in Section \ref{Assessing UP}, where we are only interested in the sign-change probabilities as a quantification of uniform precession.

Violations of the LGIs have been observed experimentally in a myriad of physical systems, such as superconducting qubits \cite{palacios-laloy_experimental_2010,PhysRevLett.111.090506}, nuclear spins \cite{athalye_investigation_2011, Souza_2011, PhysRevA.87.052102}, millimetre-scale crystals \cite{crystals}, photons \cite{Goggin_2011}, and (recently) neutron interferometry \cite{kreuzgruber_violation_2024,PhysRevA.103.032218}. The Leggett–Garg inequalities also have far-reaching applications to other fields \cite{quantum_computation, quantum_info_processing,PhysRevA.106.012214,wilde_could_2009, Li_2012}. A more exhaustive account of the experimental tests of macrorealism to date is provided in the extensive review articles Refs. \cite{Emary_2013,Vitagliano_2023}.

However, a significant issue for the Leggett-Garg scheme is that macrorealism per se is conjoined inextricably with non-invasive measurability. The ``clumsiness loophole" \cite{wilde_addressing_2012} is that any recorded violations could always be ascribed, instead, to the unintended invasiveness of the measurement process \cite{montina_dynamics_2012}. The standard way to render the measurements non-invasive is to use ideal negative result measurements \cite{dicke1981interaction, knee2012violation}. A convenient refinement of this protocol is described in Ref. \cite{PhysRevA.99.022119}, where it is shown that the correlators such as $C_{12}$ can always be measured in a context where the no-signalling-in-time condition is satisfied. This is the condition that $\langle Q_2^{(1)} \rangle$, defined in Eq.\ref{Q_2 with an earlier 1}, is the same as $\langle Q_2 \rangle$, i.e. measurements at $t_1$ have no disturbing effect on the average of $Q$ at time $t_2$. Alternative non-invasive measurement methods exist, such as the waiting detector protocol \cite{halliwell_leggett-garg_2016, PhysRevA.100.042325,mawbythesis}. If the non-invasive measurement condition is implemented convincingly, then violations of the LGIs imply the failure of \textit{macrorealism per se}.

The derivation of the Leggett–Garg inequalities solely from the postulates of macrorealism is completely independent of the system's equations of motion. It is this indifference to the underlying dynamics that necessitates sequential measurements in order to obtain a quantum–classical disparity. Alternatively, the introduction of a well-chosen and experimentally testable \textit{dynamical assumption} can supplant the requirement for non-invasive measurement. A few examples of such assumptions have appeared in the LG literature and typically consist of assuming that certain transition probabilities are zero \cite{hermens_constraints_2018,LGdecays,halliwell_leggett-garg_2016} or can be simplified in some way \cite{Emary_2013, stationarity2011,PhysRevA.52.R2497}. The key point is that the assumptions are chosen so that the correlators simplify to single-time measurement averages. Furthermore, the dynamical assumption is either implemented through the design of the experiment (e.g. the choice of Hamiltonian, where possible) or checked experimentally through separate control experiments. Dynamical assumptions are often classically motivated, and the Tsirelson scheme is an example par excellence, with its simplifying dynamical assumption of uniform precession.

The related single crossing dynamical assumption, studied in the context of a waiting detector model in Ref. \cite{halliwell_leggett-garg_2016} and Chapter 6 of Ref. \cite{mawbythesis}, assumes one crossing of the origin over a chosen time interval and would be valid for a harmonic oscillator over half a period. In fact, this is then mathematically equivalent to the assumption of uniform precession over a full period, a proof of which we provide in Appendix \ref{Single crossing assumption}.

But does a Tsirelson inequality violation truly imply genuine quantum behaviour and the failure of macrorealism, or could a stubborn macrorealist now argue that it stems purely from faulty dynamics? We saw that even trajectories with a defined position at all times can produce spurious Tsirelson inequality violations (Fig. \ref{fig:Erratic classical trajectory}). So, just as separating realism from non–invasive measurability has proved to be the greatest hassle in LG tests, do we now run into a similar problem in Tsirelson tests of separating the failure of realism from this newfound dynamical assumption?

Our strategy in Section \ref{Assessing UP} to assess the dynamical assumption  will be to use the Leggett–Garg framework to re-write the Tsirelson quantity in terms of various probabilities that can be interpreted as measures of uniform precession.

\section{Tsirelson Violations in the Quantum Harmonic Oscillator} \label{Tsirelson Violations in the QHO}

\subsection{Expansion in energy eigenbasis}

In order to discuss the Tsirelson quantity quantum-mechanically, we promote the variable $Q$ to be an operator in the Heisenberg picture, undergoing unitary time evolution as $ \hat Q(t) = U^\dagger  \hat Q U $, where $U = \text{exp}(-itH)$ with a standard QHO Hamiltonian $H$. We can write $\hat{Q}=(2{\theta}(\hat x)-1)$, where ${\theta}( \hat x) = \int_0^\infty dx |x\rangle\langle x|$ is the Heaviside operator (and $|x\rangle$ are position eigenstates). For convenience, we will work exclusively with units chosen such that $\hbar = 1$; further, it will be convenient to normalise the angular frequency of oscillations to unity, $\omega = 1$. The Tsirelson inequality is then $-1 \leq \langle A \rangle \leq 1$, where now
\begin{equation}
    \hat A \equiv \hat Q + U^\dagger \hat Q U + (U^\dagger)^2 \hat Q U^2 \ ,
\end{equation}
with $U = \text{exp}(-\frac{2\pi}{3}iH)$, recalling that, by choosing units of $\omega = 1$, one full period corresponds to $T=2\pi$. We will refer to $\hat A$ hereafter as the ``Tsirelson operator". Quantum systems \textit{can} violate this inequality—up to a maximal value of $\langle A \rangle \approx 1.26$, so Tsirelson found \cite{tsirelson_how_2006}.

We can compute the Tsirelson quantity for an arbitrary state of the QHO Hilbert space formed as a superposition of energy eigenstates, $|n\rangle$, from the ground state upwards up to a finite cut-off:
\begin{equation} \label{eq: state with cutoff}
    | \Psi \rangle = \sum_{n=0}^N c_n |n\rangle, c_n \in \mathbb{C} \ .
\end{equation}
The single-time average $\langle  Q \rangle = \langle \Psi | (2 \theta(\hat x) - 1)| \Psi \rangle$ can be written as
\begin{equation}
    \langle \Psi | (2 \theta(\hat x)-1) | \Psi \rangle = \sum_{n,k=0}^N c_n^* Q_{nk} c_k \ ,
\end{equation}
where we have defined the matrix elements $Q_{nk} = \langle n| (2 \theta(\hat x)-1) | k\rangle$. A tractable expression for $Q_{nk}$ is derived in Appendix \ref{sec: matrix elements E} and is given by 
\begin{equation}
   \frac{2\pi^{-\frac{1}{2}}}{(k-n) \sqrt{2^{n+k} n! k!}} [kH_{k-1}(0)H_n(0) - nH_{n-1}(0)H_k(0)] \ ,
\end{equation}
where $H_n(x)$ are the standard Hermite polynomials.

So, for a state in the form of Eq. \ref{eq: state with cutoff},
\begin{equation}
    \langle  Q(t)\rangle = \langle \Psi| \hat{Q} | \Psi \rangle = \sum_{n,k=0}^N c_k^* c_n e^{-i(n-k) t}Q_{nk} \ . \label{Single time average expression}
\end{equation} 
Therefore, for an arbitrary state up to an energy truncation of $N$, the Tsirelson quantity is given by
\begin{equation}
    \langle A \rangle = \sum_{n,k=0}^N c_k^* c_n (1+ e^{-i\frac{2\pi}{3}(n-k)}+e^{-i\frac{4\pi}{3}(n-k)})Q_{nk}  \ . \label{Tsirelson quantity}
\end{equation}  

Violations of the Tsirelson inequality are manifestly associated with negativity in the Wigner function. A corollary is that a single coherent state, which has a completely positive Wigner function, will not violate the Tsirelson inequality; neither will a single energy eigenstate, for it is trivially of strictly odd or even parity. A proof, in terms of probability currents, that definite-parity states do not violate the Tsirelson inequality is provided in Section \ref{Probability currents}. Perhaps more surprising is that a superposition of \textit{at least three} energy eigenstates is required to produce a violation. We provide a full proof of this fact in Appendix \ref{Energy Eigenstate Superpositions Appendix}.

\subsection{The simplest violating state}

The Tsirelson operator is Hermitian ($\hat{A}\equiv \hat{Q}_1 + \hat{Q}_2 + \hat{Q}_3 = \hat{A}^{\dag}$), and so its eigenvalues for a truncated $(N+1)$-dimensional subspace of the Hilbert space are real. The greatest eigenvalue corresponds to the largest possible Tsirelson violation, $\langle A \rangle > 1$, for a superposition $|\psi\rangle = \sum_{n=0}^N c_n|n\rangle$ of QHO energy eigenstates up to $|N\rangle$. Likewise, the lowest eigenvalue corresponds to the most severe violation of the classical lower bound, $\langle A \rangle=-1$, in the subspace. The matrix elements of $\hat A$ in the energy eigenbasis can be written in terms of the familiar $Q_{nk}$:
\begin{equation} \label{A matrix elements}
    {A}_{nk} \equiv \langle n | {A} | k \rangle = Q_{nk}(1 + e^{i\frac{2\pi}{3}(n-k)} + e^{i\frac{4\pi}{3}(n-k)}) \ ,
\end{equation}
for $n,k = 0,1,...,N$.

By explicitly diagonalising the Tsirelson operator $\hat A$ in an increasingly large subspace, which is detailed in Appendix \ref{Diagonalising appendix}, we find that the greatest Tsirelson violation in the $N=6$ subspace is given by:
\begin{equation} \label{normal Zaw}
    |\Psi_\mathcal{Z}\rangle=\frac{4}{\sqrt{42}}|0\rangle-\frac{1}{\sqrt2}|3\rangle+\sqrt{\frac{5}{42}}|6\rangle \ ,
\end{equation}
with an upper-bound Tsirelson violation $\langle A\rangle=1.1195$; or equivalently, by swapping the sign of the odd-parity eigenstate,
\begin{equation} \label{parity flipped Zaw}
    |\Psi_\mathcal{Z}'\rangle=\frac{4}{\sqrt{42}}|0\rangle+\frac{1}{\sqrt2}|3\rangle+\sqrt{\frac{5}{42}}|6\rangle \ ,
\end{equation}
with a lower-bound Tsirelson violation $\langle A \rangle=-1.1195$.
This agrees exactly with the work of Ref. \cite{PhysRevA.108.022211}, for which we label Eq. \ref{normal Zaw} the Zaw–Scarani state. We shall use this as a simple candidate state for later analysis because the $N=6$ cut-off is the first point at which violations to Tsirelson's inequality become possible (a direct consequence of Appendix \ref{Energy Eigenstate Superpositions Appendix} and Section \ref{sec: twirl}).

\begin{figure*}
    \centering
    \subfigure[]{
        \includegraphics[width=0.5\linewidth]{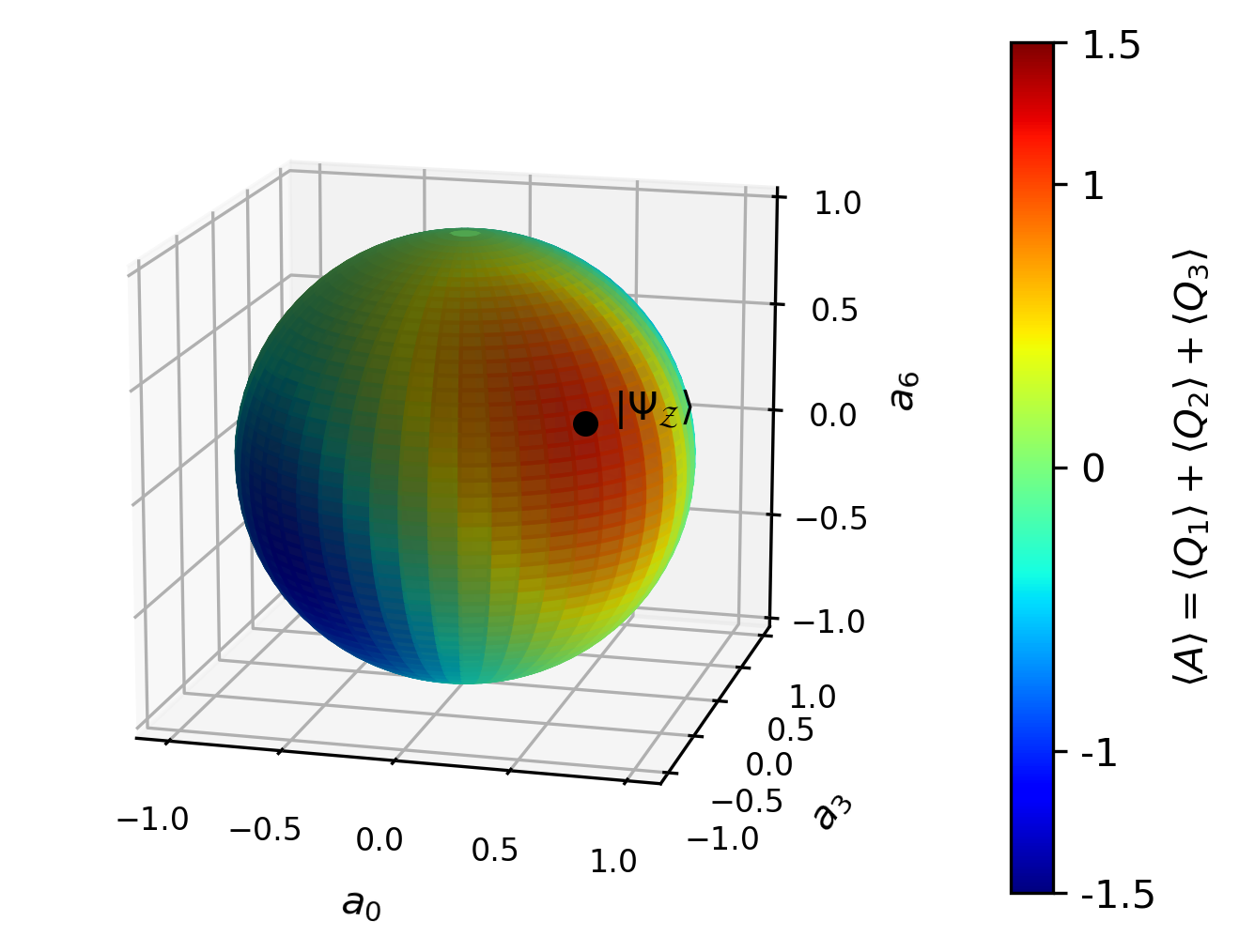}
    }
    \subfigure[]{
        \includegraphics[width=0.4\linewidth]{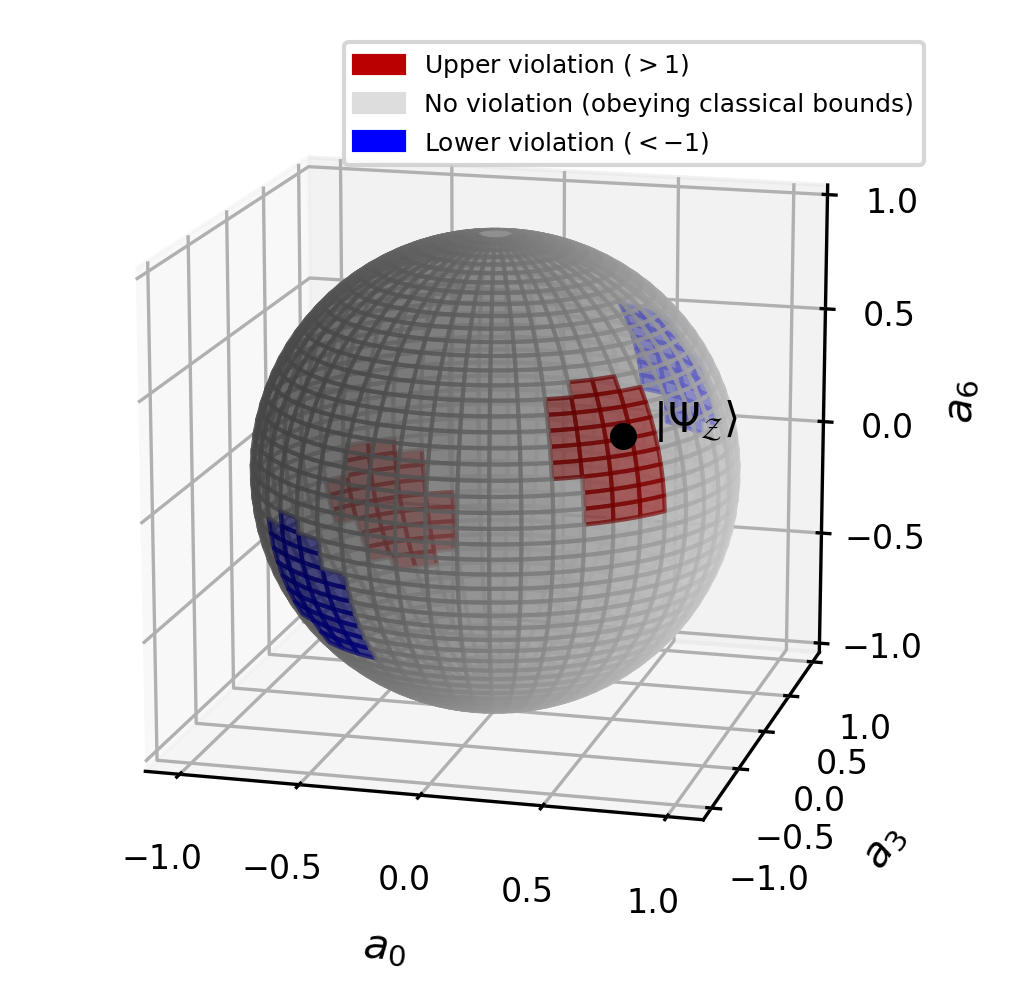}
    }
    \caption{Representation of all states in the subspace described by Eq. \ref{convenient subspace}, where the axes represent the (real) coefficients. (a) A colour map is used to represent the value of the Tsirelson quantity when each of these states is taken to be the initial state. The Zaw–Scarani state, labelled $|\Psi_Z\rangle$, has the greatest violation in this subspace and lies in the centre of the upper-bound-violating region. (b) Only the violating (non-classical) regions are highlighted, showing that each upper violation is paired with a lower violation.}
    \label{Bloch spheres}
\end{figure*}

This is not an isolated violating state. Fig. \ref{Bloch spheres} depicts a modified ``Bloch sphere" representation of the set of all possible states constructed in the subspace spanned by the $|0\rangle$, $|3\rangle$, and $|6\rangle$ energy eigenstates with real coefficients. That is,
\begin{equation} \label{convenient subspace}
    |\Psi\rangle = a_0 |0\rangle + a_3 |3\rangle + a_6 |6\rangle \ , \ \text{where} \ a_0,a_3,a_6 \in \mathbb{R} \ .
\end{equation}
Clearly, there exists a sizeable region of states over which a Tsirelson violation can be achieved to varying extent. This is good news for experimental realisations of the protocol, because it means even quantum states that are prepared imperfectly can, in principle, produce demonstrable Tsirelson violation. Another important observation is that every upper-bound-violating state is paired with a commensurate lower-bound-violating state ($\langle A \rangle \rightarrow -\langle A\rangle$), which is obtained by swapping the sign of the odd-parity coefficients (e.g. $c_3\rightarrow -c_3$).

\subsection{Structure of the Tsirelson operator} \label{sec: twirl}

This section establishes a link between the Tsirelson operator and the group $C_3$. We then apply this connection to find an alternative explanation for a result previously noted in Ref. \cite{precessions}.

The group $C_3$ has three elements which we denote $\{e, r, r^2\}$. It is an Abelian group whose generator is $r$ \cite{groupC3}. From the Cayley table of the group, outlined in Table \ref{tab: C3_cayley}, it is straightforward to determine that $\{\mathds{1}, U, U^2\}$ forms a faithful representation of $C_3$ under multiplication with $U$ acting as the generator.
\begin{table}[h!]
\centering
\begin{tabular}{|c|c|c|c|}
\hline
\textbf{} & \(e\) & \(r\) & \(r^2\)  \\
\hline
$e$    & $e$ & $r$ & $r^2$ \\
$r$    & $r$ & $r^2$ & $e$ \\
$r^2$  & $r^2$ & $e$ & $r$ \\
\hline
\end{tabular}
\caption{The Cayley table of $C_3$. The three elements of the group are $\{e,r,r^2\}$, where $e$ is the identity.}
\label{tab: C3_cayley}
\end{table}

From this observation, we can relate the Tsirelson operator to the twirl of the sign operator,
\begin{equation}
    \operatorname{Twirl}(\hat {Q}) = \frac{1}{3} \hat A = \frac{1}{3}\left[\mathds{1} \hat Q \mathds{1} + U^\dagger\hat Q U + (U^2)^\dagger \hat Q (U^2) \right] \ .
\end{equation}
The $\operatorname{Twirl}(\cdot)$ operation is the averaging of a matrix over the actions of a group, in this case $C_3$ \cite{twirling}. It is a useful operation because it extracts the component of the matrix that transforms under the identity representation of the group. Note that $U^\dagger \operatorname{Twirl}(\hat Q) U = (1)\operatorname{Twirl}(\hat Q) \implies [\operatorname{Twirl}(\hat Q), U] = 0$.

\begin{figure}[h]
    \centering
    \includegraphics[width=1\linewidth]{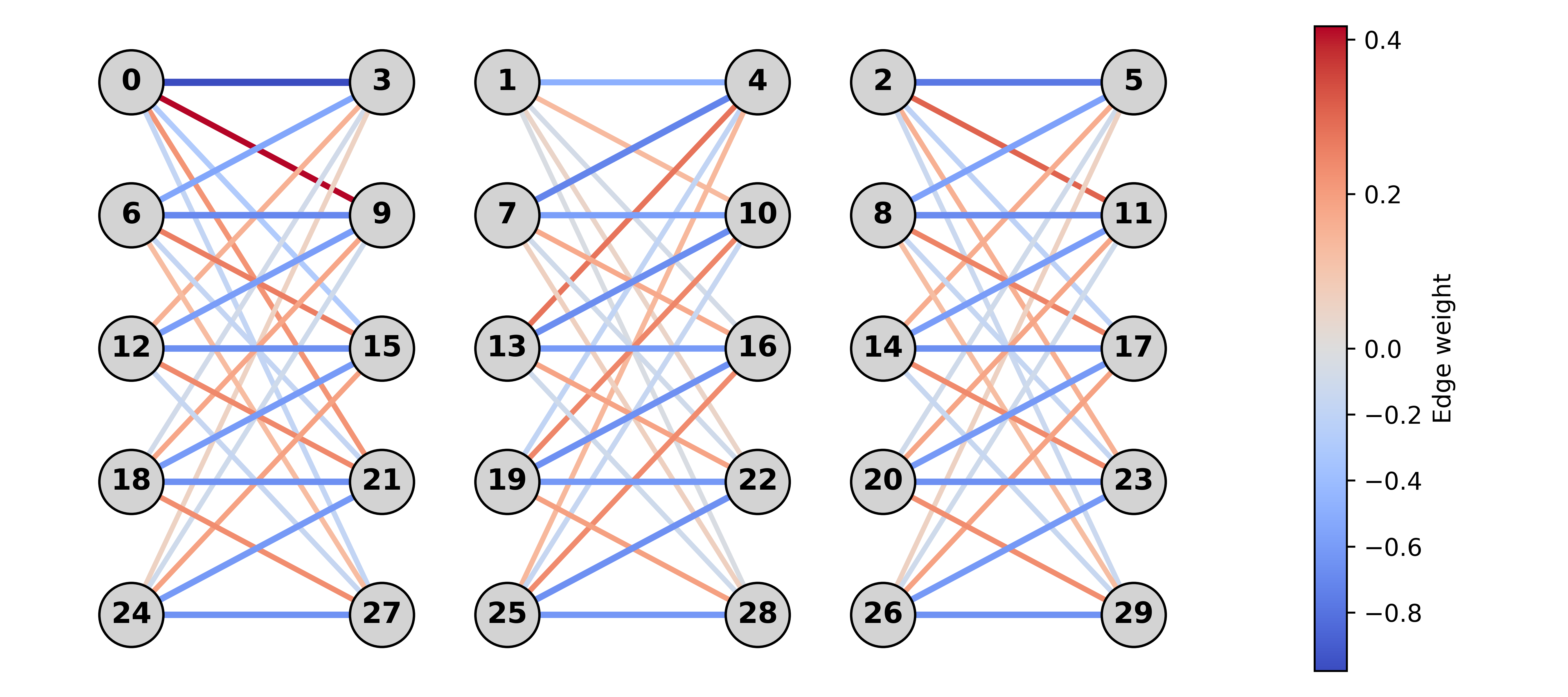}
    \caption{The Tsirelson operator in the energy eigenbasis with any finite energy cut-off can be interpreted as the adjacency matrix of a simple weighted graph. This graph is illustrated with a cut-off at $n=29$, where the nodes are energy eigenstates and the edges represent non-zero matrix elements of the Tsirelson operator. The colour represents the value of the matrix element.}
    \label{fig: graph rep}
\end{figure}

Because the two normal operators commute, they cannot mix each other's eigenspaces. The action of $U$ on the basis states $|n\rangle$ clearly partitions the Hilbert Space, $\mathcal{H}$, into 3 degenerate eigenspaces corresponding to its three eigenvalues $\lambda_j$:

\noindent Eigenspace for \( \lambda_1 = 1 \):
\begin{equation} \label{H0}
        \mathcal{H}_0 = \operatorname{span}\{|n\rangle : n \equiv 0 \mod 3\} \ .
\end{equation}
Eigenspace for \( \lambda_2 = e^{-i \frac{2\pi}{3}} \):
\begin{equation} \label{H1}
        \mathcal{H}_1 = \operatorname{span}\{|n\rangle : n \equiv 1 \mod 3\} \ .
\end{equation}
Eigenspace for \( \lambda_3 = e^{-i \frac{4\pi}{3}} \):
\begin{equation} \label{H2}
        \mathcal{H}_2 = \operatorname{span}\{|n\rangle : n \equiv 2 \mod 3\} \ .
\end{equation}
Therefore, the eigenvectors of $\operatorname{Twirl}({Q})$, and hence $A$, must be made up exclusively of basis vectors in a \textit{single} degenerate subspace. This property is illustrated in Figure \ref{fig: graph rep} by representing the Tsirelson operator as a graph, where each subgraph corresponds to one of the subspaces. The bipartite nature of the subgraphs visualises the underlying symmetric spectrum of the operator.

\section{Assessing the Dynamical Assumption} \label{Assessing UP}

\subsection{General framework}

We now come to the main issue, which is assessing the degree to which the dynamical assumptions behind the Tsirelson inequality hold. As discussed, like many other classical phenomena, uniform precession and its closely-related assumptions have no unique definition in quantum theory. We will therefore explore a number of inequivalent implementations which entail different measurement techniques to check it. Our first method
involves demonstrating an explicit connection between $\langle A \rangle$ and the probability of zero sign changes, L1 (Eqs.
\ref{L1 A} and \ref{L1 B}), which should be zero classically. Our
second method stems from the interesting observation
that a weaker condition of setting merely the difference $q(+,+,+)-q(-,-,-)$ to be zero also leads to the Tsirelson inequality, as we shall show, and thus this may also be considered a measure of uniform precession.

Specifically, we will implement the following procedure in a number of different ways:

\begin{enumerate}
        \item Declare a data set, comprising measurements of $\langle A \rangle$ (the Tsirelson test itself) as well as subsidiary quantities to gauge the dynamical assumption of uniform precession (UP).
        \item Write the Tsirelson quantity in the following form:
        \begin{multline}\label{Quantum form}
            \frac{1}{2}(1+\langle A \rangle) = (\text{Positive term}) + (\text{UP violating term}) \\ + (\text{Quantum interference term}) \ ,
        \end{multline}
        The specific form on the left-hand side is used because $\frac{1}{2}(1 \pm \langle A \rangle)$ are the probabilities for $A=\pm 1$ when it can take only values $\pm 1$, as is the case classically (since uniform precession holds). This gives a convenient scale against which to compare violations of the Tsirelson inequality and of uniform precession. 
        \item Establish whether or not the UP violating term is capable of dominating the quantum interference terms in terms of explaining the Tsirelson inequality violation. In other words, whether or not a classical model with no quantum-mechanical interferences could account for the violation. If not, the failure of macrorealism per se must also be implied.
\end{enumerate}

\subsection{Method 1: Macrorealistic description of sign changes}

We shall quantify the breaking of the dynamical assumption of uniform precession through the LG3 quantity $L_1$ (Eq. \ref{L1 A}), corresponding to the quasi-probability sum $q(+,+,+)+q(-,-,-)$,
which should be zero classically.

\textbf{Step 1:} The full data set comprises all three correlators, $C_{12},C_{23},C_{13}$, measured non-invasively, and $\langle A \rangle$ via a Tsirelson test.

\textbf{Step 2:} We find, using the probability moment expansions in Eqs. \ref{moment expansion 3-time} and \ref{projective 2-time moment expansion}, that the Tsirelson quantity can be written as
\begin{equation} \label{Key equation} 
    \frac{1}{2}(1+\langle A \rangle)=\sum_{j>i}p_{ij}(+,+) - L_1+I \ ,
\end{equation}
where the sum of two-time measurement probabilities, 
\begin{equation}
    \sum_{j>i}p_{ij} \equiv p_{12}(+,+)+p_{23}(+,+)+p_{13}(+,+) \ ,
\end{equation}
is manifestly positive, and the interference term is given by
\begin{multline}
    I=\frac{1}{4} \left( \langle Q_2 \rangle -\langle Q_2^{(1)} \rangle \right) + \frac{1}{4}\left( \langle Q_3 \rangle - \langle Q_3^{(2)} \rangle \right) \\+ \frac{1}{4}\left( \langle Q_3 \rangle - \langle Q_3^{(1)} \rangle \right) \ .
\end{multline}
This constitutes the explicit form of Eq. \ref{Quantum form}.

The candidate state we consider is the parity-flipped Zaw–Scarani state, Eq. \ref{parity flipped Zaw}. with a (re-scaled) lower-bound Tsirelson violation
\begin{equation}
    \frac{1}{2}(1+\langle A \rangle) = -0.0598<0 \ ,
\end{equation}
which forms the left-hand side of the key formula (Eq. \ref{Key equation}).

\textbf{Step 3:} Finally, we ask the question: is it possible that a non-zero no-crossing probability (that breaks uniform precession) explains the full Tsirelson violation? Referring to the crucial Eq. \ref{Key equation}, we have calculated the left-hand side term $\frac{1}{2}(1+\langle A \rangle)=-0.0598$ and the second term on the right-hand side, $L_1=0.0421$ (from Appendix \ref{Correlator appendix}). Importantly, because the first term on the right-hand side, $\sum_{j>i}p_{ij}$, is positive, then no matter its exact value, it is mathematically impossible for $(\sum_{j>i}p_{ij}-0.0421)$ to reach the value $-0.0598$. Thus, it is immediately clear that $I\neq0$ must be implied (specifically, $I<0$); that is, non-zero quantum interferences are present. We conclude that the violation of uniform precession is \textit{not} sufficient to explain the Tsirelson inequality violation. This is a crucial result, showing that this method has proven successful in identifying that uniform precession is satisfied well enough that quantum behaviour must be implied.

We also note that Eq. \ref{Key equation} can instead be written in terms of the two-time quasi-probabilities (Eq. \ref{moment expansion 2-time}) with no interferences:
\begin{equation} \label{eq: tsirelson as two-time}
    \frac{1}{2}(1+\langle A \rangle)=q_{12}(+,+)+q_{23}(+,+)+q_{13}(+,+) - L_1 \ .
\end{equation}
Each of the two-time quasi-probabilities satisfies the the L{\"u}ders bound; but, in fact, the sum $q_{12}(+,+)+q_{23}(+,+)+q_{13}(+,+)$ is itself bounded below by $-\frac{1}{8}$, as is readily shown by writing the sum purely in terms of $\langle A \rangle$ and $\langle A^2 \rangle$ using the moment expansions (Eq. \ref{moment expansion 2-time}):
\begin{equation}\label{eq: sum quasis}
    \sum_{j>i}q_{ij}(+,+) = \frac{1}{8}\langle (A+2)^2-1\rangle \ .
\end{equation}
From Eq. \ref{eq: tsirelson as two-time} and Eq. \ref{eq: sum quasis}, we see that the Tsirelson violation, $\frac{1}{2}(1+\langle A \rangle)$, will itself respect the L{\"u}ders bound if uniform precession is satisfied exactly ($L_1=0$), but in general can exceed this bound, as is the case with the maximally-violating state in the full infinite-dimensional QHO Hilbert space, which Tsirelson found to be $\frac{1}{2}(1+\langle A \rangle)=-0.13<-0.125$ \cite{tsirelson_how_2006}.

We now proceed with some further analysis on the breaking of uniform precession in this state. We already noted above that $L_1 > 0$, and we also found that $L_2 = L_3 = L_4 = 0.3193 > 0$. This complete satisfaction of the LG3s is not some esoteric feature unique to this simple candidate state we have chosen: in fact, we find in Appendix \ref{T implies LG3 satisfaction proof} that \textit{any} Tsirelson-violating state constructed in one of the three subspaces given by Eqs. \ref{H0}–\ref{H2} will satisfy all four LG3 inequalities.

Recall from Eqs. \ref{L1 B}–\ref{L4 B} that the LG3 quantities describe a probability distribution of sign changes. Therefore, there exists a valid classical (macrorealistic) description of crossing probabilities for the state. That there exists a macrorealistic description of sign-change probabilities is a crucial point: we now have the basis to claim that $L_1$ represents a valid probabilistic description of the breaking of uniform precession.

We also note, at this point, that simple superpositions of coherent states—such as the ``three-headed cat state"—will violate the Tsirelson quantity \cite{PhysRevLett.130.160201}. For example,
\begin{equation}
    |\Psi \rangle = N\left( |\sqrt{2} \ \rangle + |e^{2\pi i/3}\sqrt{2} \ \rangle + |e^{4\pi i/3} \sqrt{2} \ \rangle \right) \ ,
\end{equation}
with the appropriate normalisation factor $N$, yields a Tsirelson violation of $\langle A \rangle =-1.0795<-1$. Thus, the method of assessing UP with an LG quantity can also be used upon such states, which may be of greater interest experimentally, for Gaussian state superpositions can be easier to create in a laboratory setting \cite{Ourjoumtsev2007}.

\subsection{Method 2: Three-time sequential measurements}

Our second method uses a dynamical assumption involving $\Delta p=p(+,+,+)-p(-,-,-)$, where $p$ is a probability to be specified, which weakens the original assumption for deriving Tsirelson's inequality \cite{tsirelson_how_2006, plavala_tsirelson_2024, precessions}. This is of interest because, classically, the Tsirelson quantity $A$ can take values $\pm 3$ and $\pm 1$, but setting $\Delta p = 0$ prevents the values $\pm 3$ from contributing to the computation of the average $\langle A \rangle$, and hence trivially $-1\leq\langle A \rangle\leq 1$.

The probabilities could be specified to involve three-time projective measurements, $\Delta p_{123}$, corresponding to Eq. \ref{projective 3-time moment expansion}, or we could consider the quasi-probability $\Delta q$, corresponding to Eq. \ref{moment expansion 3-time}. We could also use the two quasi-probability distributions mentioned in the discussion surrounding Eq. \ref{Q_2 with an earlier 1}, which lie between $\Delta p_{123}$ and $\Delta q$. We shall denote these distributions $\Delta q_A$, $\Delta q_B$, which we define fully in Appendix \ref{Method 2 appendix}). Appendix \ref{Method 2 appendix} also details the computational approach to calculating all such measurement probabilities. An important point is that all of these four distributions are identical to a macrorealist but differ quantum mechanically by various interference terms.

\textbf{Step 1:} We take the data set to be $\Delta p$ measured via various three-time measurements and the Tsirelson quantity $\langle A \rangle$ measured with single-time measurements.

\textbf{Step 2:} We find an alternative way in which the Tsirelson quantity can be expressed, using this time the moment expansion Eq. \ref{projective 3-time moment expansion}:
\begin{equation}\label{method two explicit}
    \frac{1}{2}(1+\langle A \rangle) = \frac{1}{2}(1-D)+2\Delta p +I \ .
\end{equation}
Here $I$ is the interference term, and its values for each of the four cases are given below in Table \ref{table of measures of UP}. Both classically and quantum-mechanically, the triple correlator satisfies $|D|\leq 1$, which renders the first term $\frac{1}{2}(1-D)$ manifestly non-negative, so that Eq. \ref{method two explicit} is the explicit form of Eq. \ref{Quantum form} in this method.

\textbf{Step 3:} We compare the size of the UP violation, $2|\Delta p|$, with the size of the Tsirelson violation, $\frac{1}{2}|(1+\langle A \rangle)|$, noting once again that the interference terms would vanish in a purely classical model of the violation. Specifically, $2|\Delta p|=\frac{1}{2}|(1+\langle A \rangle)|$ marks the threshold that determines whether or not there is some plausible value of $D$ that can satisfy Eq. \ref{method two explicit} with $I=0$.

Table \ref{table of measures of UP} catalogues our numerical results with this approach, considering each of the four aforementioned distributions. We compare the extent of a lower-bound violation that would be possible from \textit{only} the breaking of the dynamical assumption in a classical model with the true value of the Tsirelson violation, using the same state as before (Eq. \ref{parity flipped Zaw}). With the three-time sequential measurement probability $\Delta p_{123}$ as the chosen measure of UP, we calculate that $2|\Delta p_{123}|=0.0191<0.0598$, showing once again that the breaking of the dynamical assumption falls significantly short of accounting for the full extent of the Tsirelson violation, thereby implying the presence of quantum-mechanical interference terms.

In the case of the distribution $\Delta q_B$, UP is satisfied only marginally; and when the quasi-probability distribution $\Delta q$ is instead used, which produces no interference terms, the UP violation is exactly commensurate with the Tsirelson violation. This is to be expected, since $\Delta q = 0$ is a sufficient condition to derive the Tsirelson inequality, which property follows directly from the moment expansion form of Eq. \ref{moment expansion 3-time}.

\begin{table*}[t]
    \centering
    \renewcommand{\arraystretch}{2} 
    \begin{tabular}{c|c|c|c}
        \Xhline{2\arrayrulewidth}
        Measure of UP & \makecell{Interference term, $I$} & $2\Delta p$ & \makecell{UP satisfied sufficiently? \\ ($2|\Delta p|<\frac{1}{2}|1+\langle A\rangle| = 0.0598$ ?)} \\
        \hline
        $\Delta q$ & 0 & $-0.0598$ & {\color{red}\xmark} \\
        \arrayrulecolor{gray}
        \hline
        $\Delta p_{123}$ & \makecell{$\frac{1}{2}(\langle {Q}_2 \rangle-\langle {Q}_2^{(1)} \rangle)+\frac{1}{2}(\langle {Q}_3 \rangle-\langle {Q}_3^{(12)} \rangle)$} & $-0.0191$  & {\color{darkgreen}\checkmark} \\
        \hline
        $\Delta q_A$ & $\frac{1}{2}(\langle {Q}_3 \rangle-\langle {Q}_3^{(2)} \rangle)$ & $-0.0785$ & {\color{red}\xmark} \\
        \hline
        $\Delta q_B$ & \makecell{$\frac{1}{2}(\langle {Q}_2 \rangle-\langle {Q}_2^{(1)} \rangle)+\frac{1}{2}(\langle {Q}_3 \rangle-\langle {Q}_3^{(1)} \rangle)$} & $-0.0573$ & {\color{darkgreen}\checkmark} \\
        \arrayrulecolor{black}
        \Xhline{2\arrayrulewidth}
    \end{tabular}
    \caption{The various means by which the dynamical assumption of uniform precession (UP) can be assessed through three-time probability distributions in Method 2 of Section \ref{Assessing UP}. A tick indicates that the UP violation is too small to produce the overall Tsirelson violation, which must therefore be produced, in part, by the quantum-mechanical interference terms.}
    \label{table of measures of UP}
\end{table*}

An alluring question is whether it is possible, by some variation of the parameters for the quantum state, to render $\Delta p_{123}=0$ exactly, whilst keeping within a Tsirelson violating region (such as that depicted in Fig. \ref{Bloch spheres}), which may serve the poetic justice of ``total satisfaction" of the dynamical assumption, whereby the Tsirelson violation is solely due to effects quantum-mechanical in origin. In fact—unlike in Method 1, wherein setting $L_1=0$ leads trivially to the satisfaction of the Tsirelson inequality—it \textit{is} possible in this case to set $\Delta p_{123}=0$. We find by using a constrained optimisation procedure via Lagrange multipliers that a simple state—still within the $N=6$ subspace, albeit now with complex coefficients—can be found to produce $\langle A \rangle = 1.0756$ with $\Delta p_{123}=0$. The computational procedure is documented in Appendix \ref{Constrained optimisation}.

We also note that perfect sign-operator measurements are an idealisation for any real experimental procedure and comment on the effect of smoothed projectors in Appendix \ref{Smoothed projectors}.

\section{Heuristic Notions of Uniform Precession} \label{Heuristic definitions}

In this section, we describe an alternative method of assessing uniform precession which is somewhat heuristic and indirect but wherein the measurements involved are simpler and in some cases are, appealingly,  single-time measurements. We consider measurements of two simple quantities during a given time interval: the dwell time in $x>0$ and the number of crossings of the origin. Physical measurements will reveal the quantum-mechanical values of these quantities and, we shall show, if sufficiently close to the corresponding classical values will significantly constrain possible violations of uniform precession in a trajectory model. This in turn gives heuristic indications of the smallness of the various forms of the probabilities $p(+,+,+)$ and $p(-,-,-)$ used in Methods 1 and 2 above. We also relate the Tsirelson inequality to the probability current.

\subsection{Dwell time} \label{sec: dwell time} \label{Probability currents}

We mentioned previously that dynamical assumptions are often classically motivated. One potential notion of how well the dynamical assumption of uniform precession is satisfied is the ``dwell time", which is classically given by the following integral:
\begin{equation}
    T_D(\tau) =\frac{1}{2\pi}\int_0^{\tau} \theta(x(t)) dt \ ,
\end{equation}
where we highlight the explicit dependence of $\theta$ upon the time-dependent position $x(t)$. The dwell time is a measure of the fraction of time the particle spends in $x>0$. Classically, this value should be $1/2$ over a full period ($\tau=2\pi$). This can readily be turned into a self-adjoint operator, $\hat {T}_D(\tau)$ \cite{Munoz2009DwellTime}. The average quantum-mechanical dwell time $\langle T_D(\tau)\rangle$ can be ascertained experimentally by measuring $\langle \theta(\hat x(t))\rangle$ at multiple times.

For an arbitrary state $|\psi\rangle =\sum_n c_n |n\rangle$, in terms of energy eigenstates, we can write
\begin{equation}
    \langle T_D(2\pi) \rangle =  \sum_{n}|c_n|^2 \langle n| \theta(\hat x(0))|n\rangle \ ,
\end{equation}
because $|n\rangle$ are stationary states, meaning $\langle n|  \theta(\hat x(t))|n\rangle  = \langle n| \theta(\hat x(0))|n\rangle$. Also, as energy eigenstates have definite parity, $\langle n| \theta(\hat x(0))|n\rangle=\frac{1}{2}$. Therefore,
\begin{equation} \label{Dwell time result}
    \langle T_D (2\pi)\rangle = \frac{1}{2}.
\end{equation}
We see that the dwell time, integrated over a full period, is exactly the classical value. By this measure, a classical and quantum oscillator are indistinguishable. 

The dwell time operator is diagonal over a full period (and in fact is proportional to the identity)—which can be understood as due to the fact that $\hat{T}_D(2\pi)$ is $\text{Twirl}(\theta(\hat x))$ over a representation of the group $U(1)$—and thus there is no spread in its distribution. Within the subspace $\mathcal H_0$, the dwell time operator is also diagonal when integrated to one-third or two-thirds of a period. The latter case is particularly informative, since this represents integrating between the first and final measurement times. We have found that $\langle  \hat T_D(4\pi/3)\rangle = \frac{1}{3}$ exactly—which, remarkably, also falls within the range of classically allowed values. Another observation is that looking at the dwell time over any cycle, starting at any time, yields the same results.  

It is quite striking that the quantum dwell time agrees with the classical result and yet Tsirelson violations still occur. In fact, Ref. \cite{precessions} considered alternatives to the Tsirelson scheme with a greater number of measurement times during a period and found that quantum violations faded away as the number of measurement times was increased (limiting to the dwell time result which has no quantum violations). This perspective motivates viewing the Tsirelson quantity as a Riemann sum approximation to the dwell time integral. Upon performing such an analysis (see Appendix \ref{sec: prob currents}), we have found that the Tsirelson quantity is related to the probability density current, $J(x,t)=\operatorname{Im}\left(\psi^*\partial_x \psi\right)$, at the origin:
\begin{equation} \label{eq: tsirelson current connection}
    \langle A\rangle = (6\langle T_D(2\pi)\rangle-3)+2\pi J(0,\tau) = 2\pi J(0,\tau) \ ,
\end{equation}
where $\tau$ is some time on the interval $[0,2\pi]$, as the result was obtained using the mean value theorem. Thus, we see that the dependence of $\langle A \rangle$ upon the average dwell time drops out, since $\langle  T(2\pi)\rangle$ assumes the classical value, and we see that a Tsirelson violation necessitates a `large' (i.e. $>1/2\pi$) probability current at the origin. Consequently, this relation restricts the choice of violating states. Conversely, Eq. \ref{eq: tsirelson current connection} provides a condition guaranteeing Tsirelson satisfaction. Specifically, if $|J(0,\tau)|\leq 1/2\pi$ for all $\tau\in[0,2\pi]$, then $|\langle A\rangle|\leq 1$. In Appendix \ref{sec: prob currents}, we also consider classicalisations of the quantum phase-space distribution and the resulting classical probability current. These classicalised currents were consistently lower than the corresponding quantum currents, indicating a connection to the above condition on Tsirelson satisfaction.

\subsection{Crossing number}

Another simple property of classical uniform precession, which is closely linked to Tsirelson's dynamical assumption, is that trajectories cross the origin exactly twice over a period (barring the singular case of a particle sitting at rest at the phase-space origin). Classically, we can compute the number of crossings over a time interval $\tau$ using 
\begin{equation}
    N_c(\tau) = \int_0^{\tau} \mathrm{d}t |p(t)|\delta(x(t)) \ ,
\end{equation}
where $N_c(2\pi) = 2$.

Likewise, we can define the Hermitian quantum operator
\begin{equation} \label{split into S and AS}
    \hat N_c(\tau) = \frac{1}{4} \int_0^{\tau} dt \left(  \{ |\hat p|, \delta(\hat x) \} + \text{sign}(\hat p) \{ |\hat p|, \delta(\hat x) \} \text{sign}(\hat p)  \right) \ ,
\end{equation}
suppressing the time dependence of $\hat x,\hat p$. We note that, written in this form, the first term picks up only symmetric states and the second term only anti-symmetric states when $\tau=2\pi$.

\begin{figure}
    \centering
    \includegraphics[width=0.7\linewidth]{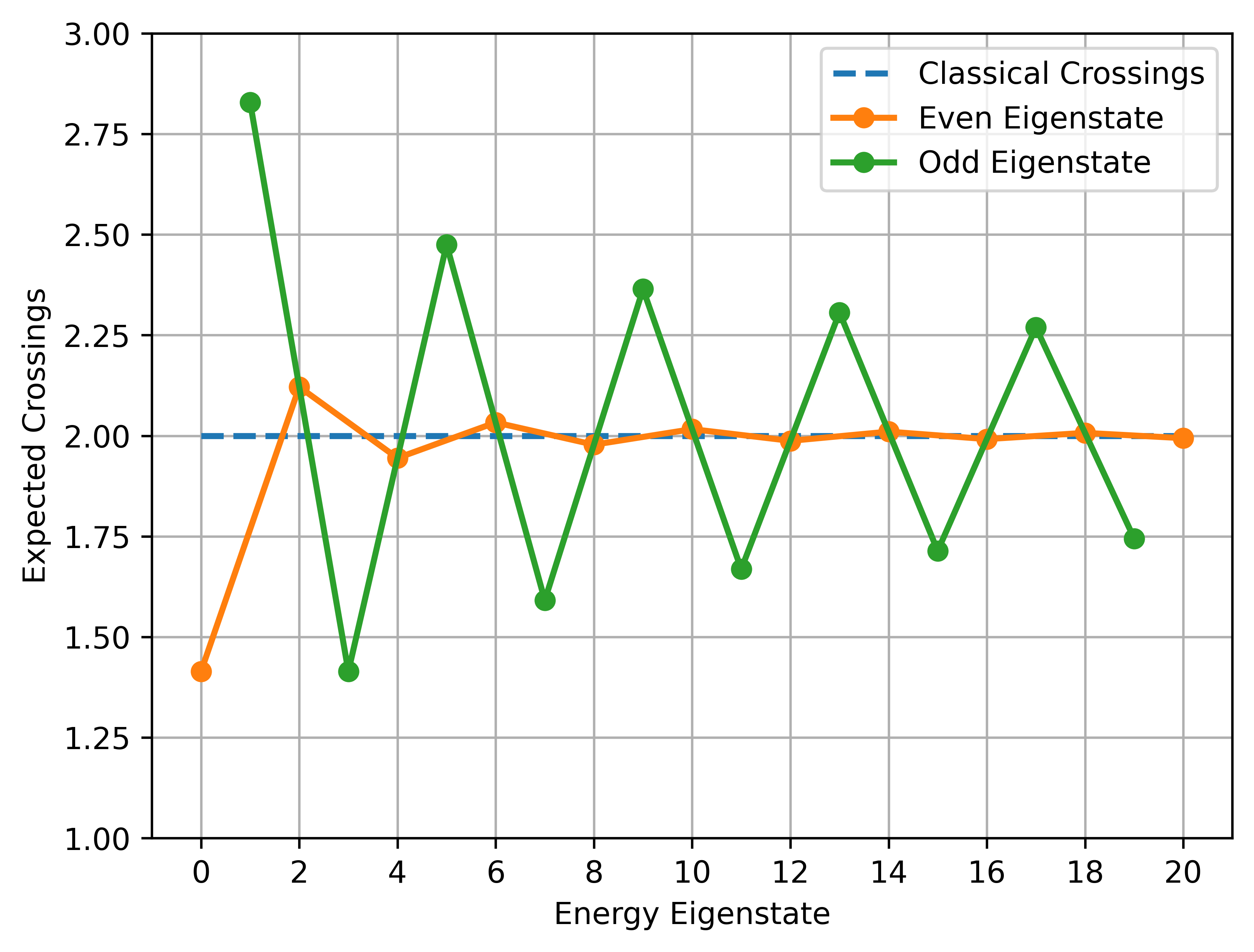}
    \caption{Plot of the average crossing number, $\langle n | N_c(2\pi)|n\rangle$, over a full period for the first 21 energy eigenstates. We see that the even eigenstates converge much more rapidly to the classical value of $2$.}
    \label{fig:crossing number for eigenstates}
\end{figure}

Explicit calculations indicate an asymptotic tendency towards the classical value of two crossings in the limit of high energy eigenstates, which is depicted in Fig. \ref{fig:crossing number for eigenstates}. Moreover, we calculate for the Zaw–Scarani state an average crossing number of $\langle  N_c(2\pi)\rangle_Z \approx 1.49$, with the possible measurement outcomes $\langle 0|\hat N_c(2\pi)|0\rangle = \langle 3|\hat N_c(2\pi)|3\rangle = \sqrt{2}$ and $\langle 6|\hat N_c(2\pi)|6\rangle \simeq 2.033$ (as the energy eigenstates are also eigenstates of $\hat N_c(2\pi)$). So, the spread of possible measurement outcomes is around $0.6$. Over a third of a period, between measurement times, $\langle  N_c(2\pi/3)\rangle_Z \approx 0.50$. These results indicate that the number of crossings in the quantum case for the Zaw–Scarani state is significantly smaller than the classical case.

A potential issue is that measuring the crossing number requires a two-time measurement, which could raise concerns of invasiveness once again. However, we note that, classically,
\begin{equation}
    |p| \delta(x) = \sqrt{p^2+x^2} \delta (x) = \sqrt{2E} \delta(x)
\end{equation}
in a fixed energy state. In turn, the first term of Eq. \ref{split into S and AS} can be taken
to be proportional to $\sqrt{2E}|\psi(0)|^2$, where we include the non-zero ground state energy contribution in $E$. This provides an alternative expression for the crossing number for fixed energy states which only involves a single-time measurement. Using this form, we find that $\langle 0|\hat N_c|0\rangle\approx1.7725$ and $\langle 6|\hat N_c|6\rangle\approx1.997$. These results are reasonably close to those obtained above and exhibit the same qualitative features. However, this method only works for the symmetric states, and we note that states like the Zaw–Scarani state contain a large anti-symmetric part.

\subsection{Implications for trajectory models}

We have seen that the quantum-mechanical properties of the dwell time are identical to its classical properties and that as for the crossing number, its quantum-mechanical properties are qualitatively the same as classical and quantitatively similar. These features suggest that measurements of the dwell time and crossing number may be used to give indirect but useful bounds on the size of uniform precession violations. These quantities are, in principle, related to the three-time probabilities used in Section \ref{Assessing UP}, but any such relation will be non-unique, and we have not been able to find a particularly useful one. We proceed using a more heuristic analysis.
 
We will suppose that there exists an underlying probability distribution on a set of configuration space trajectories $x(t)$ which are close to classical trajectories but could potentially produce a Tsirelson inequality violation by violating uniform precession.  A  trajectory model could produce non-zero values of $p(+,+,+)$ and $p(-,-,-)$ if some fraction of the trajectories are on the same side of the $x$-axis at all three measurement times. We argue that the dwell time and crossing number results limit such alternative explanations.
 
The properties of the dwell time, which hold for any time interval covering a single period and for any initial state, indicate that the underlying trajectories must spend equal times in $x>0$ and $x<0$. Hence, over two thirds of a period, each trajectory cannot spend more than half a period on one side, so must cross at least once between the first and third measurement times. This means that there can be no trajectories remaining on one side for two thirds of a period. So, the only trajectories contributing to $p(+,+,+)$ and $p(-,-,-)$ must, in order to be on the same side at all three measurement times, cross at least twice during the first third of a period, or during the second third, or both. To rule out such wiggly trajectories, we can appeal to the crossing operator results which indicate that highly erratic trajectories have very low probability.

This argument gives an independent understanding as to why uniform precession should hold reasonably well. We have not found a more precise quantitative account to compare with the results of Table \ref{table of measures of UP}.

\subsection{Classical versus quantum dynamics}

Finally, we make some general remarks about dynamics in classical versus quantum theories and why we might expect, on general grounds, that the violation of uniform precession is small.

The predictions of quantum theory are very different to the predictions of classical physics and there are a number of reasons for this. Quantum theory allows a much richer variety of initial states than does classical physics—for example, through Wigner functions that can be negative versus classical phase–space probabilities. In general, there is also a wide variety of possible single-time measurements that can be made that contribute to quantum behaviour. For two-time sequential measurements, the non-commutativity of measurements also contributes.

However, when assessing the degree to which the dynamics of a quantum-mechanical point-particle system is behaving classically, we are only interested in a very limited class of measurements, namely, coarse-grained position or phase–space measurements at different times. For such measurements, the dynamics of classical and quantum systems actually look a lot more similar. There are standard results showing that wave packets and, more generally, phase–space localised quasi-projectors evolve to a good approximation along classical paths (see for example Ref. \cite{Omnes1997} and references therein). These results can be used to show that the probabilities for histories of coarse-grained position or phase–space measurements are strongly peaked around classical paths.

This near-classical behaviour is modified if the measurements significantly chop the evolving wave function, as is the case in this paper (and, for multiple measurements, if they are applied at time intervals shorter than the energy time $\hbar/\langle E \rangle$ \cite{Omnes1997}, but this is not significant here). The magnitude of the non-classical effects producing the chopping can be seen in the specific model of Refs. \cite{MawbySHO,PhysRevA.107.032216}, which exhibited LG2 violations for measurements of $\text{sign}(x)$ for a harmonic oscillator with Gaussian initial state. This was produced  by the LG2 corresponding to the quasi-probability $q(-,+)$, which we write explicitly in the Wigner representation:
\begin{equation}
    q(-,+) = \int dp \ dq \  W_{- +} (p,q)  W_\rho (p,q) \ ,
\end{equation}
where $W_{\rho}$ is the Wigner function of the Gaussian initial state, which is clearly non-negative, and $ W_{- +} $ is the Wigner–Weyl transform of the operator  $ (1/2) \{\theta(\hat x(t_1)), \theta( \hat x(t_2) \} $, given in Refs. \cite{tsirelson_how_2006,PhysRevA.107.032216}. The classical analogue of $ W_{- +} (p,q)$ is clearly $ \theta ( x(t_1) \theta(x(t_2)) $, which is trivially non-negative; but $ W_{- +} (p,q)$ has regions of negativity which arise from the non-commutativity of the two operators, and this is the only source of non-classicality. (Physically, this is related to the diffraction-in-time effect \cite{PhysRevA.107.032216,PhysRev.88.625, moshinsky_diffraction_1976}.) Most significantly, the violation produced is only $22\%$ of the maximum possible violation (the L{\"u}ders bound). By contrast, near-maximal LG violations are possible in this system with, for example, an initial state consisting of the first excited state of the harmonic oscillator, which has a Wigner function with substantial regions of negativity. (More speculatively, we also note that the size of the uniform precession violation here is roughly $4\%$, which is peculiarly similar to the estimated bound on the so-called quantum back-flow effect \cite{bracken_probability_1994,halliwell_quantum_2013,PhysRevA.99.012124,trillo_quantum_advantage}.)

This example indicates why we might expect that, for the very limited set of measurements used to check proximity to classical behaviour, the non-classical effects arising from those measurements will generally not match the level of non-classicality arising from the choice of initial state. These assertions could be checked experimentally using control experiments in which the dynamics of a system prepared in a set of classical-like states are measured.

So, in summary, there are general physical and experimentally-testable reasons why we might expect the violation of a dynamical assumption like uniform precession to be small compared to the more significant non-classical effects that can arise from the initial state.

\section{Summary and Discussion} \label{Conclusion}

The Tsirelson inequality makes clear that, by introducing a classically-motivated dynamical assumption, the prospect of detecting quantum behaviour in temporal correlations need not be plagued by potentially invasive sequential measurements. In this paper, we have placed the Tsirelon inequality within the wider framework of Leggett-Garg tests for macrorealism, thereby yielding a method for an assessment of the dynamical assumption.

In Section \ref{Tsirelson Violations in the QHO}, we analysed violating states in an energy-truncated subspace and formalised the Tsirelson operator as the twirl of the sign operator over the group elements of the three-element cyclic group. This perspective directly explained its eigenspace structure and opens up the possibility for more tools from quantum information to provide further insight into the Tsirelson operator. We provided proofs that a superposition of at least three energy eigenstates of the QHO is required to produce a Tsirelson violation and that, in relevant subspaces, all three correlators are equal for the specific measurement time interval of two-thirds of a period. The violations possible with simple states such as the Zaw-Scarani state—and, moreover, superpositions of coherent states—are certainly promising and within the reach of experimentalists in the coming years.

But, as we noted in Section \ref{Introduction}, there is a crucial loophole in the Tsirelson inequality: particularly, that violations could ensue from the failure of the dynamical assumption of uniform precession in a classical system, meaning that a Tsirelson violation is not a sufficient condition for genuine quantum behaviour. Therefore, in this paper we have developed a subsidiary protocol to disentangle violations of macrorealism from the breaking of the dynamical assumption, to which end we have embedded the Tsirelson inequality within the wider Leggett–Garg formalism.

In Section \ref{Assessing UP}, we broke apart the Tsirelson quantity into separate terms that quantified the breaking of uniform precession and quantum interferences, such that their sizes may be compared. Recognising that uniform precession can be defined quantum-mechanically in several inequivalent ways in terms of measurable probabilities, we investigated two methods. In the first method, we proved the surprising fact that violating eigenstates of the Tsirelson operator satisfy all four LG3 inequalities, meaning that there invariably exists a macrorealistic description of sign changes, which can then be used to assess uniform precession through an LG3 test. In the second method, we considered a generalised dynamical assumption with the difference of no-sign-change probabilities (perhaps less desirable than the first method, for it involves three sequential measurements). In all cases, we found success with the protocol, demonstrating that quantum interference terms must be implied, elevating the Tsirelson test from a simple test of non-classical behaviour to a more robust demonstration of quantumness.

In Section \ref{Heuristic definitions}, by considering the heuristic notions of dwell times and the crossing number operator, we presented further qualitative evidence that uniform precession is satisfied well in states that violate the Tsirelson inequality. We found that the quantum-mechanical dwell time obeys the classical bound exactly and that the total number of origin crossings is approximately the classical value. On their own, these stand as interesting mathematical results, further elucidating the various ways in which uniform precession is and is not equivalent in the classical and quantum cases. But we have also further interpreted these quantities in the context of underlying trajectory models, noting that they greatly restrict the extent to which underlying trajectories can be erratic. We also came upon the global property that Tsirelson violations are linked to the probability currents through the origin, whereby weakly-flowing states, or the classicalised currents of quantum states, never produce a violation.

We conclude with a discussion on the key issue of whether or not the need for sequential measurements has really been eliminated—a virtue of the Tsirelson inequality we so admired in the first place—if, at the very least, a Leggett–Garg test is still required to assess uniform precession. Is non-invasive measurement then re-invoked as an untoward assumption alongside \textit{macrorealism per se} and uniform precession? At one level, this paper can be regarded as a purely quantum-mechanical analysis of Tsirelson violations, and we have found that they arise largely from the presence of interferences and a genuine failure of macrorealism, but with a small contribution stemming from UP violations.

However, an ultimate goal for Tsirelson tests will be to probe systems not knowing beforehand whether they are quantum or not. Then, to an experimentalist—who should wish to implement this protocol by collecting a data set of the Tsirelson quantity supplemented with a Leggett–Garg test for the relevant probabilities—the problem of re-invoking a Leggett–Garg test, in order that we should further conclude that \textit{macrorealism per se} has been violated, is possibly harder to justify. But we may appeal here to the heuristic results of Section \ref{Heuristic definitions}, which provide evidence that the LG3 quantity $L_1$ should be small and which support the idea that violations of uniform precession are not, in general, sufficiently large to be the primary cause of Tsirelson inequality violations. In some cases, we were partially successful in showing how these quantities may be measured through only single-time measurements. In the original Leggett–Garg scheme, there is no reason to expect \textit{a priori} that $L_1$ should be small; here, however, we have gathered evidence that it should indeed be small (and positive), and so if an experiment finds otherwise, there is reason to suspect that measurement invasiveness may be largely to blame.

The results of this paper open up several new research directions. Future work could consider an anharmonic oscillator, multiple coupled oscillators and links to field theories, entanglement, alternative measures of crossings, and stronger quantum-mechanical definitions of uniform precession, amongst other things. We have concentrated in this work upon the case of the harmonic oscillator, but it would be interesting to conduct this same analysis and assessment of uniform precession for spin systems; subsequent comparison with the results herein obtained could be insightful. We have considered the use smoothed projectors, but these order of magnitude arguments need to be bolstered by a more detailed calculation, similar to that of Ref. \cite{MawbySHO}. The analysis of Tsirelson’s inequality and the breaking of uniform precession could be extended to incorporate suspected connections with the quantum back-flow limit. This could all provide further illumination on the true nature of violations of Tsirelson’s intriguing inequality.

\begin{acknowledgments}
We are grateful to Clement Mawby, Shayan Majidy and Lin Htoo Zaw for useful discussions and comments on earlier versions of this manuscript, and the latter for bringing to our attention a result regarding superpositions of coherent states.
\end{acknowledgments}

\appendix
\onecolumngrid

\section{Tsirelson Quantity from the Wigner Function} \label{Wigner function appendix}
The Tsirelson quantity can also be written in terms of the Wigner quasiprobability distribution $W_\rho(x,p)$ as
\begin{equation}
    \langle A \rangle = \text{Tr}(\rho \hat A)=\int\int dx \ dp \ W_A(x,p)W_\rho(x,p) \ ,
\end{equation}
where $W_A(x,p)$ (and likewise $W_\rho(x,p)$) is the phase-space representation of the operator $\hat A$, obtained from its Weyl transform:
\begin{equation}
    W_A(x, p) = \int_{-\infty}^{\infty} dy\, e^{-i p y / \hbar} \left\langle x + \frac{y}{2} \middle| \hat A \middle| x - \frac{y}{2} \right\rangle \ .
\end{equation}
This formula is valid for any Hermitian operator.

\section{Single crossing dynamical assumption} \label{Single crossing assumption}

An alternative set of inequalities can be formulated under the single crossing assumption (SCA), which is an alternative dynamical assumption that instead asserts that $p(+,-,+)=p(-,+,-)=0$, forbidding that the particle cross the origin more than once. A very rapidly oscillating system could cross many times in between measurements whilst maintaining $p(+,-,+)=p(-,+,-)=0$, but we assume that the relative amplitudes of these erratic trajectories are small. In fact, similar inequalities can be obtained using the polytope method along with dynamical constraints over a full period as shown in Ref. \cite{plavala_tsirelson_2024}.

SCA can be imposed in the same way as was done uniform precession in the original Tsirelson derivation, by setting $p(+,-,+)=p(-,+,-)=0$. Using the moment expansions \ref{moment expansion 3-time}, it is straightforward to obtain
\begin{equation} \label{SCA Tsirelson}
    -1 \leq \langle Q_1 \rangle - \langle Q_2 \rangle + \langle Q_3 \rangle \equiv \langle A_\text{SCA} \rangle \leq 1 \ .
\end{equation}

One of the most interesting features of the original Tsirelson inequality, we found, was that it could be derived from the weaker condition that $p(+,+,+)=p(-,-,-)$. In fact, there is an exact analogue of this for the SCA, whereby Eq. \ref{SCA Tsirelson} follows directly from the weaker condition $p(+,-,+)=p(-,+,-)$.

Classically, one of the simplest examples of a system for which the SCA is valid would be a free particle. Acted upon by no external forces, a free particle will only ever cross a pre-defined origin at most once. In the case of a uniformly-precessing system, such as the harmonic oscillator, this assumption holds true over half a period. We choose measurement times $t_j=\frac{jT}{6}$, $j=0,1,2$, where $T$ is the time period of the oscillator.

We repeat the same analysis as before, defining an operator
\begin{equation}
    \hat A_\text{SCA} = \hat{Q}_{t=0}  -  \hat{Q}_{t=1/6}  +  \hat{Q}_{t=1/3} \ ,
\end{equation}
where we have set the time period to unity for convenience. What we will now proceed to show is that imposing SCA over half a period is mathematically identical to imposing uniform precession over a full period, in that
\begin{equation}
    \langle A_\text{SCA} \rangle = \langle A \rangle \ .
\end{equation}

Since $\langle A \rangle=\langle {Q}_{t=0} \rangle + \langle {Q}_{t=1/3} \rangle + \langle {Q}_{t=2/3} \rangle $ and $\langle A_\text{SCA} \rangle=\langle {Q}_{t=0} \rangle - \langle {Q}_{t=1/6} \rangle + \langle {Q}_{t=1/3} \rangle $, we see that these two quantities only differ by one term. If indeed it is true that $\langle A \rangle=\langle A_\text{SCA} \rangle$ then it is implied that $\langle {Q}_{t=2/3} \rangle = - \langle {Q}_{t=1/6} \rangle$. In fact, the more general result holds true that $\langle {Q}_{t} \rangle = - \langle {Q}_{t+1/2} \rangle \ \forall \ t$. That is to say, the single-time average of the dichotomic variable $Q$ merely flips sign after half a period.

We can understand why this is the case by considering what happens to the energy eigenfunctions after exactly half a period. Undergoing unitary time evolution in accordance with the QHO Hamiltonian, the eigenstate $|n\rangle$ accrues a relative phase of $e^{-i\pi n}$ after half a period. As such, the odd-$n$ eigenstates will pick up a minus sign, whilst the even-$n$ states are unaffected. In the language of wavefunctions, it is simple to see that this means a reflection across the $y$-axis after exactly half a period. Therefore, quantum mechanically, the quantities $\langle A \rangle$ and $\langle A_\text{SCA} \rangle$ are \textit{identical}.

\section{Matrix Elements in an Energy Eigenbasis} \label{sec: matrix elements E}
Let
\begin{equation}
|\Psi\rangle=\sum_{n=0}^{\infty}c_n|n\rangle,\qquad
 \hat Q = 2 \theta(\hat x)-\mathds{1} ,
\end{equation}
where $\theta(\hat x)$ is the Heaviside projector on $x>0$.
To obtain explicit matrices we truncate the Hilbert space at a finite excitation
$N$:
\begin{equation}
|\Psi\rangle=\sum_{n=0}^{N}c_n|n\rangle,\qquad
\mathbf c=(c_0,\ldots,c_N)^{\!\mathsf T}.
\end{equation}
For any operator $ \hat O$ its expectation value reduces to a quadratic form,
\begin{equation}
\langle\Psi| \hat O|\Psi\rangle
  =\sum_{n,k=0}^{N}c_n^{*}\,O_{nk}\,c_k
  =\mathbf c^{\dagger}\,O\,\mathbf c,
\qquad
O_{nk}:=\langle n|\hat O|k\rangle .
\end{equation}
Hence the task is to evaluate the matrix
$Q_{nk}=\langle n| \hat Q|k\rangle$.
Using the position representation and energy eigenstate wavefunctions $\Psi_n(x)$,
\begin{equation}
Q_{nk}
  =\!\int_{-\infty}^{\infty}\!\!\Psi_n(x)\Psi_k(x)\,
       \bigl[2\theta(\hat x)-1\bigr]\,\mathrm dx \ , 
\end{equation}
using the property that the energy eigenstates for the 1D harmonic oscillator can always be taken to be real to drop the complex conjugate on $\Psi_n(x)$. To make the integral tractable, we split the integral at the origin and use the overlap formula in the appendix of Ref.\cite{MawbySHO}, 
\begin{equation} \label{eq: Fnk}
F_{nk}\equiv\int_{0}^{\infty}\Psi_n(x)\Psi_k(x)\,\mathrm dx
       =\frac{\Psi_k'(0)\Psi_n(0)-\Psi_n'(0)\Psi_k(0)}{2(k-n)}
       \quad(n\neq k),
\end{equation}
where $F_{nk} = \langle n| \theta(\hat x)|k\rangle$ and $F_{nn}=\frac{1}{2}$. Therefore, we find
\begin{equation}
Q_{nk}=2F_{nk},\qquad Q_{nn}=0.
\end{equation}
To evaluate Eq. \ref{eq: Fnk}, we use the closed form expressions for the eigenstates of the harmonic oscillator in terms of the Hermite polynomials $H_n(x)$,
\begin{equation}
    \Psi_n(x) = \frac{\pi^{-\frac{1}{4}}}{\sqrt{2^n n!}}\exp\left(-\frac{1}{2}x^2\right)H_n(x).
\end{equation}
Consequently, using the following identities for the derivatives of Hermite polynomials,
\begin{equation}
H'_0(0)=0, \qquad H_{n}'(0)=2n\,H_{n-1}(0) \quad (n\geq 1),
\end{equation}
we find
\begin{equation} \label{eq: Qnk}
Q_{nk}=
\frac{2\pi^{-1/2}}
     {(k-n)\sqrt{2^{\,n+k}n!\,k!}}
     \bigl[k\,H_{k-1}(0)H_{n}(0)-n\,H_{n-1}(0)H_{k}(0)\bigr]
     \quad(n\neq k).
\end{equation}
Finally, since \( U|n\rangle=e^{-2\pi i n/3}|n\rangle\),
\begin{equation}
A_{nk}=Q_{nk}\Bigl[1+e^{\tfrac{2\pi i}{3}(n-k)}
                       +e^{\tfrac{4\pi i}{3}(n-k)}\Bigr].
\end{equation}
Our numerical implementation of the sign operator and Tsirelson operator on Sympy, the symbolic Python maths package, also employed the following identities:
\begin{equation} \label{eq: H identities}
H_{2m}(0)=(-2)^{m}(2m-1)!!,\quad
H_{2m+1}(0)=0.
\end{equation}
We verified this implementation by checking matrix elements of the $F_{nk}$ agreed with those cited in Ref.\cite{MawbySHO} and our implementation of the Tsirelson matrix was verified by checking some known violating states.

\section{Diagonalisation of the Tsirelson operator in an energy eigenbasis} \label{Diagonalising appendix}

Note that when $\tau=\frac{2\pi}{3}$, the term 
\begin{equation}
    1 + e^{i\tau(n-k)} + e^{2i\tau(n-k)} = 1 + \cos(\frac{2\pi}{3}(n-k)) + \cos(\frac{4\pi}{3}(n-k)) + i\sin(\frac{2\pi}{3}(n-k)) + i\sin(\frac{4\pi}{3}(n-k))
\end{equation}
in Eq. \ref{A matrix elements} vanishes unless the difference $(n-k)$ is a multiple of 3. Moreover, not only does the Zaw–Scarani state produce the largest violations state in the $N=6$ subspace, but also in the $N=7$ and $N=8$ subspaces. For $N=9$, on the other hand, we can achieve a superior violation of $\langle A \rangle\approx1.12$ with the state 
\begin{equation} \label{N=9 max. violating state}
    |\Psi_9\rangle \approx 0.6277|0\rangle -0.7064|3\rangle + 0.3255|6\rangle + 0.0304|9\rangle \ .
\end{equation}
We resort to an approximate numerical specification of the coefficients, because neat analytic forms are intractable. Importantly, any numerical errors are encapsulated in the floating point specification of these coefficients, and the remaining coefficients ($c_1,c_2,c_4,\text{etc...}$) really are exactly and identically zero.

\begin{figure}[H]
    \centering
    \includegraphics[width=0.5\linewidth]{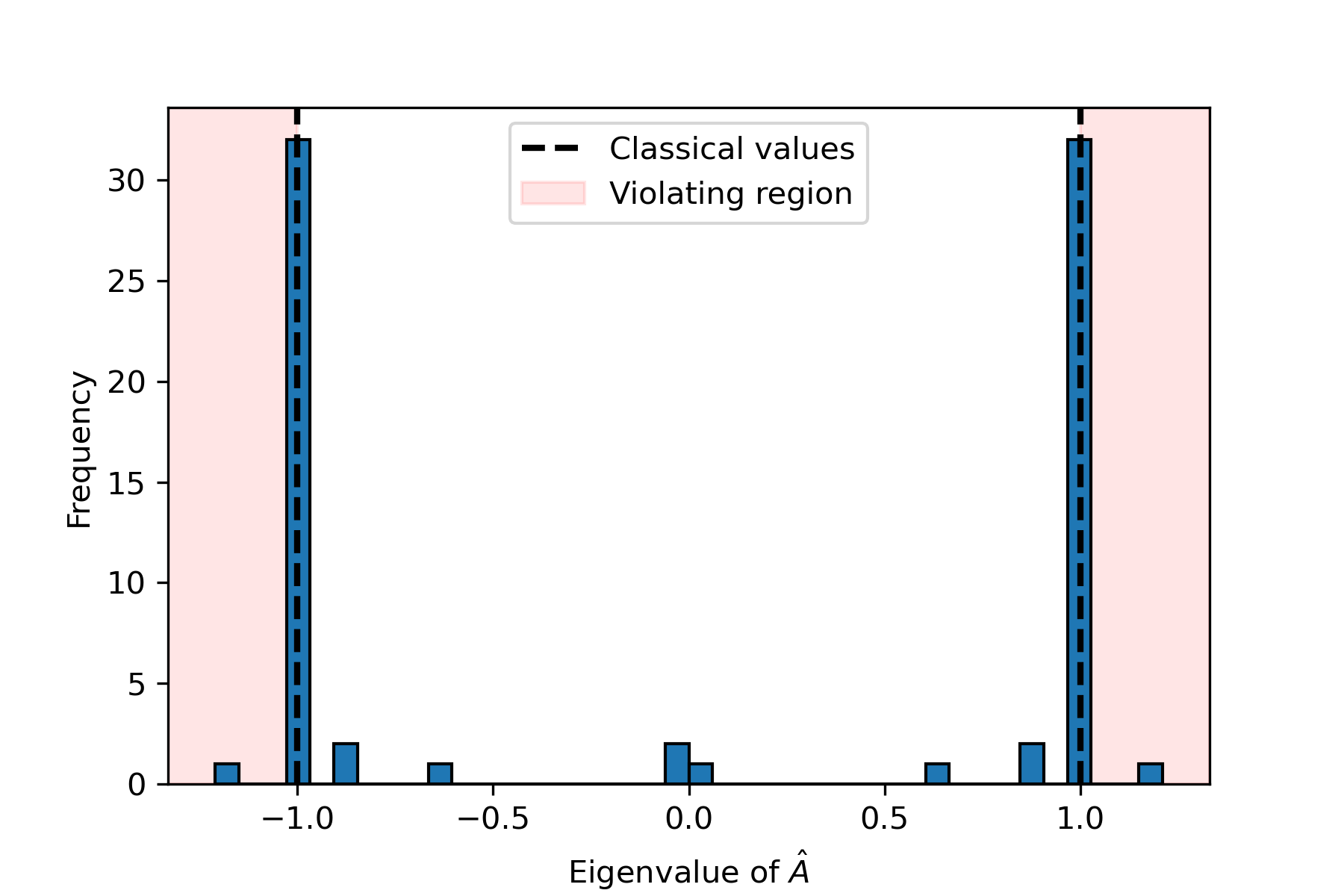}
    \caption{A histogram of the spectrum of eigenvalues of $\hat A$ is plotted by using an energy truncation of $E=70(\hbar\omega)$. The y-axis refers to the number of eigenvalues found within a given range.}
    \label{fig:eigenspectrum of A}
\end{figure}

\begin{figure}[H]
    \centering
    \includegraphics[width=0.4\linewidth]{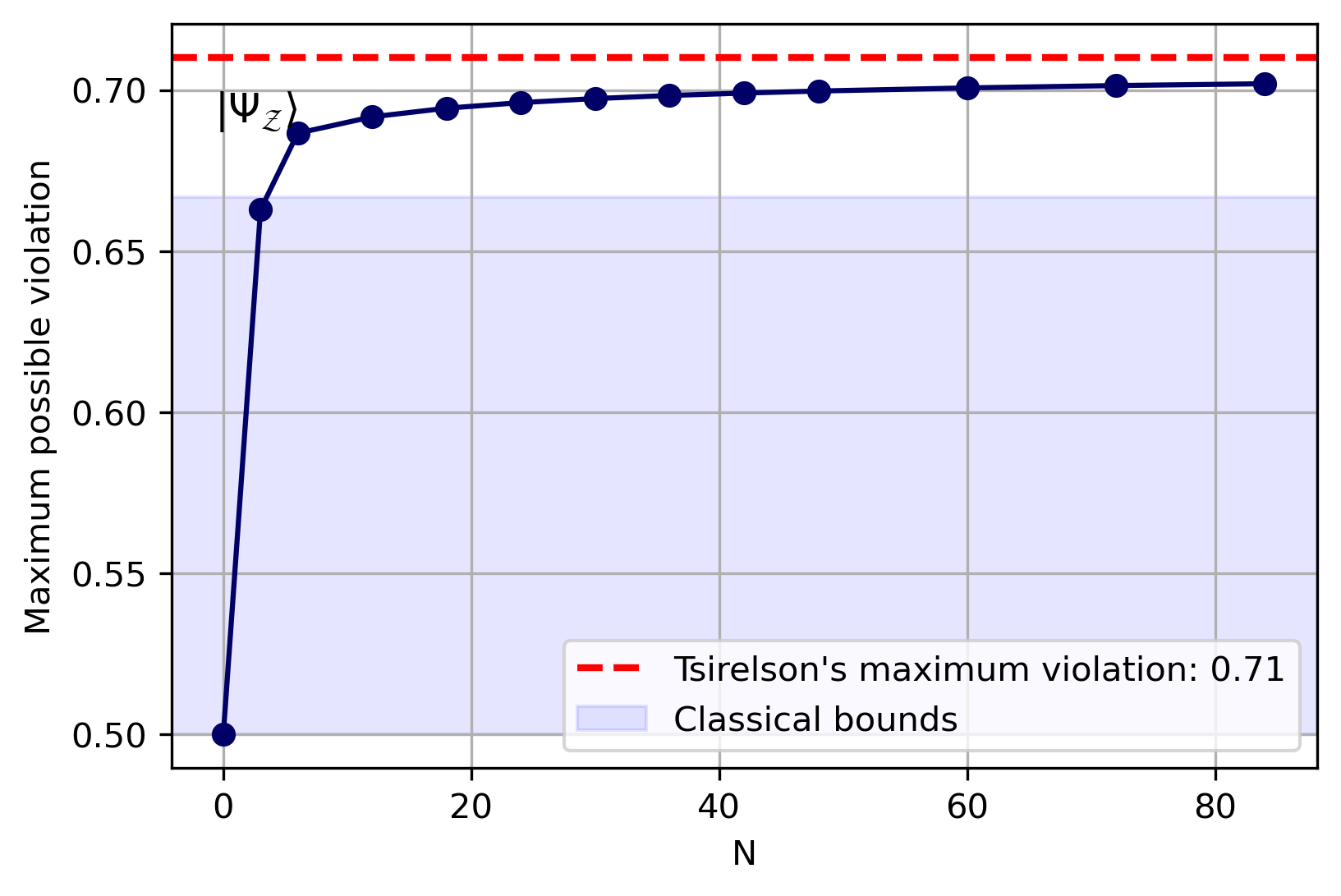}
    \caption{The computed maximal eigenvalue of $\hat A$ in an increasingly large subspace, tending towards a finite maximal violation. Note, importantly, that we re-scale the eigenvalues $\lambda$ as $\frac{1}{6}(\lambda+3)$. This is only an arbitrary scaling choice to see agreement with the definition of the operator considered by Tsirelson in Ref. \cite{plavala_tsirelson_2024}. The Zaw–Scarani state ($|\Psi_Z\rangle$) is pointed out at $N=6$.}
    \label{fig:Convergence to Tsirelson's value}
\end{figure}

Fig. \ref{fig:eigenspectrum of A} shows the results of a spectral decomposition of the Tsirelson operator in a high-energy truncated subspace. This serves as a good illustration that the classical values $A=\pm1$ are ``accumulation points" \cite{tsirelson_how_2006}, but violations past this value are indeed possible. None of our computed eigenvalues exceed $\approx 1.26$, which agrees with the value of the maximum possible violation quoted in Tsirelson's original paper \cite{tsirelson_how_2006}. A more accurate estimate of the true upper and lower bounds on the spectral norm of $\hat A$ has been obtained in the recent work of Ref. \cite{zaw2024threeanglevariantstsirelsonsprecession}. Fig. \ref{fig:Convergence to Tsirelson's value} illustrates that considering a progressively larger subspace from the low-lying energy states upwards does indeed lead to an asymptotic convergence towards Tsirelson's quoted value (up to an arbitrary scaling factor) for the maximum violation in the full QHO Hilbert space. 

We have also computed the eigenvectors of $\hat A$ in increasingly larger subspaces, catalogued in Table \ref{Table of eigenvectors}, and have found that they all obey the same pattern of being composed solely of energy eigenstates that are multiples of three.

\begin{table}[H]
    \centering
    \renewcommand{\arraystretch}{2} 
    \begin{tabular}{c|ccccccc|c} 
        \Xhline{2\arrayrulewidth}
        $N$ & $c_0$ & $c_3$ & $c_6$ & $c_9$ & $c_{12}$ & $c_{15}$ & $c_{18}$ & $\langle A \rangle$ \\
        \hline
        3 & $-\frac{1}{\sqrt{2}}$ & $ \frac{1}{\sqrt{2}} $ & 0 & 0 & 0 & 0 & 0 & 0.9772\\
        \hline
        6 & $\frac{4}{\sqrt{42}}$ & $ -\frac{1}{\sqrt2} $ & $\sqrt{\frac{5}{42}}$ & 0 & 0 & 0 & 0 & 1.1195\\
        \hline
        9 & 0.6277 & -0.7064 & 0.3255 & 0.0304 & 0 & 0 & 0 & 1.1200 \\
        \hline
        12 & -0.6529 & 0.6600 & -0.1591 & -0.2538 & 0.2200 & 0 & 0 & 1.1500\\
        \hline
        15 & -0.6509 & 0.6591 & -0.1592 & -0.2561 &  0.2259 & -0.0083 & 0 & 1.1500\\
        \hline
        18 & -0.6655 & 0.6354 & -0.0999 & -0.2771 &  0.1451 & 0.1396 & -0.1615 & 1.1661\\
        \arrayrulecolor{gray}
        \hline
        \vdots & & & & \vdots & & & & \vdots \\
        \hline
        90 & \multicolumn{7}{c|}{$|\Psi\rangle \rightarrow$ True eigenstate of ${A}$} &  1.2131\\
        \arrayrulecolor{black}
        \Xhline{2\arrayrulewidth}
    \end{tabular} 
    \caption{Explicit computation of maximally violating states in a truncated energy subspace ($|\Psi_{N}\rangle = \sum_{n=0}^N c_n |n\rangle$). The only non-zero coefficients are $c_{3j}$, where $j\in\mathbb{N}$.}
    \label{Table of eigenvectors}
\end{table}

\section{Superpositions of Two Energy Eigenstates} \label{Energy Eigenstate Superpositions Appendix}
As established in Ref.\cite{tsirelson_how_2006}, single energy eigenstates of the harmonic oscillator cannot violate Tsirelson's inequality due to their definite parity. The next logical step is to find violations produced by a superposition of two eigenstates. However, as we show now, any superposition of two energy eigenstates is insufficient to produce a violation.

\subsection{Proof Strategy}
We aim to show that the magnitude of $\langle A\rangle$ is less than $1$ for the state
\begin{equation}
    |\psi_2\rangle = \alpha |n\rangle + \beta |k\rangle,
\end{equation}
where $|\alpha|^2+|\beta|^2=1$. The expectation value of $ \hat A$ for $|\psi_2\rangle$ is 
\begin{equation}
    \langle A\rangle = |\alpha|^2\underbrace{\langle n| A |n\rangle}_{0} + |\beta|^2 \underbrace{\langle k|  A|k \rangle}_{0} + 2\operatorname{Re}(\alpha^* \beta) \langle n | A|k\rangle.  
\end{equation}
Expanding out $\langle A\rangle$ in terms of $Q_{nk}$ yields
\begin{equation}
    \langle A\rangle = 2\operatorname{Re}(\alpha^* \beta) Q_{nk} \left(1+e^{\frac{2\pi i}{3}(n-k)}+e^{\frac{4\pi i}{3}(n-k)}\right).  
\end{equation}
The sum of exponentials is 3 when $n-k = 0 \operatorname{mod} 3$ and 0 otherwise. So, $\langle A\rangle$ is clearly maximised when $(n-k)$ is a multiple of 3. Moreover, $|\alpha|^2+|\beta|^2 = 1$ puts a constraint on the maximum value of $\operatorname{Re}(\alpha^* \beta)$. Intuitively, the interference term will be maximised if the state is an equal superposition of $|n\rangle$ and $|k\rangle$. Therefore, the maximum value of $\operatorname{Re}(\alpha^* \beta)$ is $\frac{1}{2}$. To be precise, consider the following short proof: Let $\alpha=ce^{i\theta}$ and $\beta=se^{i\phi}$. Then $\operatorname{Re}(\alpha^*\beta)=cs\cos(\phi-\theta)$. $\cos(\phi-\theta)$ is maximised if $\phi=\theta$. Thus we can choose $\alpha=c$ and $\beta=s$, up to a global phase. Then, the constraint $c^2+s^2=1$ means we can parametrise $c$ and $s$ as $c=\cos(\xi)$ and $s=\sin(\xi)$. Therefore, $cs = \frac{1}{2}\sin(2\xi)$, which is maximised if $\sin(2\xi)=1$. Hence, $\operatorname{max}(cs) = \operatorname{max}\left(\operatorname{Re}(\alpha^*\beta)\right) = \frac{1}{2}$. Thus, $\langle A\rangle$ for an equal superposition of $|\psi_2\rangle = \frac{1}{\sqrt{2}}(|n\rangle +|k\rangle)$ with $n-k=0\operatorname{mod} 3$ is
\begin{equation}
    \langle A\rangle = 2 \left(\frac{1}{2}\right) Q_{nk} (3) = 3Q_{nk}.
\end{equation}
We see that the problem of showing that $|\langle A\rangle|<1$ reduces to showing that
\begin{equation}
    |Q_{nk}|<\frac{1}{3} \quad \forall n,k \quad \text{s.t} \quad k-n=3m, \text{ for }m \in \mathbb{N} \ ,
\end{equation}
where we only need to consider $m\geq 0$ because $Q_{nk}$ is symmetric. One might conjecture that the $Q_{03}$ matrix element will be the largest in magnitude that meets the $k-n=3m$ condition. If that holds then our proof would be complete, as $|Q_{03}|\approx 0.32<\frac{1}{3}$ (computed using Eq. \ref{eq: Qnk}). Therefore, the strategy we choose to take is to show two things about $Q_{nk}$:
\begin{enumerate}
    \item The largest matrix element, in magnitude, on a diagonal i.e for fixed $m$, $|Q_{n,n+2m+1}|$ is $|Q_{0,2m+1}|$.
        \begin{equation}
            Q_{nk} = 
                \begin{pmatrix}
                    0 & \fcolorbox{red}{yellow}{$Q_{01}$} & 0 & Q_{03}      & \cdots \\[6pt]
                    Q_{10} & 0 & \fcolorbox{red}{yellow}{$Q_{12}$} & 0      & \cdots \\[6pt]
                    0 & Q_{21} & 0& \fcolorbox{red}{yellow}{$Q_{23}$} & \cdots \\[6pt]
                    Q_{30} & 0 & Q_{32} & 0 & \cdots \\[6pt]
                    \vdots & \vdots & \vdots & \vdots & \ddots 
                \end{pmatrix}.
        \end{equation}
    \item The non-zero matrix elements decrease in magnitude across the 0th row i.e $|Q_{0,2m+1}|$ is a decreasing function in $m$.
        \begin{equation}
            Q_{nk} = 
                \begin{pmatrix}
                    0 & \fcolorbox{red}{yellow}{$Q_{01}$} & 0 & \fcolorbox{red}{yellow}{$Q_{03}$}     & \cdots \\[6pt]
                    Q_{10} & 0 & Q_{12} & 0 & \cdots \\[6pt]
                    0 & Q_{21} & 0& Q_{23} & \cdots \\[6pt]
                    Q_{30} & 0 & Q_{32} & 0 & \cdots \\[6pt]
                    \vdots & \vdots & \vdots & \vdots & \ddots 
                \end{pmatrix}.
        \end{equation}
\end{enumerate}
We already have established that the only non-zero matrix elements of the $Q_{nk}$ are those with opposite parity, and we ignore elements of the same parity when claiming these sequences are decreasing. Once those two statements are proven, we will have shown that $Q_{01}$ is the largest matrix element. This does not satisfy the condition $k-n=3m$, though, and so we take the next largest matrix element, $Q_{03}$, thereby completing the proof.

\subsection{Diagonals of the \texorpdfstring{$Q_{nk}$}{Qnk} matrix}
Showing the first statement is slightly complicated by the fact that the even $n$ and odd $n$ matrix elements have different formulas. When $n$ is even (and hence $k$ is odd) the relevant formula (obtained using Equations \ref{eq: Qnk} and \ref{eq: H identities}) is
\begin{equation}
    |Q^\text{even}_{nk}| =  \left(\sqrt{\frac{2}{\pi}}\frac{(n-1)!!}{\sqrt{n!}}\right) \frac{(k+1)!}{(k-n) \sqrt{k!} 2^{\frac{k+1}{2}} \left(\frac{k+1}{2}\right)!}.
\end{equation}
When $n$ is odd (and hence $k$ is even) the relevant formula is 
\begin{equation}
    |Q^\text{odd}_{nk}| = \left(\sqrt{\frac{2}{\pi}}\frac{n!!}{\sqrt{n!}}\right) \frac{k!}{(k-n) \sqrt{k!}2^{\frac{k}{2}} \left(\frac{k}{2}\right)!}.
\end{equation}
To resolve this complication we consider the odd and even indexed subsequences separately. We define the sequence $\{a_n\}_{n=0}^\infty$, where $a_n = |Q_{n,n+2m+1}|$ for fixed $m$. Then, we just need to show the even indexed subsequence is decreasing ($a_{2j+2}<a_{2j}$) and that when the next element of the sequence is odd-indexed it is less than the previous even-indexed element ($a_{2j+1}<a_{2j}$).

\paragraph{Even Subsequence is Decreasing:} Consider the fraction $a_{2j+2}/a_{2j}$ and show it is less than 1. 
\begin{equation}
    \frac{a_{2j+2}}{a_{2j}} = \left|\frac{Q_{2j+2,2j+2m+3}}{Q_{2j, 2j+2m+1}}\right|
\end{equation}
\begin{equation}
    |Q_{2j+2, 2j+2m+3}| = \frac{(2j+1)!!}{\sqrt{(2j+2)!}}\frac{(2j+2m+4)!}{(2m+1)\sqrt{(2j+2m+3)!}2^{j+m+2}(j+m+2)!}
\end{equation}
\begin{equation}
    |Q_{2j, 2j+2m+1}| = \frac{(2j-1)!!}{\sqrt{(2j)!}}\frac{(2j+2m+2)!}{(2m+1)\sqrt{(2j+2m+1)!}2^{j+m+1}(j+m+1)!}
\end{equation}
Therefore,
\begin{align}
    \frac{a_{2j+2}}{a_{2j}} &= \frac{(2j+1)}{\sqrt{(2j+2)(2j+1)}}\frac{(2j+2m+4)(2j+2m+3)}{\sqrt{(2j+2m+3)(2j+2m+2)}2(j+m+2)} \\
    &= \sqrt{\frac{(4j^2+4jm+8j)+2m+3}{(4j^2+4jm+8j)+4m+4}} \\
    \frac{a_{2j+2}}{a_{2j}}&< 1 \ . \qed
\end{align}

\paragraph{The Next Odd Element is less than the Preceding Even Element:} Using a very similar argument to above, consider the fraction $a_{2j+1}/a_{2j}$ and show it is less than 1.
\begin{equation}
    \frac{a_{2j+1}}{a_{2j}} = \left|\frac{Q_{2j+1,2j+2m+2}}{Q_{2j, 2j+2m+1}}\right|
\end{equation}
\begin{equation}
    |Q_{2j+1, 2j+2m+2}| = \frac{(2j+1)!!}{\sqrt{(2j+1)!}}\frac{(2j+2m+2)!}{(2m+1)\sqrt{(2j+2m+2)!}2^{j+m+1}(j+m+1)!}
\end{equation}
\begin{equation}
    |Q_{2j, 2j+2m+1}| = \frac{(2j-1)!!}{\sqrt{(2j)!}}\frac{(2j+2m+2)!}{(2m+1)\sqrt{(2j+2m+1)!}2^{j+m+1}(j+m+1)!}
\end{equation}
Therefore,
\begin{align}
    \frac{a_{2j+1}}{a_{2j}} &= \left(\frac{(2j+1)!!}{(2j-1)!!}\right)\sqrt{\frac{(2j)!}{(2j+1)!}} \sqrt{\frac{(2j+2m+1)!}{(2j+2m+2)!}}\\
    &= \sqrt{\frac{(2j+1)}{(2j+1)+2m+1}} \\
    \frac{a_{2j+1}}{a_{2j}}&< 1\qed
\end{align}

\subsection{0th Row of the \texorpdfstring{$Q_{nk}$}{Qnk} Matrix is Decreasing}
The second thing to show is that $Q_{0,2m+1}$ is a decreasing function in $m$. To accomplish this, consider the fraction $|Q_{0,2m+3}/Q_{0,2m+1}|$ and show it is less than 1.
\begin{equation}
    Q_{0, 2m+3}=\left(\sqrt{\frac{2}{\pi}}\frac{(-1)!!}{\sqrt{0!}}\right) \frac{(2m+4)!}{(2m+3) \sqrt{(2m+3)!} 2^{m+2} \left(m+2\right)!}
\end{equation}
\begin{equation}
    Q_{0, 2m+1}=\left(\sqrt{\frac{2}{\pi}}\frac{(-1)!!}{\sqrt{0!}}\right) \frac{(2m+2)!}{(2m+1) \sqrt{(2m+1)!} 2^{m+1} \left(m+1\right)!}
\end{equation}
where $(-1)!!=1$. Therefore,
\begin{align}
    \left|\frac{Q_{0,2m+3}}{Q_{0,2m+1}}\right|
    &= \frac{2m+1}{\sqrt{(2m+3)(2m+2)}} < \frac{2m+1}{\sqrt{(2m+1)(2m+1)}}\\
    \left|\frac{Q_{0,2m+3}}{Q_{0,2m+1}}\right|&< 1\qed
\end{align}

\section{Calculation of correlators in the quantum harmonic oscillator} \label{Correlator appendix}

Consider a general state
\begin{equation}
    |\psi\rangle=\sum_{n=0}^N c_n |n\rangle, \ c_n \in \mathbb{C} \ .
\end{equation}
The measurement operator in the Heisenberg picture is given by $\hat{Q}(t)=e^{iHt}(2{ \theta(\hat x)}-1)e^{-iHt}$, where we choose $t_2 - t_1 = \tau$ and $t_1 = 0$, such that 
\begin{equation}
    \hat {Q}_j=e^{iH(j-1)\tau}(2{\theta}(\hat x)-1)e^{-iH(j-1)\tau} \ .
\end{equation}
Then,
\begin{equation}
    C_{12} = \frac{1}{2}\langle\psi|\{ \hat {Q}_1 , \hat {Q}_2 \} | \psi \rangle \ .
\end{equation}
\begin{equation}
    \therefore C_{12} = \frac{1}{2} \sum_{n,k=0}^N c_n^{*} c_k\langle n| \hat{Q}_1 e^{i{H}\tau} \hat{Q}_1 e^{-i{H}\tau} | k \rangle + \frac{1}{2}\sum_{n,k=0}^N c_n^{*} c_k\langle n| e^{i{H}\tau} \hat{Q}_1 e^{-i{H}\tau} \hat{Q}_1 | k \rangle
\end{equation}
We can write
\begin{equation}
    e^{i{H}\tau}=\sum_{m=0}^\infty e^{im\tau} |m\rangle \langle m| \ ,
\end{equation}
so that
\begin{multline}
    C_{12} = \frac{1}{2} \sum_{n,k=0}^N c_n^{*} c_k\langle n| \hat{Q}_1 \sum_{m=0}^\infty e^{im\tau} |m\rangle \langle m| \hat{Q}_1 e^{-ik\tau} | k \rangle + \frac{1}{2}\sum_{n,k=0}^N c_n^{*} c_k\langle n| e^{in\tau} \hat{Q}_1 \sum_{m=0}^\infty e^{-im\tau} |m\rangle \langle m| \hat{Q}_1 | k \rangle \ ,
\end{multline}
which can be reduced to a form containing only the $Q_{nk}\equiv\langle n | \hat{Q}_1 | k \rangle$ matrix elements:
\begin{equation}
    \frac{1}{2} \sum_{n,k=0}^N \sum_{m=0}^\infty e^{i(m-k)\tau} c_n^{*} c_k Q_{nm} Q_{mk} + \frac{1}{2}\sum_{n,k=0}^N \sum_{m=0}^\infty e^{i(n-m)\tau} c_n^{*} c_k Q_{nm} Q_{mk} \ .
\end{equation}
The matrix $\bold{J}$ is real and symmetric; and so we can combine terms as:
\begin{equation}
    \frac{1}{2} \sum_{n,k=0}^N \sum_{m=0}^\infty e^{i(m-k)\tau} c_n^{*} c_k Q_{nm} Q_{mk} + \frac{1}{2}[\sum_{n,k=0}^N \sum_{m=0}^\infty e^{i(m-k)\tau} c_n^{*} c_k Q_{nm} Q_{mk}]^{*} \ .
\end{equation}
That is,
\begin{equation} \label{eq: C12 calculation}
    C_{12} = \mathrm{Re}\{ \sum_{n,k=0}^N \sum_{m=0}^\infty e^{i(m-k)\omega\tau} c_n^{*} c_k Q_{nm} Q_{mk} \} \ .
\end{equation}
In fact, for states like the Zaw-Scarani state with support entirely on a given $\mathcal{H}_k$, it is sufficient to just compute $C_{12}$ as, with a measurement time spacing of a third of a period, $C_{12}=C_{13}=C_{23}=C$. We prove this statement in Appendix \ref{sec: equal correlators}. For more general states, the calculation procedure for $C_{23}$ and $C_{13}$ is very similar to Eq.\ref{eq: C12 calculation}.

\section{All Three Correlators are Equal on $\mathcal{H}_k$} \label{sec: equal correlators}
For some state $|\psi_k\rangle\in \mathcal{H}_k$, the correlators $C_{ij}$ can be written as
\begin{equation}\label{eq: correlator on Hk}
    C_{ij} = \frac{1}{2} \langle \psi_k|\{ \hat Q_i,  \hat Q_j\}|\psi_k\rangle \ .
\end{equation}
Now, the expectation value $F_{ijk}=\frac{1}{2}\langle \psi_k|\{ \hat Q_i,  \hat Q_j\}|\psi_k\rangle$, $i<j$, can be expanded out as
\begin{equation}
    F_{ijk} = \operatorname{Re}\langle \psi_k|( U^{i-1})^\dagger  \hat Q_1  U^{i-1}( U^{j-1})^\dagger \hat  Q_1  U^{j-1}|\psi_k\rangle = \operatorname{Re}\langle \psi_k|\omega^{k(1-i)} \hat Q_1  U^{i-j} \hat Q_1 \omega^{k(j-1)}|\psi_k\rangle \ ,
\end{equation}
by leveraging the unitarity of $ U$ and its action on $|\psi_k\rangle$. We then group some terms in the aim of showing $F_{12k}=F_{23k}=F_{13k}$,
\begin{equation}
    F_{ijk} = \operatorname{Re} \omega^{k(j-i)}\langle \psi_k| \hat Q_1  U^{i-j} \hat Q_1 |\psi_k\rangle \ .
\end{equation}
Clearly, $F_{12k} = \operatorname{Re} \omega^{k}\langle \psi_k|\hat Q_1  U^{-1}  \hat Q_1 |\psi_k\rangle = F_{23k}$ but it also quickly follows that $F_{12k}=F_{13k}$. Note, 
\begin{equation}
    F_{13k} = \operatorname{Re} \omega^{2k}\langle \psi_k| \hat Q_1  U^{-2}  \hat Q_1 |\psi_k\rangle \ ,
\end{equation}
where $\omega^{2k} = (\omega^k)^*$, $ U^{-2} =  U$ and $\langle \psi_k| \hat Q_1  U^{-2} \hat Q_1 |\psi_k\rangle = \langle \psi_k|\hat Q_1  U^{-1}\hat Q_1 |\psi_k\rangle^*$. So, 
\begin{equation} \label{eq: equal Fs}
    F_{13k} = \operatorname{Re} \left(\omega^{k}\langle \psi_k|\hat Q_1  U^{-1}  \hat Q_1 |\psi_k\rangle \right)^* = \operatorname{Re} \omega^{k}\langle \psi_k|\hat Q_1  U^{-1} \hat Q_1 |\psi_k\rangle = F_{12k}=F_{23k}\ .
\end{equation}
Hence, by pairing Eq. \ref{eq: correlator on Hk} and Eq. \ref{eq: equal Fs}, we have shown that for a given quantum state, with support entirely on $\mathcal{H}_k$, all the correlators are equal when the measurement spacing is $\Delta t = 2\pi/3$.

\section{Calculation of sequential measurement probabilities} \label{Method 2 appendix}

We consider the initial density operator
\begin{equation}
    \rho = |\psi\rangle \langle \psi| = \sum_{n,k=0}^N c_n^{*} c_k |k\rangle \langle n |
\end{equation}
and define the Heisenberg-picture projection operators $P_{+}(t) = {\theta}(\hat x(t))$ and $P_{-}(t) = [1-{\theta}( \hat x(t))]$, in order to make more concise the notation for the time-evolved Heaviside projector onto the positive x-axis:
\begin{equation} \label{eq: Heaviside definition}
    P_{+}(t) \equiv {\theta}(\hat x(t)) = \int _0^\infty e^{-iHt}|x\rangle\langle x |e^{iHt} \mathrm{d}x \ .
\end{equation}
Then we have the simple expression
\begin{equation}
    p_{123}(s_1,s_2,s_3) = \mathrm{Tr}\left[P_{s_3}(t_3)P_{s_2}(t_2)P_{s_1}(t_1)\rho P_{s_1}(t_1)P_{s_2}(t_2)\right] \ .
\end{equation}

In order to compute these probabilities numerically, however, it is easiest to dissociate the time dependence of the operators such that we recover expressions in terms of the familiar $Q_{nk}$ matrix elements. That is,
\begin{equation}
    p_{123}(+,+,+) = \mathrm{Tr}(e^{2iH\tau}P_+(t_1)e^{-iH\tau}P_+(t_1)e^{-iH\tau} P_+(t_1) \rho P_+(t_1) e^{iH\tau}P_+(t_1)e^{-iH\tau}) \ ,
\end{equation}
recognising that $P_+(t_1)$ is simply the Heaviside operator without any time evolution. is Then, writing 
\begin{equation}
    e^{iH\tau}=\sum_{m=0}^\infty e^{im\tau}|m\rangle\langle m| \ ,
\end{equation}
setting $\omega=\hbar=1$ as always, we define the matrix elements
\begin{equation} \label{Lnk}
    L_{nk}\equiv \langle n | (\hat{Q}_1 +1) | k\rangle = Q_{nk} + \delta_{nk} \ ,
\end{equation}
where $\delta_{nk}$ is the Kronecker delta symbol, such that
\begin{equation}
    p_{123}(+,+,+) = \frac{1}{32} \sum_{m,p,q,r=0}^{\infty} \sum_{n,k=0}^{N} c_n^{*}c_ke^{i\tau(q+r-m-p)}L_{rm}L_{mp}L_{pk}L_{nq}L_{qr} \ , \label{qho p3+}
\end{equation}
recalling that $P_+(t_1)=\frac{(\hat {Q}_1+1)}{2}$. Similarly,
\begin{equation}
    p_{123}(-,-,-) = \frac{1}{32} \sum_{m,p,q,r=0}^{\infty} \sum_{n,k=0}^{N} c_n^{*}c_ke^{i\tau(q+r-m-p)}L'_{rm}L'_{mp}L'_{pk}L'_{nq}L'_{qr} \ ,
\end{equation}
where $\mathbf{L}'\equiv(\mathbf{\mathbb{1}}-\mathbf{Q})$. Computationally, we have chosen a finite cut-off value $N$ to truncate the infinite sum over $m,p,q,r$.

There are also two other quasi-probability distributions, as elucidated in the discussion surrounding Eq. \ref{Q_2 with an earlier 1}. We construct these by beginning from Eq. \ref{quasi trace expression} and either replacing $C_{13}$ by $C_{13}^{(2)}$ (the first possibility) or replacing $C_{23}$ by $C_{23}^{(1)}$ (the second). (Doing both would effectively yield Eq. \ref{projective 3-time moment expansion}.) Thus, we define
\begin{equation} \label{Cosmic penguin}
    q_A(+,+,+) = \frac{1}{2}\text{Tr}\left(P_+(t_3)P_+(t_2)\left\{P_+(t_1),\rho\right\}P_+(t_2)\right) \ .
\end{equation}
Classically, this distribution is identical to $p_{123}(+,+,+)$. In this case,
\begin{equation}
    q_{A}(+,+,+)-q_{A}(-,-,-)\equiv \Delta q_{A}=\frac{1}{4}(\langle A \rangle+D+I) \ ,
\end{equation}
where the interference terms would only stem from the second measurement: $I=\langle Q_3^{(2)} \rangle - \langle Q_3 \rangle$. Secondly,
\begin{equation}
    q_B(+,+,+) = \frac{1}{2} \text{Tr} \left( P_+(t_3) \{ P_+(t_2), P_+(t_1) \rho P_+(t_1) \} \right) \ .
\end{equation}
For reference, this is readily computed in terms of unevolved Heaviside operators:
\begin{equation} \label{weak second measurement}
    q_B(+,+,+) = \frac{1}{2} \text{Tr} \left( e^{2iH\tau} {\theta}(\hat x) e^{-2iH\tau} \{ e^{iH\tau} {\theta}(\hat x) e^{-iH\tau}, {\theta}(\hat x) \rho {\theta}(\hat x) \} \right) ,
\end{equation}
which can be computed straightforwardly in terms of the afore-defined $L_{nk}$ matrix elements; for example,
\begin{equation}
    \frac{1}{2} \text{Tr} \left( e^{2iH\tau} {\theta}(\hat x) e^{-iH\tau} {\theta}(\hat x) e^{iH\tau} {\theta}(\hat x) \rho {\theta}(\hat x) \right) =\frac{1}{32} \sum_{n,k,p,q,r} c_n^{*} c_k e^{i\tau(2p+r-q)} L_{pq} L_{qm} L_{mk} L_{np} \ .
\end{equation}

\section{Tsirelson Violation implies LG3 satisfaction on $\mathcal{H}_k$} \label{T implies LG3 satisfaction proof}
\subsection{Outline and Preliminaries}
This appendix is devoted to proving that if Tsirelson's inequality is violated by a given quantum state, with support entirely in $\mathcal{H}_k$, then that state will satisfy all four LG3s (Eqs \ref{L1 A}-\ref{L4 A}) if the measurement times $t_1, t_2, t_3$ are the same as in the Tsirelson scheme.

To show this claim, one simply needs to show two statements hold true. The conjunction of these two statements would imply that all four LG3s are greater than zero:
\begin{enumerate}
    \item A Tsirelson violation implies satisfaction of the first LG3 i.e $L_1>0$. We straightforwardly show this in Section \ref{sec: Tsirelson implies L1}. In fact, this result holds for arbitrary quantum states.
    \item For quantum states in $\mathcal{H}_0, \mathcal{H}_1, \mathcal{H}_2$, the first Leggett–Garg quantity is always the smallest of the four. Consequently, all four LG3s are positive.
\end{enumerate}

\subsection{Tsirelson Violation \texorpdfstring{$\implies L_1>0$}{Implies L1>0}}\label{sec: Tsirelson implies L1}

Consider a state that violates Tsirelson's inequality such that
\begin{equation}\label{eq:violation constraint}
    |\langle Q_1 + Q_2+Q_3\rangle| >1
\end{equation}
Now, consider the Cauchy-Schwarz inequality applied to the left hand side of Eq. \ref{eq:violation constraint}
\begin{equation}\label{eq:cauchy schwartz tsirelson}
    \langle (Q_1 + Q_2+Q_3)^2\rangle \geq (\langle Q_1 + Q_2+Q_3\rangle)^2.
\end{equation}
Using Equations \ref{eq:violation constraint} and \ref{eq:cauchy schwartz tsirelson} we can write
\begin{equation}
    \langle (Q_1 + Q_2+Q_3)^2\rangle > 1.
\end{equation}
Finally, simply expanding out the expectation value nets
\begin{equation}
    1+C_{12}+C_{13}+C_{23}>0.
\end{equation}

\subsection{\texorpdfstring{$L_1$}{L1} is the smallest LG3 on \texorpdfstring{$\mathcal H_0, \mathcal H_1$ and $\mathcal H_2$}{H0, H1 and H2}} \label{sec: subspaces proof}
Recall from Appendix \ref{sec: equal correlators} that for $|\psi_k\rangle\in \mathcal{H}_k$, $C_{12}=C_{13}=C_{23}=C$. Therefore, on these subspaces, $L_1 = 1+3C$ and the rest are $1-C$. So, by showing that $C_{12}$ is negative we will have shown that $L_1$ is the smallest LG3. Note,
\begin{equation}
    C_{12} = \frac{1}{2}\langle \psi_k|\hat Q(\omega^{-k}U+\omega^k U^
    \dagger) \hat Q|\psi_k\rangle \ ,
\end{equation}
where $\hat Q|\psi_k\rangle$ can be decomposed into $\hat Q|\psi_k\rangle = |\phi_k\rangle + |\phi_\perp\rangle$, $|\phi_k\rangle\in\mathcal{H}_k$ and $\langle \phi_k|\phi_\perp\rangle = 0$. Hence, as $|\phi_\perp\rangle$ is constructed out of orthogonal energy eigenstates to $|\phi_k\rangle$, we can write
\begin{equation} \label{eq: almost there}
    C_{12} = \frac{1}{2} \langle \phi_k|(\omega^{-k}U+\omega^k U^
    \dagger) |\phi_k\rangle + \frac{1}{2} \langle \phi_\perp|(\omega^{-k}U+\omega^k U^
    \dagger) |\phi_\perp\rangle \ .
\end{equation}
Motivated by the roots of unity representation of the cyclic group, note that
\begin{equation}
    (\mathds{1} + \omega^{-k}U+\omega^kU^\dagger)|\phi_k\rangle = 3|\phi_k\rangle \Rightarrow (\omega^{-k}U+\omega^kU^\dagger)|\phi_k\rangle = 2|\phi_k\rangle \ ,
\end{equation}
and
\begin{equation}
    (\mathds{1} + \omega^{-k}U+\omega^kU^\dagger)|\phi_\perp\rangle = 0 \Rightarrow (\omega^{-k}U+\omega^kU^\dagger)|\phi_\perp\rangle = -|\phi_\perp\rangle \ .
\end{equation}
Thus, we can write Eq. \ref{eq: almost there} as
\begin{equation} \label{eq: big condition}
    C_{12} = \langle \phi_k|\phi_k\rangle - \frac{1}{2} \langle \phi_\perp|\phi_\perp\rangle \ .
\end{equation}
By substituting in the condition $\langle \phi_\bot|\phi_\bot\rangle + \langle \phi_k|\phi_k\rangle = 1$ we find
\begin{equation}
    C_{12} = \frac{3}{2}\langle \phi_k |\phi_k\rangle - \frac{1}{2} \ .
\end{equation}
Reintroducing $|\psi_k\rangle$ nets an inequality including the spectral norm of an operator
\begin{equation}
    C_{12} \leq \frac{3}{2}||P_k \hat Q P_k||^2 - \frac{1}{2} \ ,
\end{equation}
where $P_k$ is the projection operator onto $\mathcal H_k$. We can trivially extend this upper bound by writing
\begin{equation}
    C_{12} \leq \frac{3}{2}\max_{k\in \{0,1,2\}}||P_k \hat Q P_k||^2 - \frac{1}{2} \ .
\end{equation}
But, 
\begin{equation}
    \max_{k\in \{0,1,2\}}||P_k \hat Q P_k|| = ||\operatorname{Twirl} (\hat Q)|| = \frac{1}{3}||\hat A|| \ .
\end{equation}
Therefore,
\begin{equation}
    C_{12} \leq \frac{1}{6}||\hat A||^2 - \frac{1}{2} \ .
\end{equation}
Finally, using the upper bound on $||\hat A||$ from Ref. \cite{zaw2024threeanglevariantstsirelsonsprecession},
\begin{equation}
    C_{12} \leq \frac{1}{6}(1.39)^2 - \frac{1}{2} \leq 0 \ .
\end{equation}

\section{Constrained Optimisation Procedure for Method 2} \label{Constrained optimisation}

We seek a state that violates the Tsirelson inequality but satisfies the condition $p_{123}(+,+,+) = p_{123}(-,-,-)$, or $\Delta p_{123}=0$, exactly. We can reframe this question as a constrained optimisation problem where the function we wish to maximise is 
\begin{equation}
    \langle A\rangle \equiv \langle {Q}_1 + {Q}_2 + {Q}_3 \rangle
\end{equation}
which can be written as a function of the coefficients in the energy basis expansion of the initial state:
\begin{equation}
    \langle A \rangle = f(c_n).
\end{equation}

Recall that
\begin{equation}
    p_{123}(+,+,+) = \frac{1}{32} \sum_{m,p,q,r=0}^{\infty} \sum_{n,k=0}^{N} c_n^{*}c_ke^{i\tau(q+r-m-p)}L_{rm}L_{mp}L_{pk}L_{nq}L_{qr} \ ,
\end{equation}
where the $L_{nk}$ matrix elements are defined as in Eq. \ref{Lnk}, with a similar expression for $p_{123}(-,-,-)$. Defining
\begin{equation}
    \Gamma_{nk} = L_{rm}L_{mp}L_{pk}L_{nq}L_{qr} \ ,
\end{equation}
the constraint can be rewritten in the simple form
\begin{equation}
    \Delta p_{123} \equiv 2 \sum_{n, k \text{ with opposite parity}}^N c_n^{*} c_k \Gamma_{nk} = 0 \ .
\end{equation}
or more algorithmically:
\begin{equation}
    \sum_{j=1}^{\floor*{\frac{N+1}{2}}} \sum_{i=0}^{N-(2j-1)} \text{Re}(c_i^* c_{i+2j-1}\Gamma_{i,i+2j-1}) = 0 \ . 
\end{equation}
Thus, the objective function that we will optimise is 
\begin{equation}
    L(c_n, \lambda) = f(c_n) - \lambda \left( \sum_{j=1}^{\floor*{\frac{N+1}{2}}} \sum_{i=0}^{N-(2j-1)} \text{Re}(c_i^* c_{i+2j-1}\Gamma_{i,i+2j-1})\right), 
\end{equation}
where the constraint is introduced as a Lagrange multiplier. We also introduce a constraint on the norm of the vector of being 1.

We find that the state
\begin{multline}
    (-0.281+0.52i)|0\rangle+(0.096-0.127i)|1\rangle+(-0.024+0.102i)|2\rangle \\ +(0.423-0.539i)|3\rangle+(-0.066+0.068i)|4\rangle+(0.012-0.067i)|5\rangle \\ +(-0.25 +0.262i)|6\rangle
\end{multline}
readily produces a Tsirelson violation of $\langle A \rangle=1.0756$ with $\Delta p_{123}=0$. More accurately, $\Delta p_{123}$ is not precisely and analytically zero for the quoted state, since the numerical computation was done using a truncated energy expansion. However, we find that slightly tweaking the state parameters can lead to states with $\Delta p_{123}$ slightly above or below zero, and so it is almost certain that, in principle, there exists a state with $\Delta p_{123}=0$ exactly.

\section{Smoothed Projectors} \label{Smoothed projectors}

We have modelled the measurements of position using the step function $\theta (\hat x)$, but of course realistic measurements and experimental procedures, e.g. involving a light source localized in $x>0$ (as alluded to in Section \ref{Introduction}), will not be an exact step function but rather will interpolate smoothly between $0$ and $1$ in a region of size $\epsilon$ around $x=0$. The effect of such smoothing was calculated explicitly in Ref. \cite{MawbySHO}, and it was shown that LG violations in the harmonic oscillator for measurements of $ \operatorname{sign} (x) $ are significantly reduced when the smoothing scale $\epsilon$ is of order $1$ (in the dimensionless units we are using here). This effect is related to the well-known fact that the Wigner function is rendered positive when smoothed over phase–space regions of size of order $\hbar$ (noting that when calculating probabilities in the Wigner picture, the smoothing on step functions can be switched onto the Wigner function of the initial state). We therefore also expect a reduction in the violation of the Tsirelson inequality due to smoothing if the smoothing scale $\epsilon$ approaches $1$, so we need $ \epsilon \ll 1 $ to maintain a violation.

The smoothing will also affect the probabilities $p(+,+,+)$ and $p(-,-,-)$ characterizing uniform precession and we therefore need to revisit what uniform precession means in this case. To assess this we use a simplified but instructive ``smoothing" function $ \theta_{\epsilon} (x) $, which agrees with the usual $\theta(x)$ function except that it takes the value $ \frac{1}{2}$ for $ -\epsilon / 2 < x < \epsilon / 2$. Also, for simplicity, we use a classical analysis with a state concentrated around $x=0$ and $p=0$.

Then there are two cases. If the width of the state is less than $\epsilon$, then most of the state is concentrated in the smoothing region and we have effectively  $\theta_{\epsilon} (x) = \frac{1}{2} $. We would then have $p(+,+,+)$ be of order $\frac{1}{8}$ and likewise $p(-,-,-)$, and hence $L_1$ is of order $\frac{1}{4}$. This means that any values of $L_1$ less than $\frac{1}{4}$ may be taken to mean that uniform precession is satisfied.

\begin{figure*}
    \centering
    \subfigure[]{
        \includegraphics[width=0.4\linewidth]{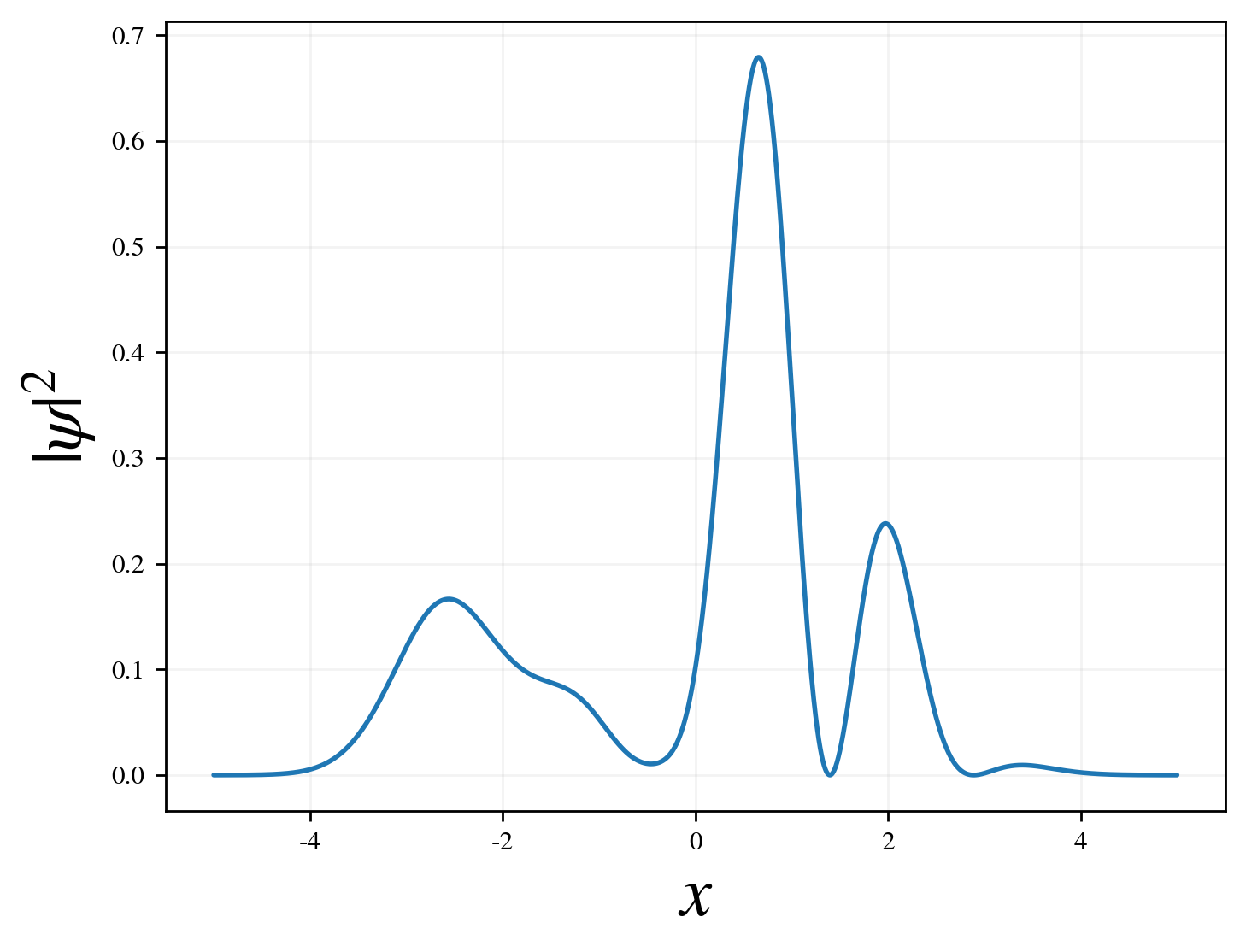}
        \label{fig: zaw plots a}
    } 
    \subfigure[]{
        \includegraphics[width=0.4\linewidth]{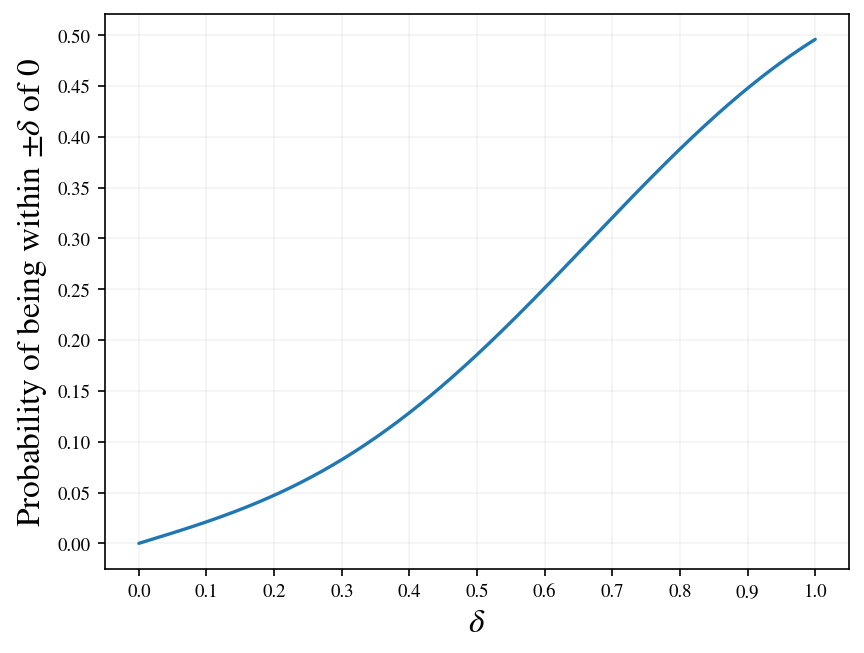}
        \label{fig: zaw plots b}
    } 
    \caption{(a) A plot of $|\psi(x)|^2$ against the dimensionless position for the Zaw-Scarani state $|\Psi_\mathcal{Z}\rangle=\frac{4}{\sqrt{42}}|0\rangle-\frac{1}{\sqrt2}|3\rangle+\sqrt{\frac{5}{42}}|6\rangle$  (b) For the Zaw-Scarani state, this plot shows the probability $p(\delta)$ of measuring the particle to be in the interval $(-\delta, \delta)$ against $\delta$, where $\delta$ is the dimensionless position.}
    \label{fig: zaw plots}
\end{figure*}

However, if the width of the state is much greater than $\epsilon$, then only a small
fraction of the state will be in the smoothing region and the above effect will be proportionately reduced. For the Zaw–Scarani state, plotted in Fig. \ref{fig: zaw plots a}, the probability $p(\delta)$ of finding the particle within any interval $[-\delta,\delta]$ around the origin is plotted in Fig. \ref{fig: zaw plots b}.  The condition for uniform precession is then that $L_1$ is less than $\frac{1}{4}p(\delta)$ where $\delta=\epsilon/2$ (so that the interval coincides with the smoothing region). As noted above, the interesting regime for a Tsirelson violation is $\epsilon \ll 1$, and from the plot Fig. \ref{fig: zaw plots b} one can see that $p(\delta)$, and hence $L_1$, will also be $\ll 1$ in this case. Hence the regime most relevant to us is actually very close to the case of exact projective measurements considered in this paper. More precise estimates are not
possible without knowledge of the experimental parameter $\epsilon$ and would require a more detailed calculations similar to those of Ref. \cite{MawbySHO}, which would also need to include an assessment of the possible reduction of the Tsirelson violation produced by the smoothing; but we will not do this here.

\section{Probability Currents} \label{sec: prob currents}
\subsection{Relating the Tsirelson quantity to the quantum current}
Consider the limit definition of the integral to write $\langle T_D(2\pi) \rangle$ as a Riemann sum,
\begin{align}
    \langle T_D(2\pi)\rangle &= \frac{1}{2\pi}\left[\lim_{N\rightarrow \infty}\frac{2\pi}{N+1}\sum_{n=0}^N \left\langle \theta \left( x\left(\frac{2\pi n}{N+1}\right)\right) \right\rangle \right] \ , \\
    &=\frac{1}{3}\sum_{n=0}^2 \left\langle\theta \left( x\left(\frac{2\pi n}{3}\right)\right) \right\rangle + \frac{\varepsilon}{2\pi} \ ,
\end{align}
where $\varepsilon$ refers to an error term. The equality certainly holds for some $\varepsilon \in \mathbb{R}$.

Assuming $\langle\theta( x(t))\rangle$ is continuously differentiable in $t$, then Taylor's theorem applies to $\langle \theta(x(t))\rangle$ and we can borrow some well-known results from numerical analysis to quantify our error term. Specifically, we notice that this amounts to a left endpoint approximation to the integral for the dwell time. Therefore we can use the formula for the error term \cite{numerical_methods},
\begin{equation}
    E_L = \frac{(2\pi)^2}{6} \frac{d\langle \theta(x)\rangle}{dt}\big|_{t=\tau}, \tau\in[0,2\pi] \ ,
\end{equation}
to state that 
\begin{align}
    \langle T_D(2\pi) \rangle &= \frac{1}{3}\sum_{n=0}^2 \left\langle\theta \left( x\left(\frac{2\pi n}{3}\right)\right) \right\rangle + \frac{E_L}{2\pi} \\
    &= \frac{1}{3}\sum_{n=0}^2 \left\langle\theta \left( x\left(\frac{2\pi n}{3}\right)\right) \right\rangle +\frac{2\pi}{6} \frac{d\langle \theta(x)\rangle}{dt}\big|_{t=\tau} \ ,
\end{align}
where $\tau\in[0,2\pi]$. The sum can be related to the Tsirelson quantity because $\left\langle \theta( x(t)) \right\rangle = \frac{\left\langle Q(t) \right\rangle+1}{2}$  and $\langle A\rangle = \sum_{n=0}^2 \left\langle Q\right\rangle $. So,
\begin{align}
    \langle T_D(2\pi) \rangle &= \frac{1}{6}\left(\sum_{n=0}^2 \left\langle Q \left(\frac{2\pi n}{3}\right) \right\rangle +3 \right) +\frac{2\pi}{6} \frac{d\langle \theta(x)\rangle}{dt}\big|_{t=\tau} \\
    &= \frac{1}{6}\left(3+\langle A\rangle\right) +\frac{2\pi}{6} \frac{d\langle \theta(x)\rangle}{dt}\big|_{t=\tau}.
\end{align}
Hence,
\begin{equation}
    \langle A\rangle = (6\langle  T_D(2\pi) \rangle-3)-2\pi\frac{d\langle \theta(x)\rangle}{dt}\big|_{t=\tau} = -2\pi\frac{d\langle \theta(x)\rangle}{dt}\big|_{t=\tau}
\end{equation}
This expression thereby provide an alternative explanation for why $\langle A\rangle=0$ for energy eigenstates (stationary states). The expectation value of the Heaviside operator has no time dependence for energy eigenstates, so those derivatives are $0$ for any value of $\tau$.

Note that because $\langle \theta (x(t+\pi))\rangle = 1 - \langle \theta(x(t))\rangle$, the derivatives of $\langle \theta(x(t))\rangle$ simply flip sign after half a period. Hence, without loss of generality, we may drop the negative signs on our expressions, since $\tau$ is some unknown time on the interval $[0,2\pi]$:
\begin{equation} \label{eq: new A expression}
    {\langle A \rangle = 2\pi \frac{d\langle \theta(x)\rangle}{dt}\big|_{t=\tau}} \ .
\end{equation}
We can rewrite Eq. \ref{eq: new A expression} as,
\begin{equation}
    \langle A\rangle = 2\pi\frac{d}{dt}\int_0^\infty |\psi(x,\tau)|^2dx.
\end{equation}
Then, using the 1D continuity equation, 
\begin{equation}
    \frac{\partial \rho}{\partial t} + \frac{\partial J}{\partial x} = 0,
\end{equation}
where $\rho = |\psi|^2$ is the probability density and $J=\operatorname{Im}\left(\psi^*\partial_x \psi\right)$ is the probability density current \cite{Sakurai_Napolitano_2020}, we can substitute the current into the Tsirelson quantity:
\begin{align}
    \langle A\rangle &= -2\pi\int_0^\infty \frac{\partial J(x,\tau)}{\partial x}dx, \\
    & = 2\pi J(0,\tau).
\end{align}
Therefore, assuming $J(\infty, \tau) \rightarrow 0$, we have directly related the Tsirelson quantity to the probability current at the origin. Specifically,
\begin{equation} \label{eq: current connection}
    \exists\tau\in[0,2\pi] \text{ }\operatorname{s.t}\text{ } \langle A\rangle = 2\pi J(0,\tau).
\end{equation}
From Eq. \ref{eq: current connection}, we can make the following claim: if $|J(0,\tau)|\leq \frac{1}{2\pi}, \forall \tau\in[0,2\pi] $, then $|\langle A\rangle|\leq 1$. That is to say, if the current at the origin stays sufficiently low, then we can guarantee the state will \textit{not} violate the Tsirelson inequality. Notably, $J(0,t)$ could be measured by taking the time derivative of measured values of $\langle \theta(x(t))\rangle$ which only requires single-time measurements.

This result also provides some further insight into why completely even or odd wavefunctions cannot violate the Tsirelson inequality. Expanding a general state as a superposition of energy eigenstate wavefunctions $\psi_n(x)$, the current at the origin can be written as
\begin{equation} \label{eq: energy current decomp}
    J(0,t) = \operatorname{Im}\sum_{n,m} c_n^*c_m \psi_n(0) \psi_m'(0) e^{i(n-m)t} \ .
\end{equation}
Even functions have $0$ slope at $x=0$ and odd functions are $0$ at $x=0$. As such, if the wavefunction in question can be decomposed into exclusively even energy eigenfunctions or exclusively odd energy eigenfunctions, then $J(0,t) = 0, \forall t\in[0,2\pi]$, so $\langle A\rangle=0$.

\subsection{A Tsirelson satisfaction condition on $\mathcal{H}_0$} \label{sec: improved current bound}
We will refer to the dwell time over $2/3$ of a period as $T_{2/3}$. Consider the following integral approximation:
\begin{align}
    T_{2/3} &= \frac{1}{6}\left(\sum_{j=1}^3 \langle \theta(x(t_j))\rangle + \langle \theta(x(t_1))\rangle\right) + \frac{1}{2\pi}\sum_{j=1}^3 E_j \\
    &= \frac{1}{12}\langle A\rangle + \frac{1}{12}\langle Q_1\rangle + \frac{1}{3} + \frac{1}{2\pi}\sum_{j=1}^3 E_j.
\end{align}
Therefore,
\begin{equation}
    \langle A\rangle = 12T_{2/3} - 4 - \langle Q_1\rangle - \frac{6}{\pi}\sum_{j=1}^3 E_j
\end{equation}
where both $E_1$ and $E_2$ are left end point errors with $E_1 = \frac{(2\pi/3)^2}{2}\frac{\mathrm{d}\langle \theta(x)\rangle}{\mathrm{d}t}\big|_{\tau_1} \ , \ \tau_1\in[0,2\pi/3]$ and $E_2 = \frac{(\pi/3)^2}{2}\frac{\mathrm{d}\langle \theta(x)\rangle}{\mathrm{d}t}\big|_{\tau_2} \ , \ \tau_2\in[2\pi/3, \pi]$. However, $E_3$ is a right endpoint error with $E_3 = -\frac{(\pi/3)^2}{2}\frac{\mathrm{d}\langle \theta(x)\rangle}{\mathrm{d}t}\big|_{\tau_3} \ , \ \tau_3\in[\pi, 4\pi/3] $. So, expanding out and simplifying the error term contribution,
\begin{equation}
    \sum_{j=1}^3 E_j = \frac{\pi^2}{18}\left[4J(\tau_1)+ J(\tau_2) - J(\tau_3)\right] \ .
\end{equation}
Therefore,
\begin{equation}
    \langle A\rangle = 12T_{2/3} - 4 - \langle Q_1\rangle - \frac{\pi}{3}\left[4J(\tau_1)+ J(\tau_2) - J(\tau_3)\right] \ .
\end{equation}
But, by noting that $-J(\tau_3) = +J(\tilde \tau _3), \ \tilde \tau_3\in [0, \pi/3]$, the sum of the three currents can be reformulated as a weighted average over the domain $[0,\pi]$. Pairing that with an application of the intermediate value theorem allows us to write
\begin{equation}
    \langle A\rangle = 12T_{2/3} - 4 - \langle Q_1\rangle - 2\pi J(\tau), \ \tau\in [0,\pi] \ .
\end{equation}
For $|\psi\rangle \in \mathcal{H}_0$, we have that $T_{2/3} = 1/3$ and $\langle Q_1\rangle = \frac{1}{3}\langle A\rangle$. Therefore, the expression for $\langle A\rangle$ simplifies to
\begin{equation} \label{eq: special case mixed rectangle 1}
    \langle A\rangle = - \frac{3\pi}{2}J(\tau), \ \tau \in [0,2\pi/3]  \ ,
\end{equation}
where the domain of $\tau$ was decreased as the current is periodic over $2\pi/3$. Once again an application of the sign flipping property of the current allows us to say
\begin{equation}
    \text{For} \ |\psi\rangle\in\mathcal{H}_0 \ , \ \exists \tau\in [0,2\pi/3] \ \text{s.t} \ \langle A\rangle = \frac{3\pi}{2}J(\tau).
\end{equation}
An immediate corollary of this is that
\begin{equation}\label{eq: H0 current bound}
    \text{For} \ |\psi\rangle\in\mathcal{H}_0 \ , \ \text{if} \ |J(\tau)|\leq \frac{2}{3\pi} \ \forall \tau\in [0,2\pi/3] \ \text{then} \ |\langle A\rangle| \leq 1 \ .
\end{equation}

\subsection{Classical and Quantum Probability Current}
\begin{figure}[H]
    \centering
    \includegraphics[width=0.5\linewidth]{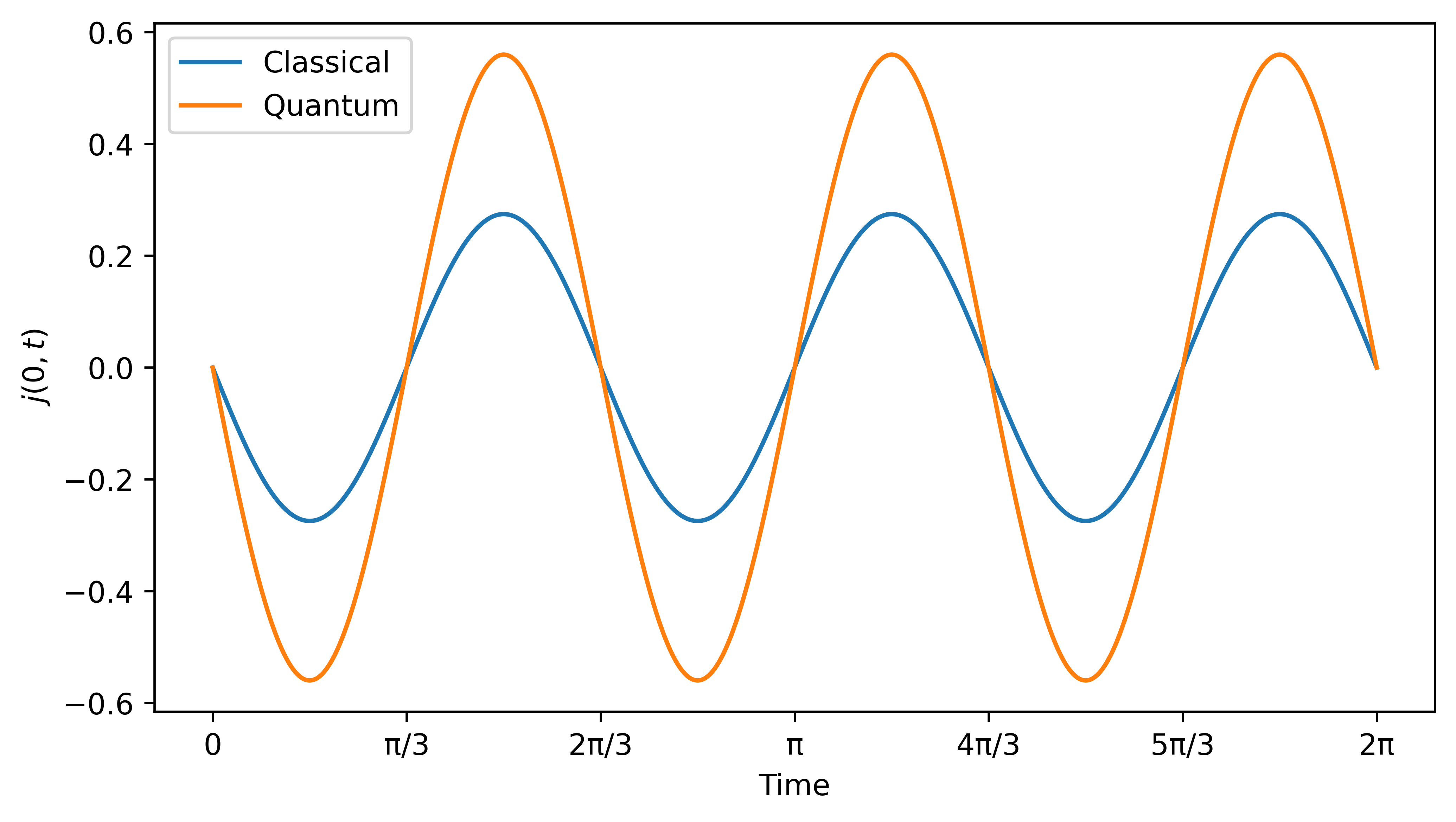}
    \caption{Plot of the probability density current for the Zaw-Scarani state at the origin over a period compared to the classical probability density current from the classical time evolution of the Husimi-Q representation of the Zaw-Scarani state. Notably, the classicalised current is consistently lower than the quantum current.}
    \label{fig: Husimi-Q vs Quantum}
\end{figure}

Peculiarly, when we computed the classical probability current, using
\begin{equation}
    j_c(0,t) = \iint \mathrm{d}x\mathrm{d}p \ p(t) \delta[x(t)]W_c(x,p) \ , 
\end{equation}
for any natural classicalisation of the phase-space distribution, $W_c(x,p)$, associated with $|\psi_Z\rangle \in \mathcal H_0$, we found the classical current was lower in magnitude than the quantum current across the entire period. Specifically, from the Wigner function of the Zaw-Scarani state, $W_Z(x,p)$, we constructed the following classical distributions: (1) $W_c(x,p)\propto |\langle \alpha|\psi_Z\rangle|^2$ (the Husimi-Q distribution), (2) $W_c(x,p)\propto|W_Z(x,p)|$, (3) $W_c(x,p)\propto W_Z(x,p)+|W_Z(x,p)|$ and (4) $W_c(x,p) =|\psi_Z(x)|^2|\phi_Z(p)|^2$ (where $\phi_Z(p)$ is the momentum-space wavefunction of the Zaw-Scarani state). In Figure \ref{fig: Husimi-Q vs Quantum}, we illustrate the resulting classical current, where the classicalisation strategy chosen is using the Husimi-Q distribution, and compare it to the quantum current. This result perhaps suggests a connection between the current based bound on the Tsirelson quantity in Eq. \ref{eq: H0 current bound} and classicalised currents of violating quantum states.

\twocolumngrid
\bibliography{references}

@article{PhysicsPhysiqueFizika.1.195,
  title = {On the {Einstein Podolsky Rosen} paradox},
  author = {Bell, J. S.},
  journal = {Physics Physique Fizika},
  volume = {1},
  issue = {3},
  pages = {195--200},
  numpages = {6},
  year = {1964},
  month = {Nov},
  publisher = {American Physical Society},
  doi = {10.1103/PhysicsPhysiqueFizika.1.195},
  url = {https://link.aps.org/doi/10.1103/PhysicsPhysiqueFizika.1.195}
}

@article{PhysRevLett.23.880,
  title = {Proposed Experiment to Test Local Hidden-Variable Theories},
  author = {Clauser, John F. and Horne, Michael A. and Shimony, Abner and Holt, Richard A.},
  journal = {Phys. Rev. Lett.},
  volume = {23},
  issue = {15},
  pages = {880--884},
  numpages = {0},
  year = {1969},
  month = {Oct},
  publisher = {American Physical Society},
  doi = {10.1103/PhysRevLett.23.880},
  url = {https://link.aps.org/doi/10.1103/PhysRevLett.23.880}
}

@article{Emary_2013,
   title={{Leggett–Garg} inequalities},
   volume={77},
   ISSN={1361-6633},
   url={http://dx.doi.org/10.1088/0034-4885/77/1/016001},
   DOI={10.1088/0034-4885/77/1/016001},
   number={1},
   journal={Reports on Progress in Physics},
   publisher={IOP Publishing},
   author={Emary, Clive and Lambert, Neill and Nori, Franco},
   year={2013},
   month=dec, pages={016001} }

@article{PhysRevLett.54.857,
  title = {Quantum mechanics versus macroscopic realism: Is the flux there when nobody looks?},
  author = {Leggett, A. J. and Garg, Anupam},
  journal = {Phys. Rev. Lett.},
  volume = {54},
  issue = {9},
  pages = {857--860},
  numpages = {0},
  year = {1985},
  month = {Mar},
  publisher = {American Physical Society},
  doi = {10.1103/PhysRevLett.54.857},
  url = {https://link.aps.org/doi/10.1103/PhysRevLett.54.857}
}

@article{Leggett2008,
	title = {Realism and the physical world},
	volume = {71},
	url = {https://dx.doi.org/10.1088/0034-4885/71/2/022001},
	doi = {10.1088/0034-4885/71/2/022001},
	pages = {022001},
	number = {2},
	journal = {Reports on Progress in Physics},
        year = 2008,
	author = {Leggett, A J},
	date = {2008-01},
}

@article{Vitagliano_2023,
   title={{Leggett–Garg} macrorealism and temporal correlations},
   volume={107},
   pages = {040101},
   ISSN={2469-9934},
   url={http://dx.doi.org/10.1103/PhysRevA.107.040101},
   DOI={10.1103/physreva.107.040101},
   number={4},
   journal={Physical Review A},
   publisher={American Physical Society (APS)},
   author={Vitagliano, Giuseppe and Budroni, Costantino},
   year={2023},
   month=apr }

@article{dicke1981interaction,
  title={Interaction-free quantum measurements: A paradox?},
  author={Dicke, RH},
  journal={Am. J. Phys},
  volume={49},
  number={10},
  doi = {https://doi.org/10.1119/1.12592},
  pages={925--930},
  year={1981}
}

@article{budroni2014temporal,
  title={Temporal quantum correlations and Leggett–Garg inequalities in multilevel systems},
  author={Budroni, Costantino and Emary, Clive},
  journal={Physical review letters},
  volume={113},
  number={5},
  pages={050401},
  year={2014},
  publisher={APS}
}

@article{knee2012violation,
  title={Violation of a {Leggett–Garg} inequality with ideal non-invasive measurements},
  author={Knee, George C and Simmons, Stephanie and Gauger, Erik M and Morton, John JL and Riemann, Helge and Abrosimov, Nikolai V and Becker, Peter and Pohl, Hans-Joachim and Itoh, Kohei M and Thewalt, Mike LW and others},
  journal={Nature communications},
  volume={3},
  number={1},
  pages={606},
  year={2012},
  doi = {https://doi.org/10.1038/ncomms1614},
  publisher={Nature Publishing Group UK London}
}

@article{palacios-laloy_experimental_2010,
   author = {Agustin Palacios-Laloy and François Mallet and François Nguyen and Patrice Bertet and Denis Vion and Daniel Esteve and Alexander N. Korotkov},
   doi = {10.1038/nphys1641},
   issn = {17452481},
   issue = {6},
   journal = {Nature Physics},
   title = {Experimental violation of a {Bell's} inequality in time with weak measurement},
   volume = {6},
   pages = {10.1038},
   year = {2010}
}

@article{montina_dynamics_2012,
  title = {Dynamics of a Qubit as a Classical Stochastic Process with Time-Correlated Noise: Minimal Measurement Invasiveness},
  author = {Montina, Alberto},
  journal = {Phys. Rev. Lett.},
  volume = {108},
  issue = {16},
  pages = {160501},
  numpages = {5},
  year = {2012},
  month = {Apr},
  publisher = {American Physical Society},
  doi = {10.1103/PhysRevLett.108.160501},
  url = {https://link.aps.org/doi/10.1103/PhysRevLett.108.160501}
}

@article{wilde_could_2009,
author = {Wilde, Mark M.  and McCracken, James M.  and Mizel, Ari },
title = {Could light harvesting complexes exhibit non-classical effects at room temperature?},
journal = {Proceedings of the Royal Society A: Mathematical, Physical and Engineering Sciences},
volume = {466},
number = {2117},
pages = {1347-1363},
year = {2010},
doi = {10.1098/rspa.2009.0575},
URL = {https://royalsocietypublishing.org/doi/abs/10.1098/rspa.2009.0575},
}

@article{athalye_investigation_2011,
  title = {Investigation of the {Leggett–Garg} Inequality for Precessing Nuclear Spins},
  author = {Athalye, Vikram and Roy, Soumya Singha and Mahesh, T. S.},
  journal = {Phys. Rev. Lett.},
  volume = {107},
  issue = {13},
  pages = {130402},
  numpages = {5},
  year = {2011},
  month = {Sep},
  publisher = {American Physical Society},
  doi = {10.1103/PhysRevLett.107.130402},
  url = {https://link.aps.org/doi/10.1103/PhysRevLett.107.130402}
}

@article{PhysRevA.106.012214,
  title = {{Leggett–Garg} inequalities for testing quantumness of gravity},
  author = {Matsumura, Akira and Nambu, Yasusada and Yamamoto, Kazuhiro},
  journal = {Phys. Rev. A},
  volume = {106},
  issue = {1},
  pages = {012214},
  numpages = {11},
  year = {2022},
  month = {Jul},
  publisher = {American Physical Society},
  doi = {10.1103/PhysRevA.106.012214},
  url = {https://link.aps.org/doi/10.1103/PhysRevA.106.012214}
}

@article{kreuzgruber_violation_2024,
  title = {Violation of a {Leggett–Garg} Inequality Using Ideal Negative Measurements in Neutron Interferometry},
  author = {Kreuzgruber, Elisabeth and Wagner, Richard and Geerits, Niels and Lemmel, Hartmut and Sponar, Stephan},
  journal = {Phys. Rev. Lett.},
  volume = {132},
  issue = {26},
  pages = {260201},
  numpages = {6},
  year = {2024},
  month = {Jun},
  publisher = {American Physical Society},
  doi = {10.1103/PhysRevLett.132.260201},
  url = {https://link.aps.org/doi/10.1103/PhysRevLett.132.260201}
}

@article{wilde_addressing_2012,
	title = {Addressing the Clumsiness Loophole in a {Leggett–Garg} Test of Macrorealism},
	volume = {42},
	issn = {1572-9516},
	url = {https://doi.org/10.1007/s10701-011-9598-4},
	doi = {10.1007/s10701-011-9598-4},
	pages = {256--265},
	number = {2},
        year = 2012,
	journal = {Foundations of Physics},
	shortjournal = {Foundations of Physics},
	author = {Wilde, Mark M. and Mizel, Ari},
	date = {2012-02-01},
}

@article{halliwell_comparing_2017,
  title = {Comparing conditions for macrorealism: {Leggett–Garg} inequalities versus no-signaling in time},
  author = {Halliwell, J. J.},
  journal = {Phys. Rev. A},
  volume = {96},
  issue = {1},
  pages = {012121},
  numpages = {11},
  year = {2017},
  month = {Jul},
  publisher = {American Physical Society},
  doi = {10.1103/PhysRevA.96.012121},
  url = {https://link.aps.org/doi/10.1103/PhysRevA.96.012121}
}

@article{halliwell_leggett-garg_2016,
  title = {{Leggett–Garg} correlation functions from a noninvasive velocity measurement continuous in time},
  author = {Halliwell, J. J.},
  journal = {Phys. Rev. A},
  volume = {94},
  issue = {5},
  pages = {052114},
  numpages = {8},
  year = {2016},
  month = {Nov},
  publisher = {American Physical Society},
  doi = {10.1103/PhysRevA.94.052114},
  url = {https://link.aps.org/doi/10.1103/PhysRevA.94.052114}
}

@article{nieto_quantum_1993,
	title = {Quantum phase and quantum phase operators: some physics and some history},
	volume = {1993},
	url = {https://dx.doi.org/10.1088/0031-8949/1993/T48/001},
	doi = {10.1088/0031-8949/1993/T48/001},
	pages = {5},
	issue = {T48},
	journal = {Physica Scripta},
	author = {Nieto, Michael Martin},
	year = {1993},
}

@article{PhysRevA.103.032218,
  title = {{Leggett–Garg} tests for macrorealism: Interference experiments and the simple harmonic oscillator},
  author = {Halliwell, J. J. and Bhatnagar, A. and Ireland, E. and Nadeem, H. and Wimalaweera, V.},
  journal = {Phys. Rev. A},
  volume = {103},
  issue = {3},
  pages = {032218},
  numpages = {11},
  year = {2021},
  month = {Mar},
  publisher = {American Physical Society},
  doi = {10.1103/PhysRevA.103.032218},
  url = {https://link.aps.org/doi/10.1103/PhysRevA.103.032218}
}

@article{halliwell_necessary_2019,
	title = {Necessary and sufficient conditions for macrorealism using two- and three-time {Leggett–Garg} inequalities},
	volume = {1275},
	url = {https://dx.doi.org/10.1088/1742-6596/1275/1/012008},
	doi = {10.1088/1742-6596/1275/1/012008},
	pages = {012008},
	number = {1},
	journal = {Journal of Physics: Conference Series},
	author = {Halliwell, J. J.},
        year = 2019,
	date = {2019-09},
}

@article{PhysRevA.100.042103,
  title = {Fine's theorem for {Leggett–Garg} tests with an arbitrary number of measurement times},
  author = {Halliwell, J. J. and Mawby, C.},
  journal = {Phys. Rev. A},
  volume = {100},
  issue = {4},
  pages = {042103},
  numpages = {12},
  year = {2019},
  month = {Oct},
  publisher = {American Physical Society},
  doi = {10.1103/PhysRevA.100.042103},
  url = {https://link.aps.org/doi/10.1103/PhysRevA.100.042103}
}

@misc{tsirelson_how_2006,
      title={How often is the coordinate of a harmonic oscillator positive?}, 
      author={Boris Tsirelson},
      eprint={quant-ph/0611147},
      archivePrefix={arXiv},
}

@article{plavala_tsirelson_2024,
	title = {Tsirelson inequalities: Detecting cheating and quantumness in a single framework},
	volume = {109},
	url = {https://link.aps.org/doi/10.1103/PhysRevA.109.062216},
	doi = {10.1103/PhysRevA.109.062216},
	pages = {062216},
	number = {6},
	journal = {Phys. Rev. A},
	author = {Plávala, Martin and Heinosaari, Teiko and Nimmrichter, Stefan and Gühne, Otfried},
	date = {2024-06},
        year = 2024,
}

@article{PhysRevA.52.R2497,
  title = {Proposed test for realist theories using {Rydberg} atoms coupled to a high-{$Q$} resonator},
  author = {Huelga, S. F. and Marshall, T. W. and Santos, E.},
  journal = {Phys. Rev. A},
  volume = {52},
  issue = {4},
  pages = {R2497--R2500},
  numpages = {0},
  year = {1995},
  month = {Oct},
  publisher = {American Physical Society},
  doi = {10.1103/PhysRevA.52.R2497},
  url = {https://link.aps.org/doi/10.1103/PhysRevA.52.R2497}
}

@article{halliwell_quasi,
	title = {Leggett–Garg inequalities and no-signaling in time: A quasiprobability approach},
	volume = {93},
	url = {https://link.aps.org/doi/10.1103/PhysRevA.93.022123},
	doi = {10.1103/PhysRevA.93.022123},
	pages = {022123},
	number = {2},
	journal = {Phys. Rev. A},
	author = {Halliwell, J. J.},
	date = {2016-02},
	note = {Publisher: American Physical Society},
}

@article{PhysRevA.103.062212,
  title = {Detecting violations of macrorealism when the original {Leggett–Garg} inequalities are satisfied},
  author = {Majidy, Shayan and Halliwell, Jonathan J. and Laflamme, Raymond},
  journal = {Phys. Rev. A},
  volume = {103},
  issue = {6},
  pages = {062212},
  numpages = {14},
  year = {2021},
  month = {Jun},
  publisher = {American Physical Society},
  doi = {10.1103/PhysRevA.103.062212},
  url = {https://link.aps.org/doi/10.1103/PhysRevA.103.062212}
}

@article{quantum_computation,
  title = {Information-theoretic temporal {Bell} inequality and quantum computation},
  author = {Morikoshi, Fumiaki},
  journal = {Phys. Rev. A},
  volume = {73},
  issue = {5},
  pages = {052308},
  numpages = {5},
  year = {2006},
  month = {May},
  publisher = {American Physical Society},
  doi = {10.1103/PhysRevA.73.052308},
  url = {https://link.aps.org/doi/10.1103/PhysRevA.73.052308}
}

@article{klyshko_bell_1996,
	title = {The {Bell} theorem and the problem of moments},
	volume = {218},
	issn = {0375-9601},
	url = {https://www.sciencedirect.com/science/article/pii/0375960196004446},
	doi = {https://doi.org/10.1016/0375-9601(96)00444-6},
	pages = {119--127},
	number = {3},
	journal = {Physics Letters A},
	author = {Klyshko, D. N.},
        year = 1996,
	date = {1996},
}

@article{PhysRevA.87.022114,
  title = {Negative probabilities, {Fine's} theorem, and linear positivity},
  author = {Halliwell, J. J. and Yearsley, J. M.},
  journal = {Phys. Rev. A},
  volume = {87},
  issue = {2},
  pages = {022114},
  numpages = {8},
  year = {2013},
  month = {Feb},
  publisher = {American Physical Society},
  doi = {10.1103/PhysRevA.87.022114},
  url = {https://link.aps.org/doi/10.1103/PhysRevA.87.022114}
}

@misc{quantum_info_processing,
      title={Temporal {Leggett–Garg-Bell} inequalities for sequential multi-time actions in quantum information processing, and a re-definition of Macroscopic Realism}, 
      author={Marek Zukowski},
      eprint={1009.1749},
      archivePrefix={arXiv}, 
}

@article{crystals,
  title = {Experimental Detection of Quantum Coherent Evolution through the Violation of {Leggett–Garg-type} Inequalities},
  author = {Zhou, Zong-Quan and Huelga, Susana F. and Li, Chuan-Feng and Guo, Guang-Can},
  journal = {Phys. Rev. Lett.},
  volume = {115},
  issue = {11},
  pages = {113002},
  numpages = {5},
  year = {2015},
  month = {Sep},
  publisher = {American Physical Society},
  doi = {10.1103/PhysRevLett.115.113002},
  url = {https://link.aps.org/doi/10.1103/PhysRevLett.115.113002}
}

@article{PhysRevLett.111.090506,
  title = {Partial-Measurement Backaction and Nonclassical Weak Values in a Superconducting Circuit},
  author = {Groen, J. P. and Rist\`e, D. and Tornberg, L. and Cramer, J. and de Groot, P. C. and Picot, T. and Johansson, G. and DiCarlo, L.},
  journal = {Phys. Rev. Lett.},
  volume = {111},
  issue = {9},
  pages = {090506},
  numpages = {5},
  year = {2013},
  month = {Aug},
  publisher = {American Physical Society},
  doi = {10.1103/PhysRevLett.111.090506},
  url = {https://link.aps.org/doi/10.1103/PhysRevLett.111.090506}
}

@article{Souza_2011,
   title={A scattering quantum circuit for measuring {Bell’s} time inequality: a nuclear magnetic resonance demonstration using maximally mixed states},
   volume={13},
   ISSN={1367-2630},
   url={http://dx.doi.org/10.1088/1367-2630/13/5/053023},
   DOI={10.1088/1367-2630/13/5/053023},
   number={5},
   journal={New Journal of Physics},
   publisher={IOP Publishing},
   author={Souza, A M and Oliveira, I S and Sarthour, R S},
   year={2011},
   month=may, pages={053023} }

@article{PhysRevA.87.052102,
  title = {Violation of entropic {Leggett–Garg} inequality in nuclear spins},
  author = {Katiyar, Hemant and Shukla, Abhishek and Rao, K. Rama Koteswara and Mahesh, T. S.},
  journal = {Phys. Rev. A},
  volume = {87},
  issue = {5},
  pages = {052102},
  numpages = {5},
  year = {2013},
  month = {May},
  publisher = {American Physical Society},
  doi = {10.1103/PhysRevA.87.052102},
  url = {https://link.aps.org/doi/10.1103/PhysRevA.87.052102}
}

@article{Goggin_2011,
   title={Violation of the {Leggett–Garg} inequality with weak measurements of photons},
   volume={108},
   ISSN={1091-6490},
   url={http://dx.doi.org/10.1073/pnas.1005774108},
   DOI={10.1073/pnas.1005774108},
   number={4},
   journal={Proceedings of the National Academy of Sciences},
   publisher={Proceedings of the National Academy of Sciences},
   author={Goggin, M. E. and Almeida, M. P. and Barbieri, M. and Lanyon, B. P. and O’Brien, J. L. and White, A. G. and Pryde, G. J.},
   year={2011},
   month=jan, pages={1256–1261} }

@article{Li_2012,
   title={Witnessing Quantum Coherence: from solid-state to biological systems},
   volume={2},
   pages = {885},
   ISSN={2045-2322},
   url={http://dx.doi.org/10.1038/srep00885},
   DOI={10.1038/srep00885},
   number={1},
   journal={Scientific Reports},
   publisher={Springer Science and Business Media LLC},
   author={Li, Che-Ming and Lambert, Neill and Chen, Yueh-Nan and Chen, Guang-Yin and Nori, Franco},
   year={2012},
   month=nov }

@article{PhysRevA.106.032222,
  title = {Detecting quantumness in uniform precessions},
  author = {Zaw, Lin Htoo and Aw, Clive Cenxin and Lasmar, Zakarya and Scarani, Valerio},
  journal = {Phys. Rev. A},
  volume = {106},
  issue = {3},
  pages = {032222},
  numpages = {16},
  year = {2022},
  month = {Sep},
  publisher = {American Physical Society},
  doi = {10.1103/PhysRevA.106.032222},
  url = {https://link.aps.org/doi/10.1103/PhysRevA.106.032222}
}

@article{PhysRevLett.130.160201,
  title = {Dynamics-Based Entanglement Witnesses for Non-Gaussian States of Harmonic Oscillators},
  author = {Jayachandran, Pooja and Zaw, Lin Htoo and Scarani, Valerio},
  journal = {Phys. Rev. Lett.},
  volume = {130},
  issue = {16},
  pages = {160201},
  numpages = {7},
  year = {2023},
  month = {Apr},
  publisher = {American Physical Society},
  doi = {10.1103/PhysRevLett.130.160201},
  url = {https://link.aps.org/doi/10.1103/PhysRevLett.130.160201}
}

@article{PhysRevA.108.022211,
  title = {Dynamics-based quantumness certification of continuous variables using time-independent Hamiltonians with one degree of freedom},
  author = {Zaw, Lin Htoo and Scarani, Valerio},
  journal = {Phys. Rev. A},
  volume = {108},
  issue = {2},
  pages = {022211},
  numpages = {19},
  year = {2023},
  month = {Aug},
  publisher = {American Physical Society},
  doi = {10.1103/PhysRevA.108.022211},
  url = {https://link.aps.org/doi/10.1103/PhysRevA.108.022211}
}

@article{PhysRevA.109.042402,
  title = {Certification of genuine multipartite entanglement in spin ensembles with measurements of total angular momentum},
  author = {Huynh-Vu, Khoi-Nguyen and Zaw, Lin Htoo and Scarani, Valerio},
  journal = {Phys. Rev. A},
  volume = {109},
  issue = {4},
  pages = {042402},
  numpages = {22},
  year = {2024},
  month = {Apr},
  publisher = {American Physical Society},
  doi = {10.1103/PhysRevA.109.042402},
  url = {https://link.aps.org/doi/10.1103/PhysRevA.109.042402}
}

@misc{zaw2024threeanglevariantstsirelsonsprecession,
      title={All three-angle variants of {Tsirelson's} precession protocol, and improved bounds for wedge integrals of {Wigner} functions}, 
      author={Lin Htoo Zaw and Valerio Scarani},
      eprint={2411.03132},
      archivePrefix={arXiv}, 
}

@article{PhysRevA.110.062408,
  title = {Even-parity precession protocol for detecting nonclassicality and entanglement},
  author = {Chen, Jinyan and Tiong, Jackson and Zaw, Lin Htoo and Scarani, Valerio},
  journal = {Phys. Rev. A},
  volume = {110},
  issue = {6},
  pages = {062408},
  numpages = {11},
  year = {2024},
  month = {Dec},
  publisher = {American Physical Society},
  doi = {10.1103/PhysRevA.110.062408},
  url = {https://link.aps.org/doi/10.1103/PhysRevA.110.062408}
}

@article{vaartjes_certifying_2025,
	title = {Certifying the quantumness of a nuclear spin qudit through its uniform precession},
	volume = {1},
        pages = {10.1016},
	issn = {2950-6360},
	url = {https://doi.org/10.1016/j.newton.2025.100017},
	number = {1},
	journal = {Newton},
	author = {Vaartjes, Arjen and Nurizzo, Martin and Zaw, Lin Htoo and Wilhelm, Benjamin and Yu, Xi and Holmes, Danielle and Schwienbacher, Daniel and Kringhøj, Anders and van Blankenstein, Mark R. and Jakob, Alexander M. and Hudson, Fay E. and Itoh, Kohei M. and Murray, Riley J. and Blume-Kohout, Robin and Anand, Namit and Dzurak, Andrew S. and Jamieson, David N. and Scarani, Valerio and Morello, Andrea},
	urldate = {2025-04-18},
	date = {2025-03-03},
        year = 2025,
}

@article{MawbySHO,
  title = {{Leggett–Garg} tests for macrorealism in the quantum harmonic oscillator and more general bound systems},
  author = {Mawby, C. and Halliwell, J. J.},
  journal = {Phys. Rev. A},
  volume = {105},
  issue = {2},
  pages = {022221},
  numpages = {15},
  year = {2022},
  month = {Feb},
  publisher = {American Physical Society},
  doi = {10.1103/PhysRevA.105.022221},
  url = {https://link.aps.org/doi/10.1103/PhysRevA.105.022221}
}

@article{PhysRevA.107.032216,
  title = {{Leggett–Garg} violations for continuous-variable systems with Gaussian states},
  author = {Mawby, C. and Halliwell, J. J.},
  journal = {Phys. Rev. A},
  volume = {107},
  issue = {3},
  pages = {032216},
  numpages = {17},
  year = {2023},
  month = {Mar},
  publisher = {American Physical Society},
  doi = {10.1103/PhysRevA.107.032216},
  url = {https://link.aps.org/doi/10.1103/PhysRevA.107.032216}
}

@article{PhysRev.88.625,
  title = {Diffraction in Time},
  author = {Moshinsky, Marcos},
  journal = {Phys. Rev.},
  volume = {88},
  issue = {3},
  pages = {625--631},
  numpages = {0},
  year = {1952},
  month = {Nov},
  publisher = {American Physical Society},
  doi = {10.1103/PhysRev.88.625},
  url = {https://link.aps.org/doi/10.1103/PhysRev.88.625}
}

@article{moshinsky_diffraction_1976,
	title = {Diffraction in time and the time–energy uncertainty relation},
	volume = {44},
	issn = {0002-9505},
	url = {https://doi.org/10.1119/1.10581},
	doi = {10.1119/1.10581},
	pages = {1037--1042},
	number = {11},
	journal = {American Journal of Physics},
	author = {Moshinsky, M.},
	date = {1976-11},
        year = 1976,
}

@book{Sakurai_Napolitano_2020, place={Cambridge}, edition={3}, title={Modern Quantum Mechanics}, publisher={Cambridge University Press}, author={Sakurai, J. J. and Napolitano, Jim}, year={2020}}

@book{numerical_methods,
author = {Cheney, Ward. and Kincaid, David (David Ronald)},
title = {Numerical mathematics and computing},
year = {2013},
booktitle = {Numerical mathematics and computing},
edition = {7th ed., International ed.},
isbn = {9781133491811},
language = {english},
publisher = {Brooks/Cole Cengage Learning},
}

@article{precessions,
  title = {Detecting quantumness in uniform precessions},
  author = {Zaw, Lin Htoo and Aw, Clive Cenxin and Lasmar, Zakarya and Scarani, Valerio},
  journal = {Phys. Rev. A},
  volume = {106},
  issue = {3},
  pages = {032222},
  numpages = {16},
  year = {2022},
  month = {Sep},
  publisher = {American Physical Society},
  doi = {10.1103/PhysRevA.106.032222},
  url = {https://link.aps.org/doi/10.1103/PhysRevA.106.032222}
}

@article{twirling,
  title = {Group twirling and noise tailoring for multiqubit controlled phase gates},
  author = {Liu, Guoding and Xie, Ziyi and Xu, Zitai and Ma, Xiongfeng},
  journal = {Phys. Rev. Res.},
  volume = {6},
  issue = {4},
  pages = {043221},
  numpages = {38},
  year = {2024},
  month = {Dec},
  publisher = {American Physical Society},
  doi = {10.1103/PhysRevResearch.6.043221},
  url = {https://link.aps.org/doi/10.1103/PhysRevResearch.6.043221}
}

@article{LGdecays,
  title = {{Leggett–Garg} inequalities and decays of unstable systems},
  author = {Giacosa, Francesco and Pagliara, Giuseppe},
  journal = {Phys. Rev. A},
  volume = {104},
  issue = {5},
  pages = {052225},
  numpages = {7},
  year = {2021},
  month = {Nov},
  publisher = {American Physical Society},
  doi = {10.1103/PhysRevA.104.052225},
  url = {https://link.aps.org/doi/10.1103/PhysRevA.104.052225}
}

@article{stationarity2011,
  title = {Violation of a Temporal {Bell} Inequality for Single Spins in a Diamond Defect Center},
  author = {Waldherr, G. and Neumann, P. and Huelga, S. F. and Jelezko, F. and Wrachtrup, J.},
  journal = {Phys. Rev. Lett.},
  volume = {107},
  issue = {9},
  pages = {090401},
  numpages = {4},
  year = {2011},
  month = {Aug},
  publisher = {American Physical Society},
  doi = {10.1103/PhysRevLett.107.090401},
  url = {https://link.aps.org/doi/10.1103/PhysRevLett.107.090401}
}

@article{Wigner_function,
  title = {On the Quantum Correction For Thermodynamic Equilibrium},
  author = {Wigner, E.},
  journal = {Phys. Rev.},
  volume = {40},
  issue = {5},
  pages = {749--759},
  numpages = {0},
  year = {1932},
  month = {Jun},
  publisher = {American Physical Society},
  doi = {10.1103/PhysRev.40.749},
  url = {https://link.aps.org/doi/10.1103/PhysRev.40.749}
}

@article{Arrival_times,
  title = {Arrival times, complex potentials, and decoherent histories},
  author = {Halliwell, J. J. and Yearsley, J. M.},
  journal = {Phys. Rev. A},
  volume = {79},
  issue = {6},
  pages = {062101},
  numpages = {17},
  year = {2009},
  month = {Jun},
  publisher = {American Physical Society},
  doi = {10.1103/PhysRevA.79.062101},
  url = {https://link.aps.org/doi/10.1103/PhysRevA.79.062101}
}

@article{PhysRevA.99.012124,
  title = {Quasiprobability for the arrival-time problem with links to backflow and the {Leggett-Garg} inequalities},
  author = {Halliwell, J. J. and Beck, H. and Lee, B. K. B. and O'Brien, S.},
  journal = {Phys. Rev. A},
  volume = {99},
  issue = {1},
  pages = {012124},
  numpages = {12},
  year = {2019},
  month = {Jan},
  publisher = {American Physical Society},
  doi = {10.1103/PhysRevA.99.012124},
  url = {https://link.aps.org/doi/10.1103/PhysRevA.99.012124}
}

@article{halliwell_quantum_2013,
	title = {Quantum backflow states from eigenstates of the regularized current operator},
	volume = {46},
	url = {https://dx.doi.org/10.1088/1751-8113/46/47/475303},
	doi = {10.1088/1751-8113/46/47/475303},
	pages = {475303},
	number = {47},
	journal = {Journal of Physics A: Mathematical and Theoretical},
	author = {Halliwell, J J and Gillman, E and Lennon, O and Patel, M and Ramirez, I},
	date = {2013-11},
        year = 2013,
}

@article{PhysRevLett.120.210402,
  title = {Nonclassicality of the Harmonic-Oscillator Coherent State Persisting up to the Macroscopic Domain},
  author = {Bose, S. and Home, D. and Mal, S.},
  journal = {Phys. Rev. Lett.},
  volume = {120},
  issue = {21},
  pages = {210402},
  numpages = {6},
  year = {2018},
  month = {May},
  publisher = {American Physical Society},
  doi = {10.1103/PhysRevLett.120.210402},
  url = {https://link.aps.org/doi/10.1103/PhysRevLett.120.210402}
}

@book{zachos2005quantum,
  author    = {C. K. Zachos and D. B. Fairlie and T. L. Curtright},
  title     = {Quantum Mechanics in Phase Space: An Overview with Selected Papers},
  publisher = {World Scientific},
  year      = {2005},
  address   = {Singapore},
  isbn      = {9789812383846}
}

@article{PhysRevLett.134.190201,
  title = {Tsirelson's Inequality for the Precession Protocol Is Maximally Violated by Quantum Theory},
  author = {Zaw, Lin Htoo and Weilenmann, Mirjam and Scarani, Valerio},
  journal = {Phys. Rev. Lett.},
  volume = {134},
  issue = {19},
  pages = {190201},
  numpages = {8},
  year = {2025},
  month = {May},
  publisher = {American Physical Society},
  doi = {10.1103/PhysRevLett.134.190201},
  url = {https://link.aps.org/doi/10.1103/PhysRevLett.134.190201}
}

@book{groupC3,
author = {Joshi, A. W.},
title = {Elements of group theory for physicists },
year = {1985 - 1982},
address = {New Delhi},
booktitle = {Elements of group theory for physicists},
edition = {3rd ed},
isbn = {0852264488},
language = {english},
publisher = {Wiley Eastern},
}

@article{hermens_constraints_2018,
	title = {Constraints on macroscopic realism without assuming non-invasive measurability},
	volume = {63},
	issn = {1355-2198},
	url = {https://www.sciencedirect.com/science/article/pii/S1355219817300990},
	doi = {https://doi.org/10.1016/j.shpsb.2017.11.003},
	pages = {50--64},
	journal = {Studies in History and Philosophy of Science Part B: Studies in History and Philosophy of Modern Physics},
	author = {Hermens, R. and Maroney, O. J. E.},
	year = {2018},
}

@article{trillo_quantum_advantage,
  title={Quantum advantages for transportation tasks – projectiles, rockets, and quantum backflow},
  author={Trillo, David and Le, Thinh P and Navascu{\'e}s, Miguel},
  journal={npj Quantum Information},
  volume={9},
  number={1},
  pages={69},
  year={2023},
  doi = {https://doi.org/10.1038/s41534-023-00739-z},
  publisher={Nature Publishing Group UK London}
}

@article{bracken_probability_1994,
	title = {Probability backflow and a new dimensionless quantum number},
	volume = {27},
	url = {https://dx.doi.org/10.1088/0305-4470/27/6/040},
	doi = {10.1088/0305-4470/27/6/040},
	pages = {2197},
	number = {6},
	journal = {Journal of Physics A: Mathematical and General},
	author = {Bracken, A. J. and Melloy, G. F.},
	date = {1994-03},
        year = 1994,
}

@misc{mawbythesis,
      title={Tests of Macrorealism in Discrete and Continuous Variable Systems}, 
      author={Clement Mawby},
      year={2024},
      eprint={2402.16537},
      archivePrefix={arXiv},
}

@book{muga2008time,
  title     = {Time in Quantum Mechanics},
  editor    = {J. G. Muga and R. Sala Mayato and I. L. Egusquiza},
  series    = {Lecture Notes in Physics},
  volume    = {734},
  publisher = {Springer},
  year      = {2008},
  address   = {Berlin},
  isbn      = {978-3-540-73472-7},
  doi       = {10.1007/978-3-540-73473-4}
}

@incollection{pauli1958handbuch,
  author    = {W. Pauli},
  title     = {General Principles of Quantum Mechanics},
  booktitle = {Handbuch der Physik},
  volume    = {5},
  pages     = {1--168},
  publisher = {Springer},
  year      = {1958},
  address   = {Berlin},
  note      = {{Pauli's} objection to a time operator appears in a footnote},
}

@article{Hillery1984,
  author       = {M. Hillery and R. F. O’Connell and M. O. Scully and E. P. Wigner},
  title        = {Distribution functions in physics: Fundamentals},
  journal      = {Physics Reports},
  volume       = {106},
  pages        = {121--167},
  year         = {1984},
  doi          = {10.1016/0370-1573(84)90160-1}
}

@article{Tatarskii1983,
  author       = {V. I. Tatarskii},
  title        = {The {Wigner} representation of quantum mechanics},
  journal      = {Soviet Physics Uspekhi},
  volume       = {26},
  number       = {4},
  pages        = {311--327},
  year         = {1983},
  doi          = {10.1070/PU1983v026n04ABEH004354}
}

@article{Case2008,
  author       = {W. B. Case},
  title        = {Wigner functions and {Weyl} transforms for pedestrians},
  journal      = {American Journal of Physics},
  volume       = {76},
  number       = {10},
  pages        = {937--946},
  year         = {2008},
  doi          = {10.1119/1.2957889}
}

@article{PhysRevA.99.022119,
  title = {{Leggett-Garg} tests of macrorealism: Checks for noninvasiveness and generalizations to higher-order correlators},
  author = {Halliwell, J. J.},
  journal = {Phys. Rev. A},
  volume = {99},
  issue = {2},
  pages = {022119},
  numpages = {8},
  year = {2019},
  month = {Feb},
  publisher = {American Physical Society},
  doi = {10.1103/PhysRevA.99.022119},
  url = {https://link.aps.org/doi/10.1103/PhysRevA.99.022119}
}

@article{Ourjoumtsev2007,
  author    = {Ourjoumtsev, Alexei and Jeong, Hyunseok and Tualle-Brouri, Rosa and Grangier, Philippe},
  title     = {Generation of optical 'Schr{\"o}dinger cats' from photon number states},
  journal   = {Nature},
  year      = {2007},
  volume    = {448},
  number    = {7155},
  pages     = {784--786},
  doi       = {10.1038/nature06054}
}

@incollection{Munoz2009DwellTime,
  author       = {Muñoz, José and Egusquiza, Iñigo L. and del Campo, Adolfo and Seidel, Dirk and Muga, J. Gonzalo},
  title        = {Dwell-Time Distributions in Quantum Mechanics},
  booktitle    = {Time in Quantum Mechanics -- Vol.\ 2},
  editor       = {Muga, J. Gonzalo and Ruschhaupt, A. and del Campo, Adolfo},
  series       = {Lecture Notes in Physics},
  volume       = {789},
  pages        = {97--125},
  year         = {2009},
  publisher    = {Springer, Berlin, Heidelberg},
  doi          = {10.1007/978-3-642-03174-8_5},
}

@article{Omnes1997,
  author       = {Roland Omnès},
  title        = {Quantum-classical correspondence using projection operators},
  journal      = {Journal of Mathematical Physics},
  volume       = {38},
  number       = {2},
  pages        = {697--707},
  year         = {1997},
  doi          = {10.1063/1.531847}
}

@article{PhysRevA.100.042325,
  title = {Exploration of an augmented set of {Leggett-Garg} inequalities using a noninvasive continuous-in-time velocity measurement},
  author = {Majidy, Shayan-Shawn and Katiyar, Hemant and Anikeeva, Galit and Halliwell, Jonathan and Laflamme, Raymond},
  journal = {Phys. Rev. A},
  volume = {100},
  issue = {4},
  pages = {042325},
  numpages = {17},
  year = {2019},
  month = {Oct},
  publisher = {American Physical Society},
  doi = {10.1103/PhysRevA.100.042325},
  url = {https://link.aps.org/doi/10.1103/PhysRevA.100.042325}
}

\end{document}